%% file: ANA-JETM-2018-02-PAPER.tex
\newcommand*{\ATLASLATEXPATH}{}
\documentclass[PAPER, coverpage=false, cernpreprint=true, texlive=2016, subfigure, UKenglish]{\ATLASLATEXPATH atlasdoc}

\usepackage{\ATLASLATEXPATH atlaspackage}
\usepackage{\ATLASLATEXPATH atlasbiblatex}
 
\usepackage{\ATLASLATEXPATH atlascontribute}
 
\usepackage{\ATLASLATEXPATH atlasphysics}
 
\graphicspath{{logos/}{figures/}}
 
\usepackage{ANA-JETM-2018-02-PAPER-defs}
\usepackage{atlasjetetmiss}
 
\usepackage[percent]{overpic}
\usepackage{epstopdf}
\usepackage{multirow}
 
\renewcommand*{\insitu}{\text{\emph{in~situ}}\xspace}
\renewcommand*{\Insitu}{\text{\emph{In~situ}}\xspace}

\newcommand*{\smallR}{small-\ensuremath{R}\xspace}

\AtlasTitle{\emph{In situ} calibration of large-radius jet energy and mass in 13~TeV~proton--proton collisions with the ATLAS detector}

\AtlasJournalRef{Eur. Phys. J. C 79 (2019) 135}
\AtlasDOI{DOI:10.1140/epjc/s10052-019-6632-8}

\AtlasAbstract{

The response of the ATLAS detector to large-radius jets is measured \emph{in situ} using 36.2~fb$^{-1}$~of $\sqrt{s}=13$~TeV~proton--proton collisions provided by the LHC and recorded by the ATLAS experiment during 2015 and 2016. The jet energy scale is measured in events where the jet recoils against a reference object, which can be either a calibrated photon, a reconstructed $Z$ boson, or a system of well-measured small-radius jets. The jet energy resolution and a calibration of forward jets are derived using dijet balance measurements. The jet mass response is measured with two methods: using mass peaks formed by $W$ bosons and top quarks with large transverse momenta and by comparing the jet mass measured using the energy deposited in the calorimeter with that using the momenta of charged-particle tracks. The transverse momentum and mass responses in simulations are found to be about 2--3\% higher than in data. This difference is adjusted for with a correction factor. The results of the different methods are combined to yield a calibration over a large range of transverse momenta ($p_{\mathrm{T}}$). The precision of the relative jet energy scale is 1--2\% for 200~GeV~<~$p_{\mathrm{T}}$~<~2~TeV, while that of the mass scale is 2--10\%. The ratio of the energy resolutions in data and simulation is measured to a precision of 10--15\% over the same $p_{\mathrm{T}}$ range.
}
 
\author{The ATLAS Collaboration}
 
\AtlasRefCode{JETM-2018-02}
 
\PreprintIdNumber{CERN-EP-2018-191}
\hypersetup{pdftitle={CERN-EP-2018-191},pdfauthor={The ATLAS Collaboration}}
 
\begin{document}
 
\maketitle

\section{Introduction} 
\begin{figure}[b]\centering
\includegraphics[width=0.95\textwidth]{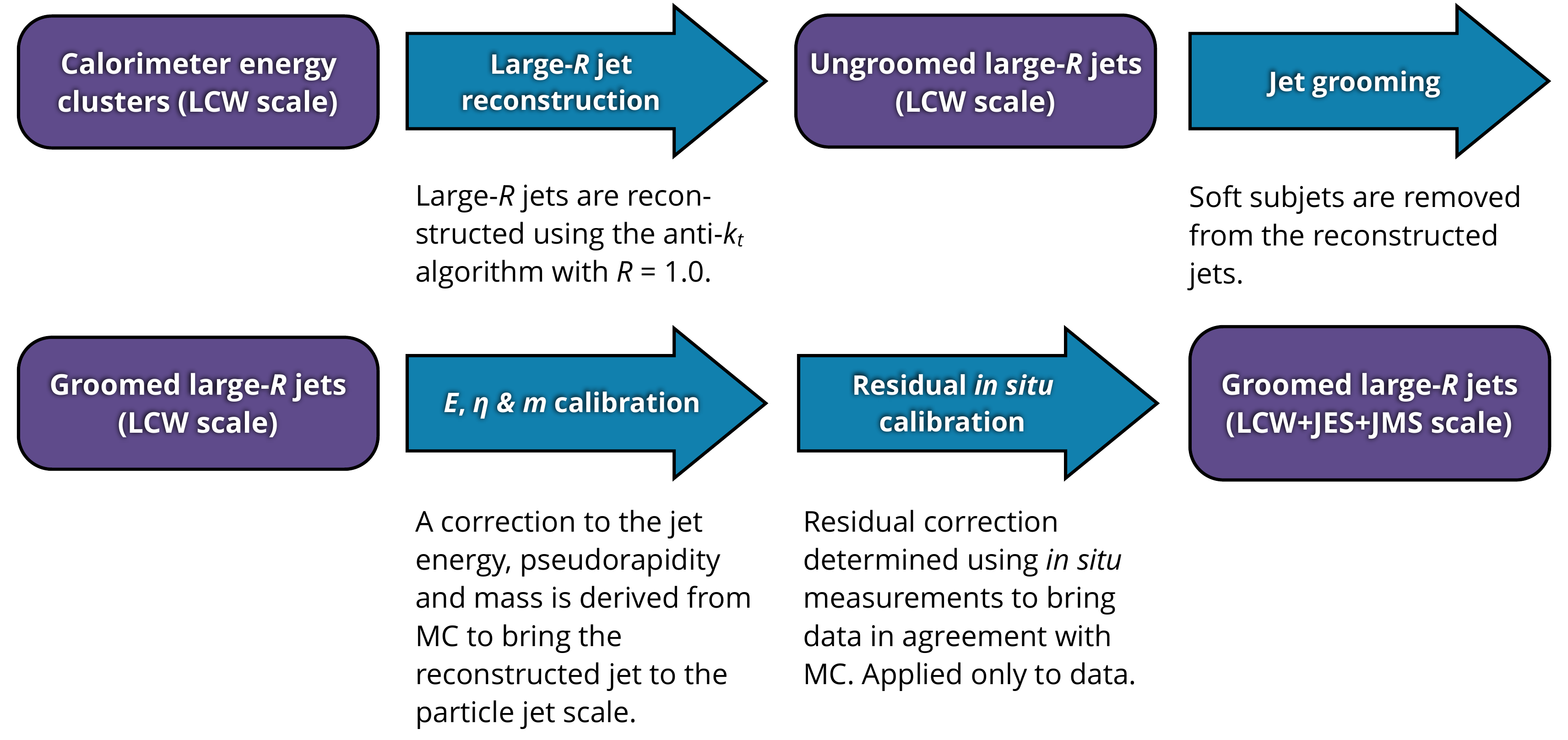}
\caption{Overview of the \largeR jet reconstruction and calibration procedure described in this paper. The calorimeter energy clusters from which jets are reconstructed have already been adjusted to point at the event's primary hard-scatter vertex.}\label{fig:flowchart}
\end{figure}
 
Signatures with high transverse momentum, $p_{\mathrm{T}}$, massive particles such as Higgs bosons, top quarks, and $W$ or $Z$ bosons have become ubiquitous during Run 2 of the Large Hadron Collider (LHC). These particles most often decay hadronically. Due to their large transverse momentum, the decay products become collimated and may be reconstructed as a single jet with large radius parameter $R$~\cite{Larkoski:2017jix,Asquith:2018igt} (a `\largeR' jet). The sensitivity of searches and measurements that use \largeR jets depends on an accurate knowledge of the transverse momentum \pt\ and mass $m$ responses of the detector~\cite{PERF-2012-02}. A calibration of the \largeR energy and mass scales derived using Monte Carlo simulation yields uncertainties as large as 10\%. The calibration described in this paper results in a reduction of these uncertainties by more than a factor of three.
 
In this paper, a suite of \insitu calibration techniques is described which measure the response in proton--proton ($pp$) collision data at $\sqrt{s}=$ 13~\tev. The results of several methods are combined to provide a calibration that defines the nominal \largeR jet energy scale (JES) and the jet mass scale (JMS). These measurements provide a significant increase in the precision with which the \largeR jet \pt and mass scales are known across most of the kinematically accessible phase space. The jet energy and mass resolutions (JER, JMR) are also measured \insitu and compared with the predictions of Monte Carlo simulations (MC). Additional uncertainties on jet substructure observables used to identify boosted objects are derived from data in Ref.~\cite{JETM-2018-03}.
 
Jet reconstruction starts with clusters of topologically connected calorimeter cell signals. These topological clusters, or `topo-clusters', are brought to the hadronic scale using the local hadronic cell weighting scheme (LCW)~\cite{PERF-2014-07}. \LargeR jets are reconstructed with the anti-$k_{t}$ algorithm~\cite{Cacciari:2008gp} using a radius parameter $R = 1.0$. The jets are groomed with the `trimming' algorithm of Ref.~\cite{Krohn:2009th}, which removes regions of the jet with a small relative contribution to the jet transverse momentum. This procedure reduces the impact from additional $pp$ interactions in the event and from the underlying event, improving the energy and mass resolution.
 
The several stages of the ATLAS \largeR jet calibration procedure are illustrated in Figure~\ref{fig:flowchart}. The trimmed \largeR jets are calibrated to the energy scale of stable final-state particles using corrections based on simulations. This jet-level correction is referred to as the simulation-based calibration and includes a correction to the jet mass~\cite{STDM-2011-19}. Finally, the jets are calibrated \insitu using response measurements in $pp$ collision data. A correction based on a statistical combination of data-to-simulation ratios of these response measurements is applied only to data and adjusts for the residual (typically 2--3\%) mismodelling of the response. Uncertainties in the JES and JMS are derived by propagating uncertainties from the individual \insitu response measurements through the statistical combination.

\begin{figure}[h!]\centering
\subfigure[dijet event]{\includegraphics[width=0.25\textwidth,page=2]{fig_02a.pdf}}
\hfill
\subfigure[$Z$+jet or $\gamma$+jet event]{\includegraphics[width=0.30\textwidth,page=1]{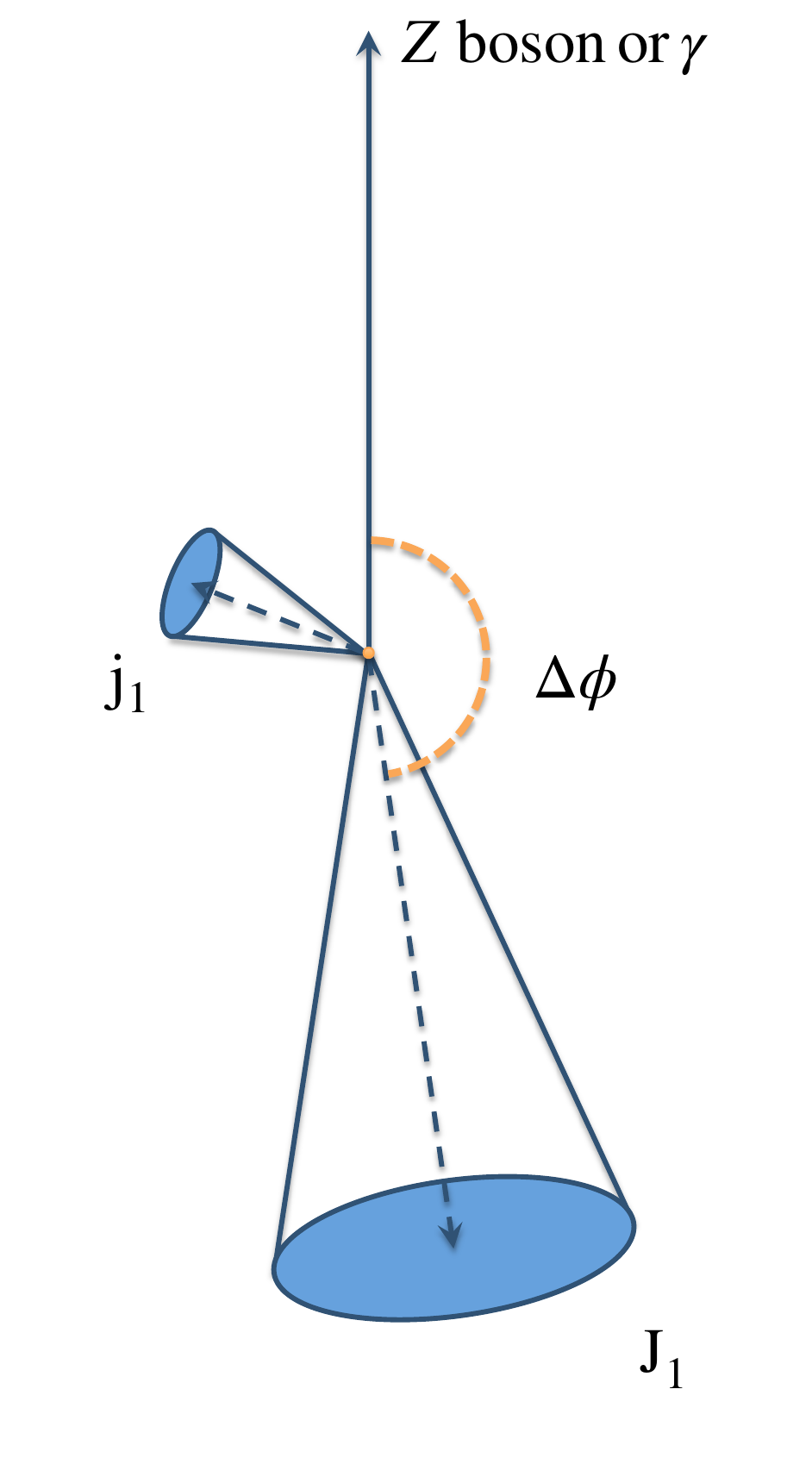}}
\hfill
\subfigure[multijet event]{\includegraphics[width=0.24\textwidth,page=3]{fig_02a.pdf}}
\caption{Schematic representation of the events used to measure the JES and JER: (a) a dijet event, (b) a $Z$+jet or $\gamma$+jet event and (c) a multijet event with several jets recoiling against the leading (\largeR) jet. The labels J$_{i}$ refer to the $i$th leading \largeR jet, while j$_i$ refers to the $i$th leading small-$R$ jet that fulfils $\DeltaR(\mathrm{J}_1,\mathrm{j}) > 1.4$. $\Delta\phi$ is the difference between the azimuthal angle of the jet and the reference object, while $\Delta\alpha$ is
the difference between the azimuthal angle of the jet and the vectorial sum of the recoil system momenta.}
\label{fig:balance_fig}
\end{figure}

The \insitu calibration is determined in two separate steps. In the first step, the JES is measured with the same methods  used to calibrate \smallR jets~\cite{PERF-2016-04}. These techniques rely on the transverse momentum balance in a variety of final states, illustrated in Figure~\ref{fig:balance_fig}. The JES correction factor is a product of two terms. The absolute calibration is derived from a statistical combination of three measurements from $Z$+jet, $\gamma$+jet, and multijet events in the central region of the detector. A relative intercalibration, derived using dijet events, propagates the well-measured central JES into the forward region of the detector. The \insitu calibration accounts for detector effects which are not captured by simulation. The JES correction is applied as a four-momentum scale factor to jets in data; therefore, it also affects the jet mass calibration.
 
In the second step of the \insitu calibration, the jet mass response is measured using two methods following the application of the \insitu JES correction. The mass response is measured in lepton+jets top quark pair production (\ttbar production)~\cite{ATLAS-CONF-2016-008} with a fit to the peaks in the jet mass distribution formed by high-\pt $W$ bosons and top quarks decaying into fully hadronic final states. 
A second measurement is performed with the $R_{\mathrm{trk}}$ method~\cite{PERF-2012-02}, which takes advantage of the independent measurements by the calorimeter and the inner tracker. This method provides a calibration for the calorimeter jet mass measurement over a broad \pt range. The results from the two methods are combined as a smooth function of \pt in two mass bins, which could be applied to data as an \insitu correction as outlined in Section~\ref{sec:combination}.
 
The JER and JMR are also measured \insitu and compared with the prediction of the simulation. The dijet balance method takes advantage of the transverse momentum balance in dijet events to extract the JER. The JMR is obtained from fits to the top quark and $W$ boson mass peaks in high-\pt lepton+jets $t\bar{t}$ events.
 
Sections~\ref{sec:detector} and~\ref{sec:simulation} provide overviews of the ATLAS detector, the data set studied, and the simulations used in this paper. Section~\ref{sec:objects} describes the reconstruction of  \largeR jets in ATLAS. The following section presents the results of the balance methods that measure the jet energy scale: the intercalibration, which uses dijet events to ensure a uniform response over the central and forward regions of the detector in Section~\ref{sec:etacal}, the $Z$+jet balance method in Section~\ref{sec:zjetbalance}, the $\gamma$+jet balance method in Section~\ref{sec:gjetbalance}, and the multijet balance method in Section~\ref{sec:mjbalance}. Section~\ref{sec:insitumass} presents the methods that are used to measure the jet mass response: the $R_{\mathrm{trk}}$ method and its results for the energy and mass scale in Section~\ref{sec:rtrk} and the fits to the $W$ boson and top quark mass peaks in high-\pt lepton+jets \ttbar events in Section~\ref{sec:forwardfolding}, which are also used to measure the JMR. The measurement of the JER in dijet events is discussed in Section~\ref{sec:djjer}. The methodology of the combination procedure is presented in Section~\ref{sec:combination}, as well as the resultant combined \insitu calibration of the JES and JMS. Section~\ref{sec:conclusion} summarizes the results.
 
\section{The ATLAS detector and data set} \label{sec:detector}
The ATLAS experiment consists of three major sub-detectors: the inner detector, the calorimeters, and the muon spectrometer. The inner detector, closest to the interaction point, is used to track charged particles in a 2~T axial magnetic field produced by a thin superconducting solenoid. It consists of a pixel detector, a silicon tracker equipped with micro-strip detectors, and a transition radiation tracker that provides a large number of space points in the outermost layers of the tracker. It covers the pseudorapidity\footnote{The ATLAS reference system is a Cartesian right-handed coordinate system, with the nominal collision point at the origin. The anticlockwise beam direction defines the positive $z$-axis, while the positive $x$-axis is defined as pointing from the collision point to the centre of the LHC ring and the positive $y$-axis points upwards. The azimuthal angle $\phi$ is measured around the beam axis, and the polar angle $\theta$ is measured relative to the $z$-axis. Pseudorapidity is defined as $\eta = - \ln[\tan(\theta/2)]$, and transverse energy is defined as $E_{\mathrm{T}} = E \sin \theta$.} range $|\eta| <$ 2.5. Surrounding the tracker and solenoid, a sampling calorimeter measures the energy of particles produced in the collisions with $|\eta| <$ 4.9. The energies of electrons and photons are measured precisely in a high-granularity liquid-argon electromagnetic calorimeter. The cylindrical ``barrel'' covers $|\eta|<1.475$, and the ``endcaps'' on either end of the detector cover $1.375 < |\eta|<3.2$. An iron/scintillator tile calorimeter measures the energy of hadrons in  the  central rapidity range, $|\eta|<1.7$, and a liquid-argon hadronic endcap calorimeter provides coverage for $1.5<|\eta|<3.2$. The forward liquid-argon calorimeter measures electrons, photons, and hadrons for $3.2 < |\eta|< 4.9$. Finally, a muon spectrometer in the magnetic field of a system of superconducting air-core toroid magnets identifies muons in the range $|\eta| < 2.7$ and measures their transverse momenta. The ATLAS trigger system consists of a hardware-based first-level trigger
followed by a software-based high-level trigger, which apply a real-time selection to reduce the up to 40~MHz LHC collision rate to an average rate of events written to storage of 1~kHz~\cite{TRIG-2016-01}. A detailed description of the ATLAS experiment is given in Ref.~\cite{PERF-2007-01}.
 
The data set used in this analysis consists of $pp$ collisions delivered
by the LHC at a centre-of-mass energy of $\sqrt{s}=$ 13~\TeV~during 2015 and 2016. The specific trigger requirements vary among the various \insitu analyses and are described in the relevant sections.
All data are required to meet ATLAS standard quality criteria. Data taken during periods in which detector subsystems were not fully
functional are discarded. Data quality criteria also reject events that have significant contamination
from detector noise or with issues in the read-out. The remaining data correspond to an integrated
luminosity of $36.2~\ifb$.
 
Due to the high luminosity of the LHC, multiple $pp$ collisions occur during
each bunch crossing. Interactions which occur within the bunch crossing of
interest (in-time pile-up) or in neighbouring bunch crossings (out-of-time
pile-up) may alter the measured energy or mass scale of jets or lead to the
reconstruction of additional `stochastic' jets, seeded by upwards
fluctuations in the local pile-up energy density. The average number of additional
$pp$ collisions per bunch crossing is 24 in the Run 2 data
from 2015 and 2016 analysed here.
 
\section{Simulations} \label{sec:simulation}
The data are compared with detailed simulations of the ATLAS detector response~\cite{SOFT-2010-01}
based on the \texttt{Geant4}~\cite{Agostinelli:2002hh} toolkit. Hard-scatter events for all processes studied were simulated with several different event generators to assess possible systematic effects due to limitations in the physics modelling.
Several different simulation packages were also used to hadronize final-state quarks and gluons in order to compare the impact of various models of hadronization and parton showering on the measurements.
 
Dijet events were generated using several different generator configurations. Depending on the analysis, nominal dijet samples were generated using either \PYTHIA8 (v8.186)~\cite{Sjostrand:2014zea} or \POWHEGBOX 2.0~\cite{Alioli:2010xd,Frixione:2007vw,Nason:2004rx} interfaced with \PYTHIA8. These samples were generated with the A14 set of tuned  parameters~\cite{ATL-PHYS-PUB-2014-021} and the NNPDF2.3 LO parton distribution function (PDF) set~\cite{Ball:2014uwa}. Samples generated with \HERWIG7~\cite{Bahr:2008pv} and \SHERPA v2.1~\cite{Gleisberg:2008ta} were used for comparison. The \HERWIG7 sample used the UE-EE-5 set of tuned parameters~\cite{Gieseke:2012ft} and CTEQ6L1 PDF set~\cite{Pumplin:2002vw}. The \SHERPA leading-order multileg generator includes $2\rightarrow2$ and $2\rightarrow3$ processes at matrix element level, combined using the CKKW prescription~\cite{Catani:2001ccx}.

$Z$+jets events are generated using \POWHEGBOX 2.0 interfaced to the Pythia 8.186 parton shower model. The CT10 PDF set is used in the matrix element~\cite{Lai:2010vv}. The AZNLO set of tuned parameters~\cite{STDM-2012-23} is used, with PDF set CTEQ6L1, for the modelling of non-perturbative effects. The EvtGen 1.2.0 program~\cite{EvtGen} is used for the properties of $b$- and $c$-hadron decays. Photos++ 3.52~\cite{Davidson:2010ew} is used for QED emissions from electroweak vertices and charged leptons. Samples of $Z$+jet events are compared to a second sample generated using \SHERPA 2.2.1. Matrix elements are calculated for up to 2 partons at NLO and 4 partons at LO using Comix~\cite{Gleisberg:2008fv} and OpenLoops~\cite{Cascioli:2011va} and merged with the \SHERPA parton shower~\cite{Schumann:2007mg} according to the ME+PS@NLO prescription~\cite{Hoeche:2012yf}. The NNPDF30nnlo PDF set is used in conjunction with dedicated parton shower tuning developed by the \SHERPA authors. $\gamma$+jets events are compared to a sample generated with the \SHERPA 2.1.1 event generator. Matrix elements are calculated with up to 3 or 4 partons at LO and merged with the \SHERPA parton shower according to the ME+PS@LO prescription. The CT10 PDF set is used in conjunction with dedicated parton shower tuning developed by the \SHERPA authors.
$Z$+jets events are generated using \POWHEGBOX 2.0 interfaced to the Pythia 8.186 parton shower model. The CT10 PDF set is used in the matrix element~\cite{Lai:2010vv}. The AZNLO set of tuned parameters~\cite{STDM-2012-23} is used, with PDF set CTEQ6L1, for the modelling of non-perturbative effects. The EvtGen 1.2.0 program~\cite{EvtGen} is used for the properties of $b$- and $c$-hadron decays. Photos++ 3.52~\cite{Davidson:2010ew} is used for QED emissions from electroweak vertices and charged leptons. Samples of $Z$+jet events are compared to a second sample generated using \SHERPA 2.2.1. Matrix elements are calculated for up to 2 partons at NLO and 4 partons at LO using Comix~\cite{Gleisberg:2008fv} and OpenLoops~\cite{Cascioli:2011va} and merged with the \SHERPA parton shower~\cite{Schumann:2007mg} according to the ME+PS@NLO prescription~\cite{Hoeche:2012yf}. The NNPDF30nnlo PDF set is used in conjunction with dedicated parton shower tuning developed by the \SHERPA authors. $\gamma$+jets events are compared to a sample generated with the \SHERPA 2.1.1 event generator. Matrix elements are calculated with up to 3 or 4 partons at LO and merged with the \SHERPA parton shower according to the ME+PS@LO prescription. The CT10 PDF set is used in conjunction with dedicated parton shower tuning developed by the \SHERPA authors.
 
For $\gamma$+jet events, \PYTHIA 8 was used as the nominal generator, where the $2\rightarrow2$ matrix element is convolved with the NNPDF2.3LO PDF set. The A14 event tune was used. These events are compared to a sample generated with \SHERPA v2.1.1, which includes up to four jets in the matrix element. These events were generated using the default \SHERPA tune and the CT10 PDF set.
 
Top quark pair production and single top production in the $s$-channel and $Wt$ final state were simulated at NLO accuracy with \POWHEGBOX v2~\cite{Frixione:2007nw} and the CT10 PDF set. For electroweak $t$-channel single top quark production, \POWHEGBOX v1 was used, which utilizes the four-flavour scheme for NLO matrix element calculations together with the fixed four-flavour PDF set CT10f4. In all cases, the nominal sample was interfaced with \PYTHIA 8 with the CTEQ6L1 PDF set, which simulates the parton shower, fragmentation, and underlying event. The $h_\text{damp}$ parameter in \POWHEG, which regulates the \pt of the first additional emission beyond the Born level and thus the \pt of the recoil emission against the \ttbar system, was set to the mass of the top quark ($172.5$~\GeV). Systematic uncertainties in the modelling of hadronization were evaluated using a \POWHEG sample interfaced to \HERWIG 7. $W$+jet events, simulated in \SHERPA v2.2.0, are considered as a background to \ttbar production.    
 
The effect of pile-up on reconstructed jets was modelled by overlaying multiple simulated minimum-bias inelastic $pp$ events on the signal event. These additional events were generated with \PYTHIA 8, using the A2 set of tuned parameters~\cite{ATL-PHYS-PUB-2012-003} and MSTW2008LO PDF set~\cite{Martin:2009iq}. The distribution of the average number of interactions per bunch crossing in simulated samples is reweighted to match that of the analyzed dataset.
 
 
\section{\LargeR jet reconstruction and simulation calibration} \label{sec:objects}
This section describes the reconstruction of \largeR jets and the grooming procedure. Three classes of jets are used: calorimeter jets, particle-level (or `truth') jets, and track jets. The \largeR jets considered in this paper are reconstructed using the \antikt algorithm~\cite{Cacciari:2008gp} with a radius parameter $R = 1.0$. For balancing and veto purposes, jets reconstructed with radius parameter
$R = 0.4$ (`\smallR jets') are used in some parts of the analysis with their own calibration procedures applied~\cite{PERF-2016-04}.
The specific implementation of the jet clustering algorithm used is taken from the \textsc{FastJet} package~\cite{fastjet,Cacciari:2005hq}.
 
\subsection{\LargeR jets}
 
Calorimeter jets are formed from topological clusters of calorimeter cells. The clusters are seeded by cells with an energy significantly above the calorimeter noise. The \largeR jets used in this paper are reconstructed using topological clusters that are calibrated to correct for response differences between energy deposition from electromagnetic particles (electrons and photons) and hadrons with the LCW scheme of Ref.~\cite{PERF-2014-07}. Small-$R$ jets reconstructed from ``electromagnetic scale'' topo-clusters are used as a reference system in the multijet balance method of Section~\ref{sec:mjbalance}. Results are labelled with ``LCW'' or ``EM'' to indicate the calibration of the clusters. Topological clusters are defined to be massless. The four-momenta of these topo-clusters, initially defined as pointing to the geometrical centre of the ATLAS detector, are adjusted to point towards the hard-scatter primary vertex of the event, which is defined as the primary vertex with the largest associated sum of track $\pt^2$.
 
To reduce the effects of pile-up, soft emissions, and the underlying event on jet substructure measurement, the trimming algorithm is applied to the jets. Trimming reclusters the jet constituents of each $R = 1.0$ jet using the \kt algorithm~\cite{Ellis:1993tq} and $R_{\mathrm{sub}}=$ 0.2, producing a collection of subjets for each jet. Subjets with $\pt^{\mathrm{subjet}} / \pt^{\mathrm{jet}} < 0.05$ are removed, and the jet four-momentum is recalculated from the remaining constituents.
 
In this paper, trimmed \largeR jets with \pt{} $>$ 200~\gev{} and $|\eta| < 2.5$ are studied.
 
\subsection{Particle-level jets and the simulation-based jet calibration}
 
The reference for the simulation-based jet calibration is formed by particle-level jets. These are created by clustering stable particles originating from the hard-scatter interaction in the simulation event record which have a lifetime $\tau$ in the laboratory frame such that $c\tau > 10$~mm. Particles that do not leave significant energy deposition in the calorimeter (i.e.\ muons and neutrinos) are excluded. Particle-level jets are reconstructed and trimmed using the same algorithms as those applied to \largeR jets built from topological clusters, incorporating the grooming procedure within the jet definition.
 
After reconstruction of the calorimeter jets, a correction derived from a sample of simulated
dijet events is applied to restore the average reconstructed calorimeter jet energy scale to that of
particle-level jets. A correction is also applied to the $\eta$ of the reconstructed jet to correct for a bias
relative to particle-level jets in certain regions of the  detector~\cite{ATLAS-CONF-2015-037}. Both corrections
are applied as a function of the reconstructed jet energy and the detector pseudorapidity, $\eta_{\mathrm{det}}$,
defined as the pseudorapidity calculated relative to the geometrical centre of the ATLAS detector. This yields a
better location of the energy-weighted centroid of the jet than the use of the pseudorapidity calculated relative
to the hard-scatter primary vertex.
 
Reconstructed jets are matched to particle-level jets using an angular matching procedure that minimizes the distance $\Delta R = \sqrt{(\Delta \phi)^2 + (\Delta \eta)^2}$. The energy response is defined as $E_{\mathrm{reco}}/E_{\mathrm{truth}}$, where $E_{\mathrm{reco}}$ is the reconstructed jet energy prior to any calibration (later denoted $E_0$) and $E_{\mathrm{truth}}$ is the energy of the corresponding particle-level jets. The mass response is defined as $m_{\mathrm{reco}}/m_{\mathrm{truth}}$, where $m_{\mathrm{reco}}$ and $m_{\mathrm{truth}}$ represent the jet mass of the matched detector-level and particle-level jets, respectively. The average response is determined in a Gaussian fit to the core of the response distribution. The parameterization of the average jet energy response $R_E = \langle E_\text{reco}/E_\text{truth} \rangle$ used for the simulation calibration is presented as a function of \etadet and for several values of the truth jet energy in Figure~\ref{fig:mccalibration_energy_mass}(a). The correction is typically 5--10\%, with a weak dependence on the jet energy and a characteristic structure in \etadet that reflects the calorimeter geometry.
 
The simulation-based JES correction factor $c_{\mathrm{JES}}$ is determined as a function of the jet energy and pseudorapidity \etadet. It is applied to the jet four-momentum as a multiplicative scale factor. The pseudorapidity correction $\Delta\eta$ only
changes the direction. This means that the reconstructed \largeR{} jet energy, mass, $\eta$, and \pt{} become
\begin{equation*}
E_{\mathrm{reco}} = c_\text{JES}\,E_0,~~~m_{\mathrm{reco}} = c_\text{JES}\,m_0,~~~\eta_{\mathrm{reco}} = \eta_0+\Delta\eta,~~~\pt^{\mathrm{reco}} = c_\text{JES}\,|\vec{p}_0| / \cosh{(\eta_0+\Delta\eta)},
\end{equation*}
where the quantities $E_0$, $m_0$, $\eta_0$, and $\vec{p}_0$ refer to the jet properties prior to any calibration, as determined by the trimming algorithm. The quantities $c_\text{JES}$ and $\Delta\eta$ are smooth functions of the \largeR jet kinematics. None of the calibration steps affect the azimuthal angle $\phi$ of the jet.
 
The \largeR jet invariant mass is calibrated in a final step. This is important when using the jet mass in physics analyses, because the jet mass is more sensitive than the transverse momentum to soft, wide-angle contributions and to cluster merging and splitting, as well as to the calorimeter geometry. For the mass correction the jet mass response $R_m = \langle m_\text{reco}/m_\text{truth} \rangle$ is determined using the same procedure as
for the jet energy calibration. The mass calibration is applied after the standard JES calibration.
The mass response is presented in Figure~\ref{fig:mccalibration_energy_mass} for three representative values of the truth
jet mass: 40~\gev{} in panel (b), the $W$ boson mass in panel (c), and the top quark mass
in panel (d). The mass response is close to unity for jets with \pt{} between 200 and 800~\gev{} and as large as 1.5 for very energetic jets with relatively low mass. Several effects can impact the jet mass response. The reconstructed mass can be artificially increased by the splitting of \topos{} during their creation. This effect is particularly important for jets with small particle-level mass relative to their \pt ($m/\pt\lessapprox 0.05$). Similarly, when several particles form one \topo{}, or when particles fail to produce any \topo{}, the mass response is decreased. This effect is significant for jets with large particle-level mass relative to their \pt ($m/\pt\gtrapprox 0.5$).
 
The simulation-based correction to the \largeR jet mass $c_{\mathrm{JMS}}$ is applied as a function of the jet $E_{\mathrm{reco}}$, \etadet, and $\log(m_{\mathrm{reco}}/E_{\mathrm{reco}})$, keeping the \largeR jet energy fixed and thus allowing the $\pt$ to vary~\cite{ATLAS-CONF-2015-037}. This factor is also a smooth function of the \largeR jet kinematics.
This has the following impact on the reconstructed jet kinematics:
\begin{eqnarray*}
E_{\mathrm{reco}} = c_\text{JES}\,E_0,~~~m_{\mathrm{reco}} = c_\text{JES}\,c_\text{JMS}\,m_0,~~~\eta_{\mathrm{reco}} = \eta_0+\Delta\eta,~~~\pt^{\mathrm{reco}} = c_\text{JES} \sqrt{E_0^2-c^2_\text{JMS}\,m_0^2} / \cosh{(\eta_0+\Delta\eta)}.
\end{eqnarray*}
All results that correspond to jets that are brought to the particle-level with the simulation-based calibration are labelled with ``JES+JMS''.

\begin{figure}[tbh]\centering
\subfigure[]{
\includegraphics[width=0.49\textwidth]{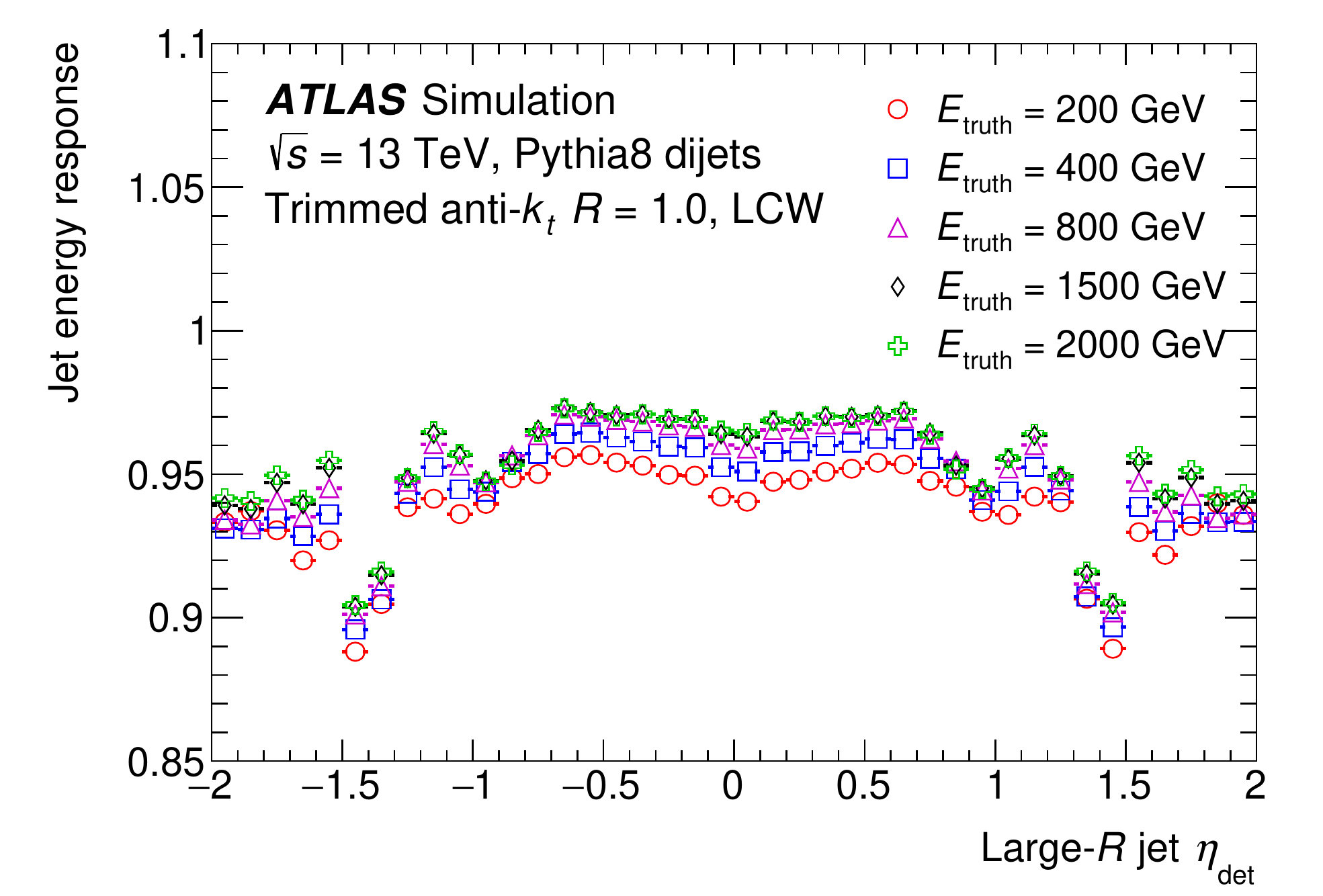}}
\subfigure[]{
\includegraphics[width=0.49\textwidth]{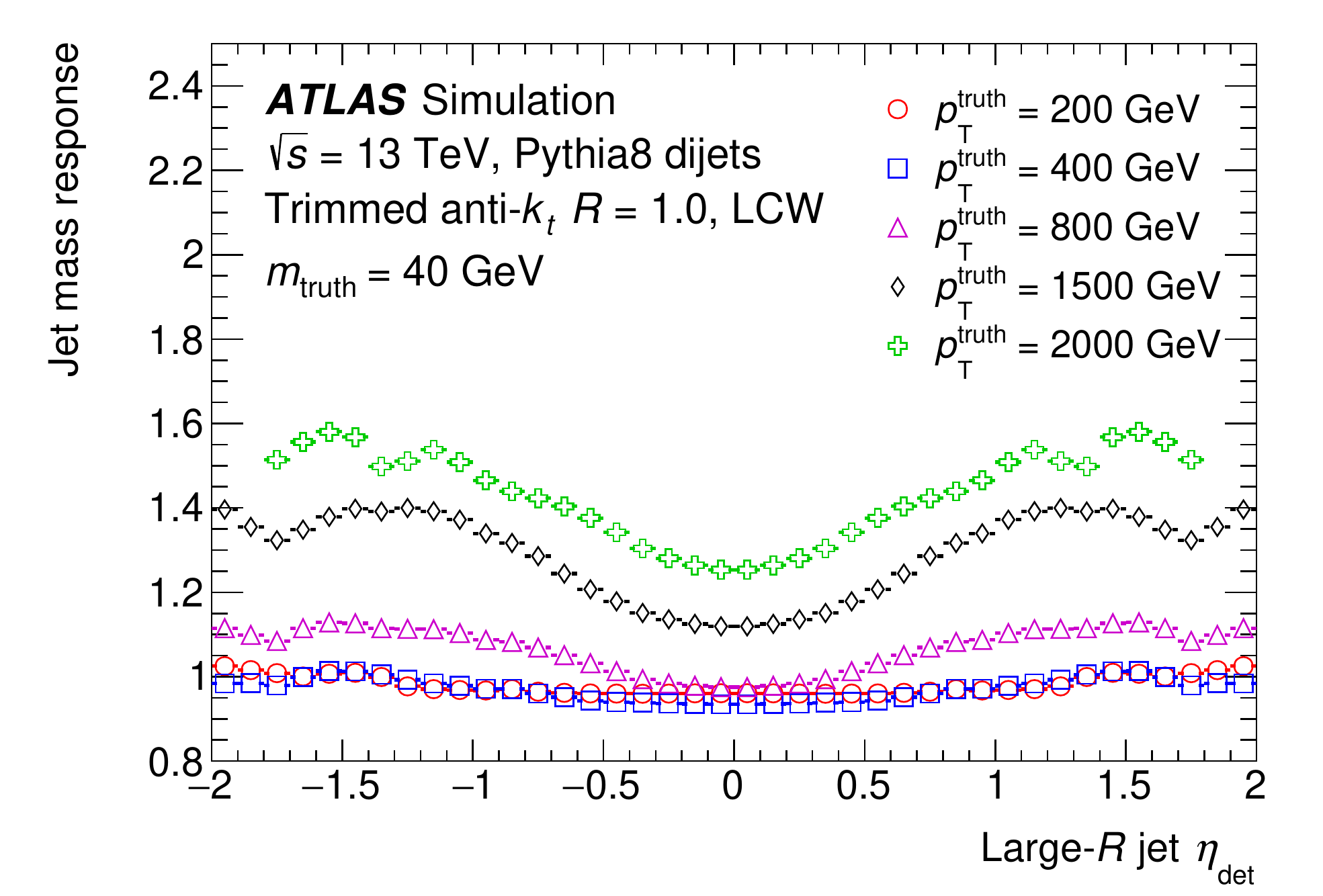}} \\
\subfigure[]{
\includegraphics[width=0.49\textwidth]{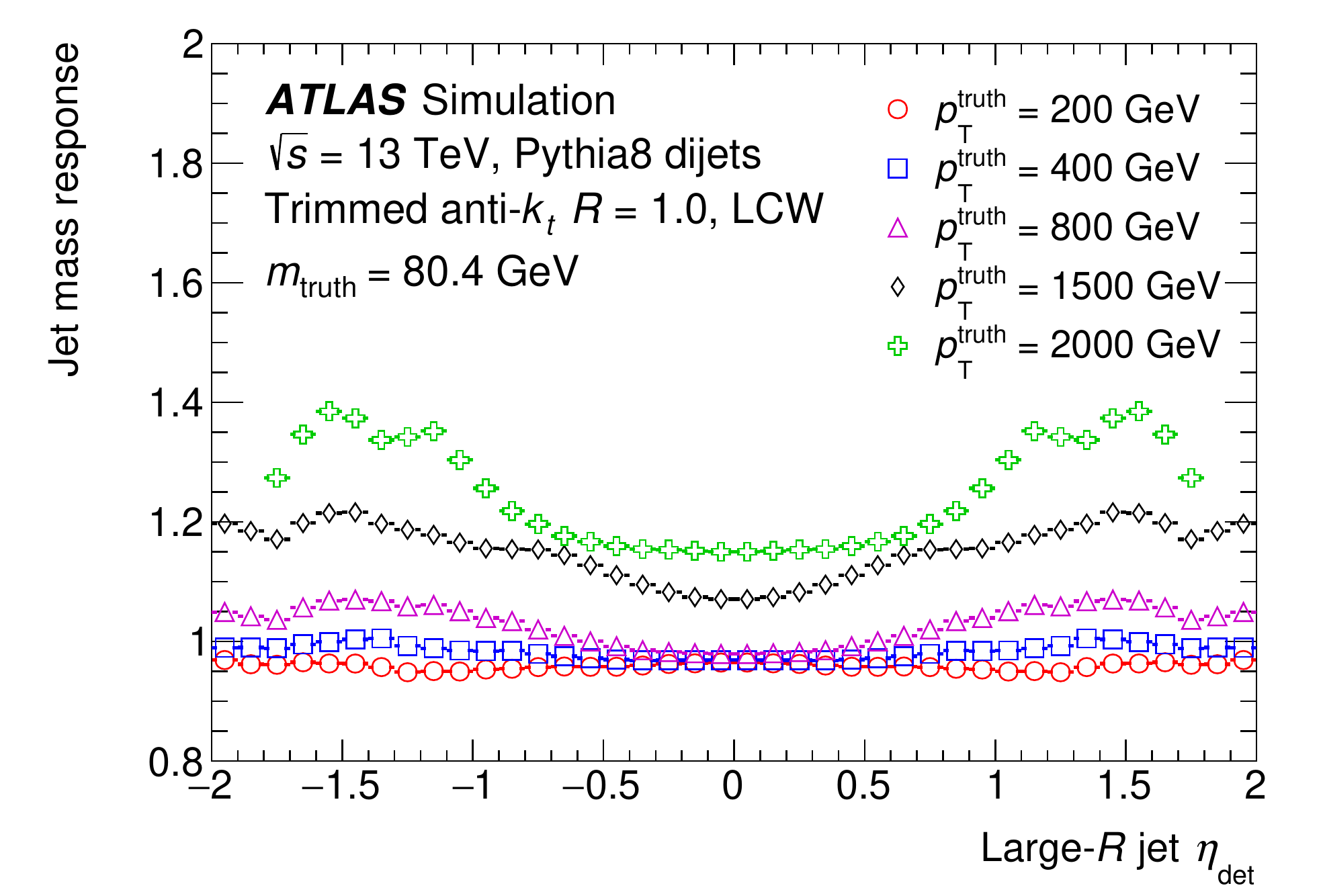}}
\subfigure[]{
\includegraphics[width=0.49\textwidth]{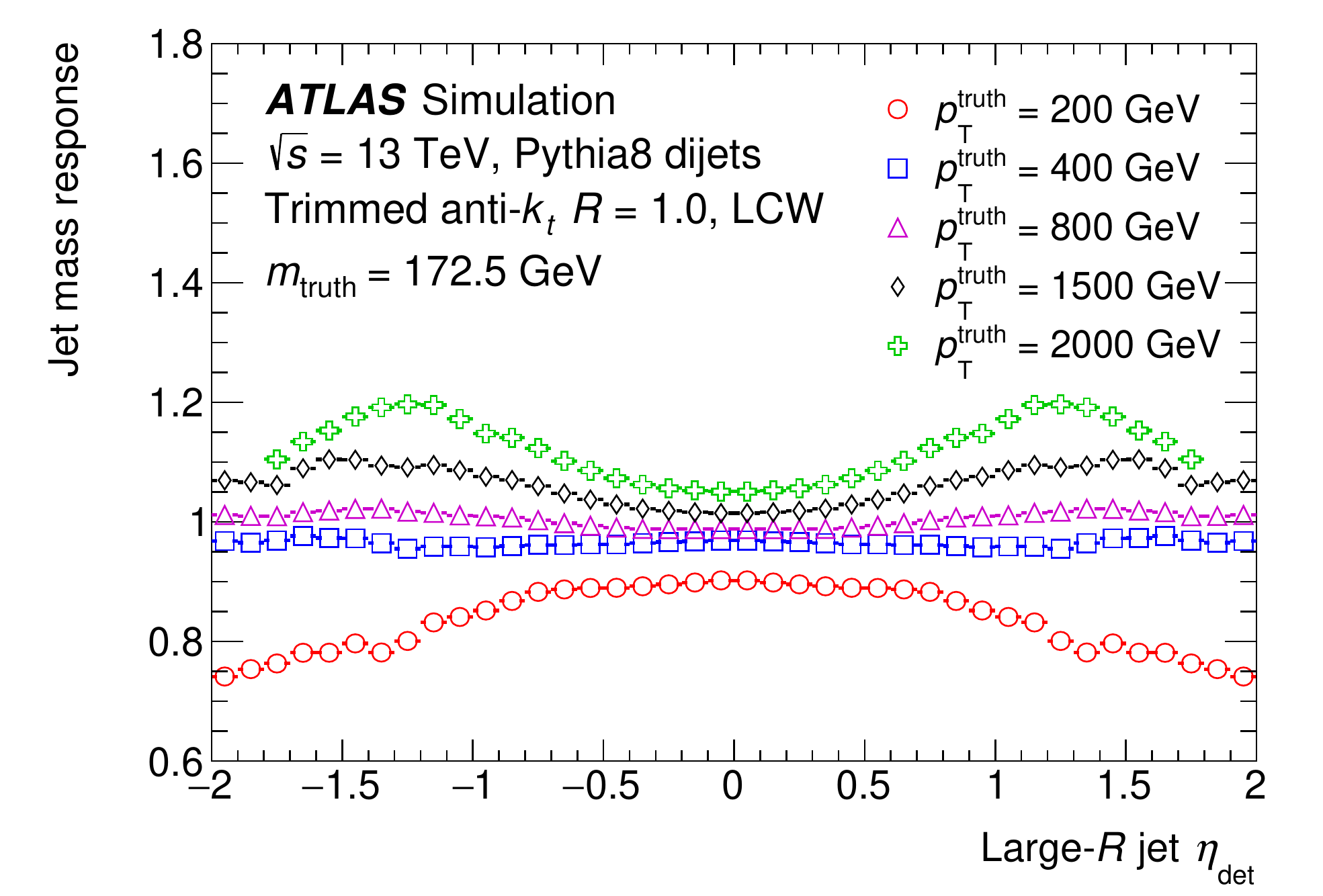}}
\caption{The response for (a) the jet energy and (b, c, d) the jet mass of \largeR{} jets. The jet energy response is presented as a function of jet detector pseudorapidity $\etadet$ for several values of the truth jet energy, ranging from 200~\gev{} to 2~\tev. The jet mass response is presented as a function of jet pseudorapidity for several values of the jet transverse momentum from 200~\gev{} to 2~\tev{} and for three representative values of the truth jet mass: (b) 40~\gev{}, representing a typical value for quark or gluon jets, (c) the $W$ boson mass, and (d) the top quark mass. The response is determined in simulation of dijet events as the ratio of the reconstructed jet mass to the mass of the corresponding particle-level jet. These results are used to define the jet-level mass correction applied in the simulation calibration.}
\label{fig:mccalibration_energy_mass}
\end{figure}
 
\subsection{Tracks and track jets}
 
Tracks are reconstructed from the hits generated by charged particles passing through the inner tracking detector (ID). They are required to have $p_{\mathrm{T}} >$ 500~\MeV. To reduce fake tracks, candidate tracks must be composed of at least one pixel detector hit and at least six hits in the silicon tracker. The track transverse impact parameter $|d_0|$ relative to the primary vertex must be less than 1.5~mm and the longitudinal impact parameter $|z_0|$ multiplied by $\sin{\theta}$ relative to the primary vertex must be less than 3~mm~\cite{ATLAS-CONF-2012-042,ATL-PHYS-PUB-2015-051}.
 
Jets reconstructed from charged-particle tracks are used as a reference in calibration and uncertainty studies, taking advantage of the independence of instrumental systematic effects between the ID and the calorimeter. Track jets are reconstructed by applying the same jet reconstruction procedure to tracks as those used when constructing the topo-cluster jets described above, including the jet trimming algorithm. Track jets are not calibrated.
 
\subsection{The combined jet mass}
 
The jet mass resolution is improved by combining the jet mass
measurement in the calorimeter with the measurement of the charged component
of the jet within the ID~\cite{ATLAS-CONF-2016-035,Son:2012mb,Katz:2010mr,Schaetzel:2013vka,Plehn:2010st,Plehn:2009rk,Larkoski:2015yqa,Bressler:2015uma,Elder:2018mcr}. A track jet is reconstructed from ID tracks with $\pt > 500~\MeV$~which are ghost-associated~\cite{Cacciari:2008gn} to the topo-cluster \largeR jet. The measurement of this track jet's mass is multiplied by the
ratio of the transverse momenta of the calorimeter jet and the track jet to obtain the track-assisted mass:
\begin{equation}
m^{\mathrm{TA}} = m^{\mathrm{track}} \, \frac{p_{\mathrm{T}}^{\mathrm{calo}}}{p_{\mathrm{T}}^{\mathrm{track}}}.
\label{eq:track_assisted}
\end{equation}
where $m^{\mathrm{TA}}$ is the track-assisted mass, $m^{\mathrm{track}}$ the mass obtained from the tracker, and $\pt^{\mathrm{calo}}$ and $\pt^{\mathrm{track}}$ are the transverse momenta measured respectively by the calorimeter and tracker.
This alternative mass measurement has better resolution for high-\pt jets with low values of $m/\pt$. A weighted least-squares combination of the mass measurements is subsequently performed with weights:
\begin{equation*}
m^{\mathrm{comb}} = w_{\textrm{calo}}\,m^{\mathrm{calo}} + w_{\textrm{TA}}\,m^{\mathrm{TA}},
\end{equation*}
where $w_{\textrm{calo}}$ and $w_{\textrm{TA}}$ are determined by the expected mass resolutions $\sigma_{\mathrm{calo}}$ and $\sigma_{\mathrm{\mathrm{TA}}}$ of the calorimeter and track-assisted measurements, using the central 68\% inter-quantile range of the jet mass response distribution in dijet events:
\begin{equation*}
w_{\textrm{calo}} = \frac{\sigma_{\mathrm{calo}}^{-2}}{\sigma_{\mathrm{calo}}^{-2} + \sigma_{\mathrm{TA}}^{-2}}, \,\, w_{\textrm{TA}} = \frac{\sigma_{\mathrm{TA}}^{-2}}{\sigma_{\mathrm{calo}}^{-2} + \sigma_{\mathrm{TA}}^{-2}},
\end{equation*}
such that the resolution of the combined mass measurement is always better than either of the two inputs within
the sample from which the weights are derived. In this paper, \insitu measurements are presented for the jet mass
reconstructed from topo-clusters and for the track-assisted mass. The constraint $ w_{\textrm{calo}} +  w_{\textrm{TA}} = 1$
ensures that the combined mass is calibrated, if the scales of both mass definitions are fixed.
 
\section{\Insitu \pt{} response measurements}\label{sec:jes}
\label{sec:balancept}
 
In this section, the methods used to derive the \insitu
calibration for the energy (or transverse momentum) response are presented.
These methods use \pt conservation in events where a \largeR jet
recoils against a well-measured reference object. The first
method is based on the \pt balance in dijet events with a central ($|\etadet| \leq 0.8$)
and a forward ($|\etadet| > 0.8$) jet. It is applied after the simulation calibration described in
Section~\ref{sec:objects}. The $\eta$-intercalibration corrects the \pt of
forward jets to make the jet energy response uniform as a function of
pseudorapidity. After the $\eta$-intercalibration procedure,
three further balance methods are used to provide an absolute \pt scale
calibration. In the $Z$+jet balance method, the recoiling system is a
reconstructed $Z \rightarrow \mu^+\mu^-$ or
$Z \rightarrow e^+e^-$ decay, in the $\gamma$+jet balance method it is
a photon, and in the multijet balance method the system is formed
by several calibrated \smallR jets with low \pt. These three methods offer
complementary coverage over a broad \pt range. The $Z$+jet balance method
provides the most precise results in the low-\pt interval
between 200 and 500~\gev, the $\gamma$+jet balance between
500~\gev~and~1~\tev, and the multijet balance extends to 2.5~\tev.
Results of the three methods are presented in this section and are
combined into a global constraint on the JES in
Section~\ref{sec:combination}.
 
\subsection{Dijet $\eta$-intercalibration} \label{sec:etacal}
 
The relative $\eta$-intercalibration extends the jet calibration to the forward detector region, $0.8 < |\eta| < 2.5$. It is derived from the differences in the \pt{} balance between a central reference and a forward jet in data and simulations. The $\eta$-intercalibration is determined in dijet events using a procedure similar to that used for \smallR jets~\cite{PERF-2012-01}. The \pt balance of the dijet system is characterized by its asymmetry $\mathcal{A}$, defined in terms of the forward (probe) and central (reference) jet \pt ($\pt^{\mathrm{probe}}$ and $\pt^{\mathrm{ref}}$) as
\begin{equation*}
\mathcal{A} = \frac{\pt^{\mathrm{probe}} - \pt^{\mathrm{ref}}}{\pt^{\mathrm{avg}}},
\label{eq:asymmetry_dijet}
\end{equation*}
where $\pt^{\mathrm{avg}} = (\pt^{\mathrm{probe}} + \pt^{\mathrm{ref}})/2$. The central reference jets are required to be within $|\eta|<0.8$. The balancing probe jet $\eta_\text{det}$ defines the detector region whose response is being probed. The asymmetry distribution is studied in bins of $\pt^{\mathrm{avg}}$ and the probe jet \etadet. In each bin, the relative response difference between the central and forward jets is
\begin{equation}
\Rrel = \left\langle \frac{\pt^{\mathrm{probe}}}{\pt^{\mathrm{ref}}} \right\rangle = \frac{2+\langle\mathcal{A}\rangle}{2-\langle\mathcal{A}\rangle},
\end{equation}
where $\langle\mathcal{A}\rangle$ is the mean value of the asymmetry. The asymmetry distribution is approximately Gaussian, and the mean value is extracted using a Gaussian fit to the core of the distribution.
 
\LargeR jets with \pt from 180~\GeV~to~2~\TeV~within $|\eta|<2.5$ are considered. Dijet events in data are selected using several dedicated single-jet triggers based on \smallR jets. Their efficiency has been evaluated for \largeR jets and each trigger is used in its region of full efficiency for those jets.  These triggers provide enough events for this technique to be used over a wide range of \pt. To ensure a $2\rightarrow2$ body topology, events with energetic additional radiation are vetoed with an upper cut on the transverse momentum of the third jet $J_3$, and the leading two jets are required to satisfy a minimum angular separation in azimuth. Both of these requirements are varied in order to derive systematic uncertainties accounting for their impact on the response measurements.  These selections and systematic variations are summarized in Table~\ref{tab:etacalvars}. No pile-up jet tagging employing the Jet Vertex Tagger likelihood measure (JVT)~\cite{PERF-2014-03,ATLAS-CONF-2014-018} is applied for \largeR jets, since in this kinematic region the contamination by pile-up jets is negligible.
 
\begin{table}[h!]
\centering
\caption{\label{tab:etacalvars} Summary of the dijet topology selection and systematic variations considered for the $\eta$-intercalibration analysis. The label $\text{J}_{3}$ refers to the third trimmed $R=1.0$ jet in the event after ordering the jets in \pt.}
\begin{tabular}{l c c c}
\hline
Variable & Nominal Selection & Up Variation & Down Variation \\ \hline
$\pt^{\mathrm{J_3}} / \ptavg$ & $< 0.4$ & $<0.5$ & $<0.3$ \\
$\Delta\phi(\mathrm{ref,probe})$ & $>2.5$ & $>2.8$ & $>2.2$ \\
\hline
\end{tabular}
\end{table}
 
The relative jet-\pt{} response \Rrel is shown in Figure~\ref{fig:etacal:RvsEta} as a function of the \largeR jet pseudorapidity for data, \POWPYTHIA 8, and \SHERPA for two \pt intervals. The relative jet response as a function of the \largeR jet \pt is shown in Figure~\ref{fig:etacal:RvsPt} for two pseudorapidity ranges of the probe jet. In the central region, the relative responses of all three samples agree by design. The relative response in data increases in the forward region due to features of the experimental response which are not well-reproduced in the simulation and hence not accounted for in the simulation-based JES calibration factor $c_\text{JES}$. Compared to the measured response, the prediction remains relatively constant around unity. The difference between the simulated and measured responses reaches about 5\% around $|\eta| = 2.5$. Similar trends are observed for $R=0.4$ jets in Ref.~\cite{PERF-2016-04}. In the lower panel of Figure~\ref{fig:etacal:RvsEta}~and~Figure~\ref{fig:etacal:RvsPt}, the ratio of simulation to data is shown. An interpolation using a filter with a sliding Gaussian kernel across \etadet yields a smooth function of jet \pt{} and \etadet. The inverse of this smooth function is taken as the $\eta$-intercalibration correction factor $c_\text{rel} (\pt,\etadet)$, which is applied as a jet four-momentum scale factor.
 
The uncertainties associated with the $\eta$-intercalibration are shown in Figure~\ref{fig:etacal:unc} for two representative \pt bins. The uncertainties associated with the veto on additional radiation and the $\Delta\phi$ requirement placed on the dijet topology are derived by varying these selection criteria to the values listed in Table~\ref{tab:etacalvars} and re-deriving the calibration. An additional systematic uncertainty accounts for the choice of event generator and parton shower models. The simulation uncertainty is derived by comparing the relative jet-\pt{} response for two event generators: \POWPYTHIA 8 and \SHERPA. In general, the uncertainties associated with the derived calibration are small, amounting to a $\sim$1\% uncertainty within the region of interest for \largeR jets ($|\eta|<2.0$). Uncertainties originating from the kinematic requirements made to select events are typically negligible, except in the highest $\pt^{\mathrm{avg}}$ bins.

\begin{figure}[tbh]\centering
\subfigure[]{
\includegraphics[width=0.48\textwidth]{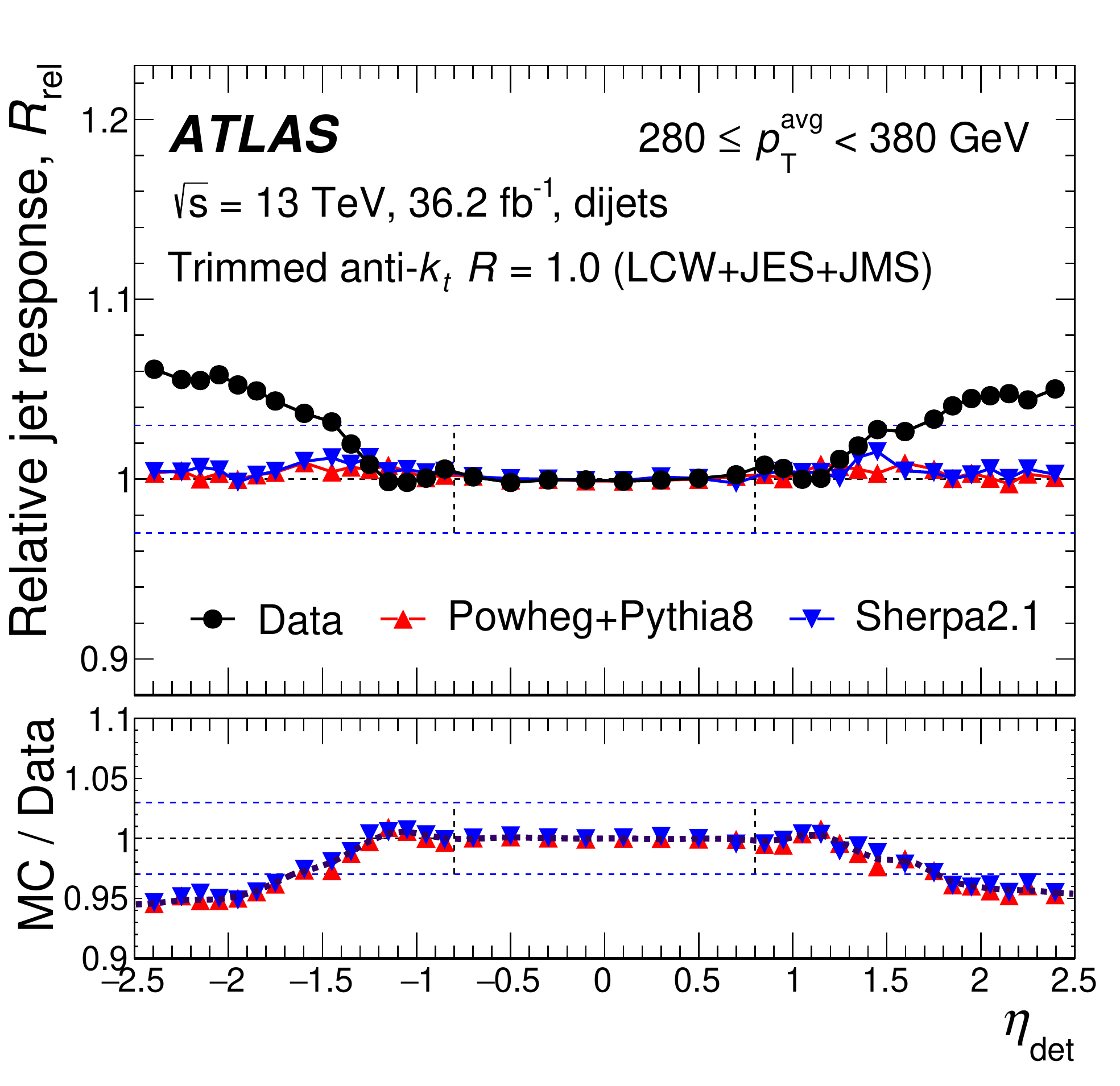}
}
\subfigure[]{
\includegraphics[width=0.48\textwidth]{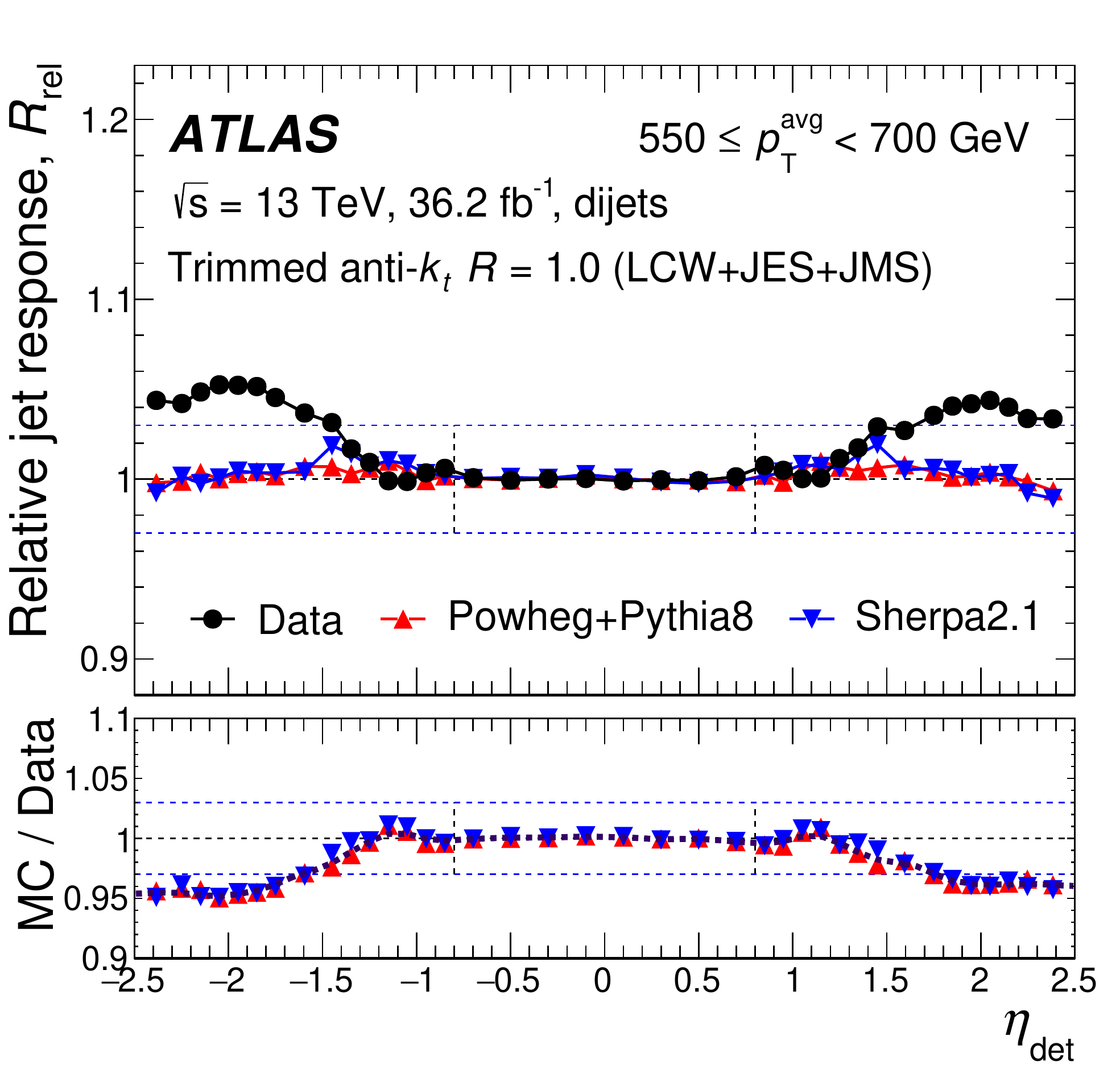}
}
\caption{The relative \largeR jet response \Rrel as a function of the \largeR jet detector pseudorapidity \etadet in two representative average transverse momentum $\pt^{\mathrm{avg}}$ bins (a) $280~\GeV < \pt^{\mathrm{avg}}~< 380~\GeV$ and (b) $550~\GeV < \pt^{\mathrm{avg}} < 700~\GeV$. The average response within the reference region $|\etadet|<0.8$ is unity by construction. In the lower panels, the dotted lines interpolating between \POWPYTHIA markers are obtained by smoothing with a filter using a sliding Gaussian kernel. }
\label{fig:etacal:RvsEta}
\end{figure}
 
\begin{figure}[tbh]\centering
\subfigure[]{
\includegraphics[width=0.48\textwidth]{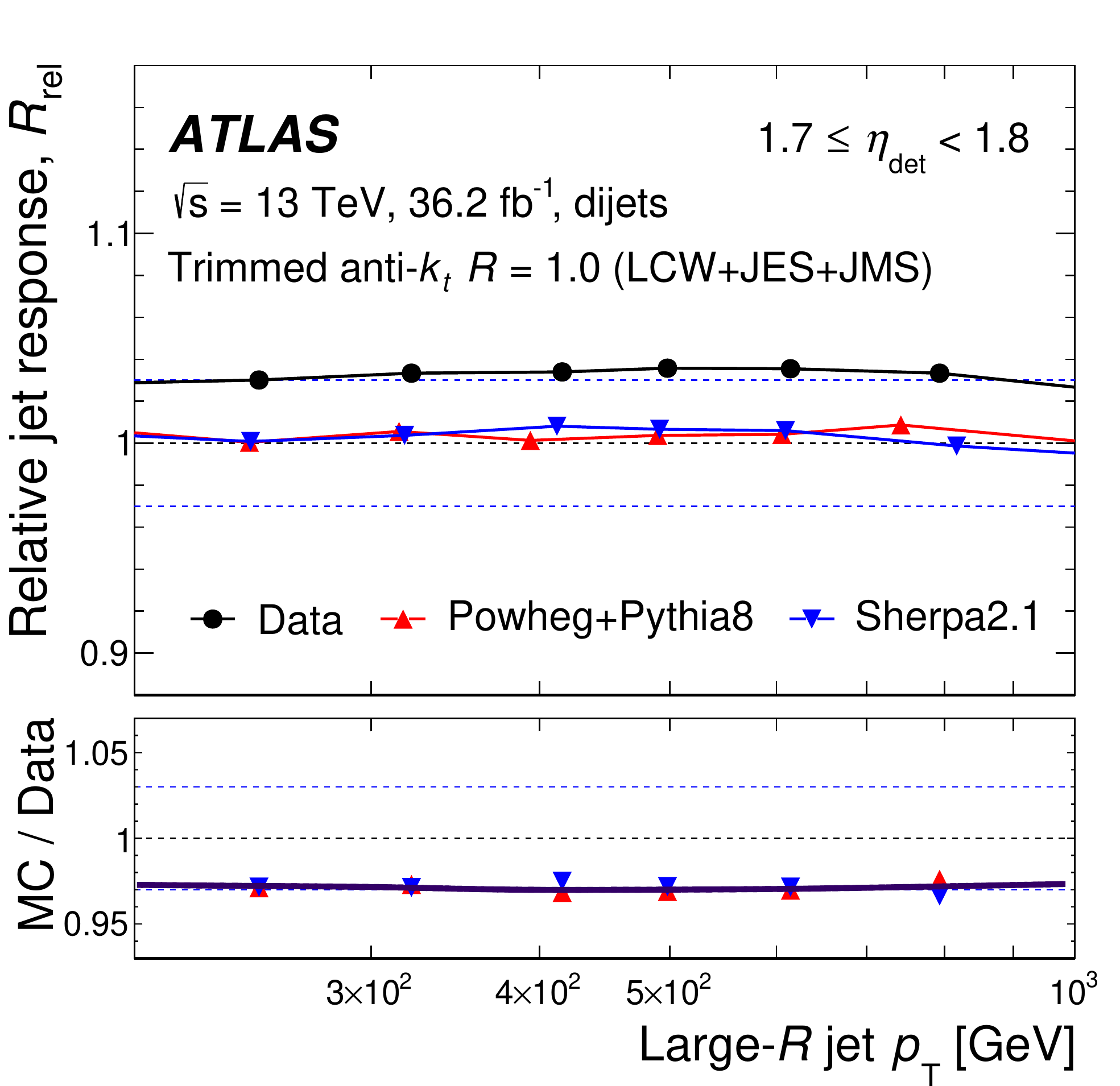}
}
\subfigure[]{
\includegraphics[width=0.48\textwidth]{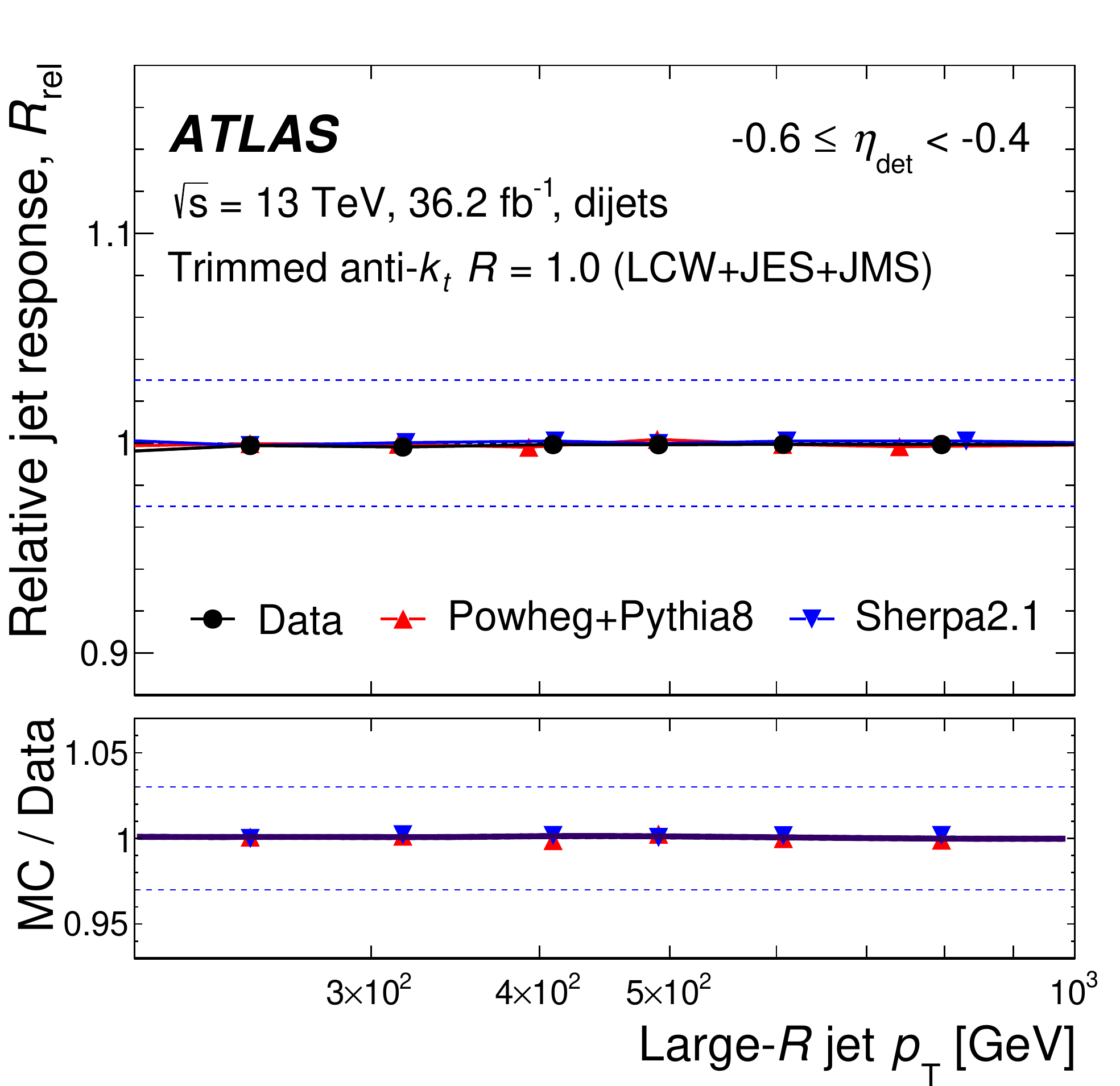}
}
\caption{The relative \largeR jet response \Rrel as a function of the \largeR jet \pt in two representative detector pseudorapidity $\etadet$ bins in the forward and central reference regions (a) $1.7 < \etadet < 1.8$ and (b) $-0.6 < \etadet < -0.4$. In the lower panels, the lines interpolating between \POWPYTHIA markers are obtained by smoothing with a filter using a sliding Gaussian kernel.} 
\label{fig:etacal:RvsPt}
\end{figure}
 
\begin{figure}[tbh]\centering
\subfigure[]{
\includegraphics[width=0.45\textwidth]{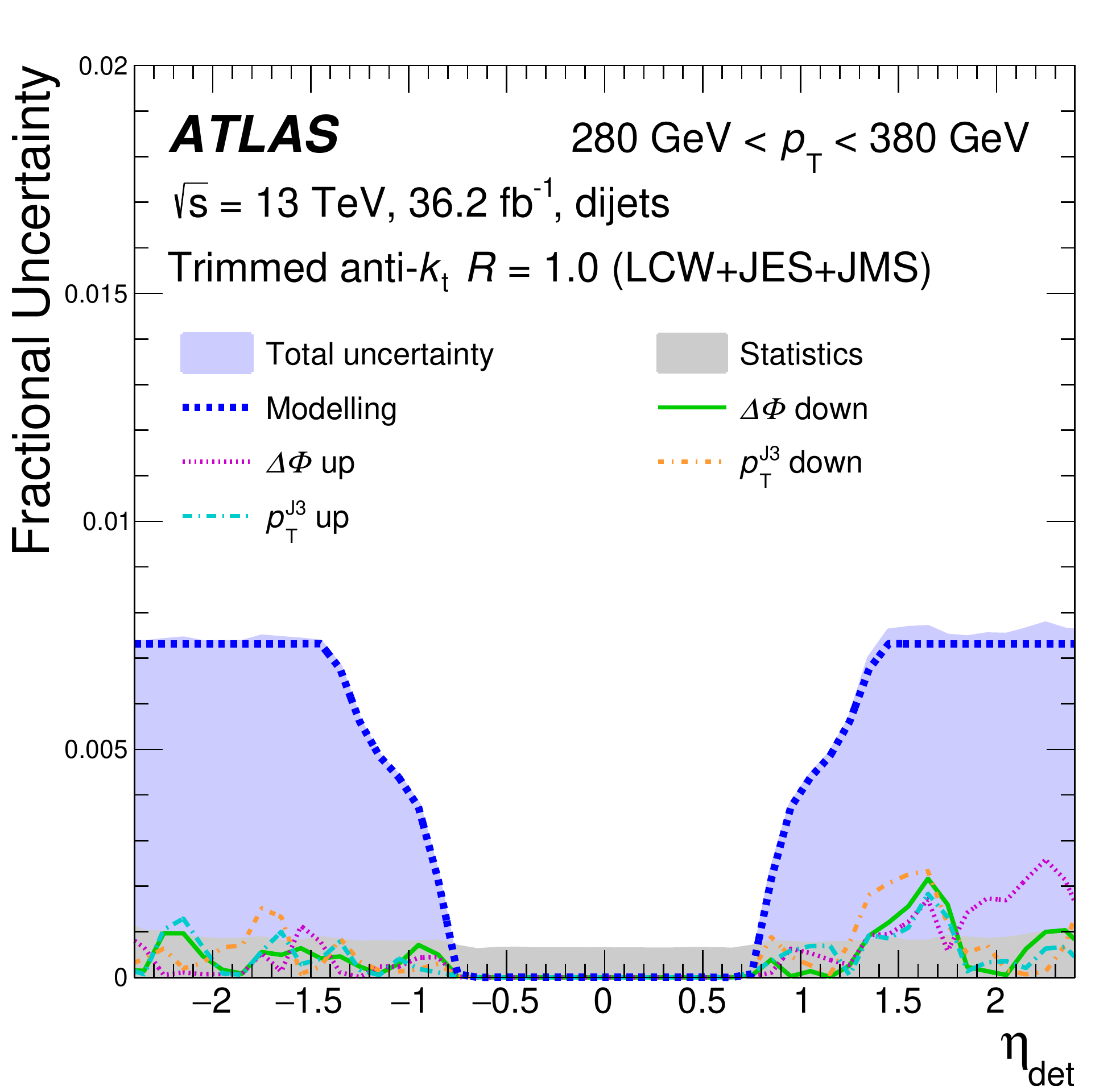}
}
\subfigure[]{
\includegraphics[width=0.45\textwidth]{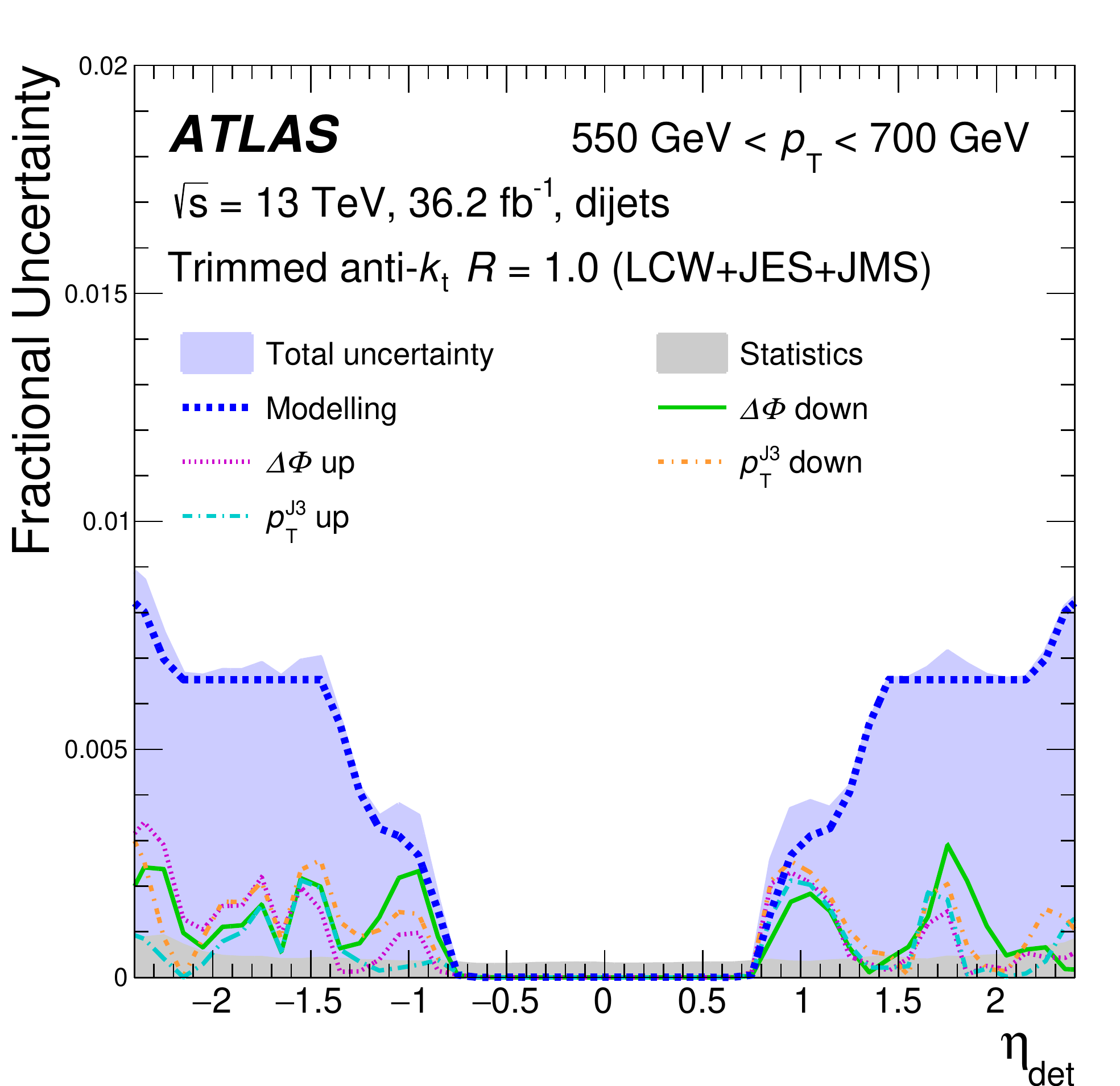}
}
\caption{Uncertainties associated with the \largeR jet $\eta$-intercalibration as a function of detector pseudorapidity $\etadet$ in two representative average transverse momentum $\pt^{\mathrm{avg}}$ bins (a) $280~\GeV < \pt^{\mathrm{avg}} < 380~\GeV$ and (b) $550~\GeV < \pt^{\mathrm{avg}} < 700~\GeV$. The uncertainties evaluated using variations of the dijet topology selection are negligible relative to the simulation modelling uncertainty, which typically amounts to a 1\% uncertainty for \largeR jets within $0.8 < |\etadet| < 2.0$. }
\label{fig:etacal:unc}
\end{figure}
\clearpage
 
\subsection{$Z$+jet balance} \label{sec:zjetbalance}
For \largeR jets within $|\etadet|<0.8$, an \insitu calibration is derived by examining the \pt balance of a \largeR jet and a leptonically decaying $Z$ boson, either $Z\rightarrow e^+e^-$ or $Z\rightarrow \mu^+\mu^-$ (Figure~\ref{fig:balance_fig}(b)). Both of these channels provide a precise, independent reference measurement of the jet energy, either from the inner detector and muon spectrometer tracks used to reconstruct muons or from the well-measured electromagnetic showers and inner detector tracks used to reconstruct electrons. The applicable range of this calibration is limited by the kinematic range where $Z$ boson production is relatively abundant, that is, up to a $Z$ boson \pt of about 500~\GeV. Electrons used to reconstruct the $Z$ boson are required to pass `medium likelihood identificiation' quality and `Loose' isolation requirements and must be reconstructed within $|\eta|<2.47$ (excluding the transition region $1.36 < |\eta| < 1.52$ between the barrel and endcap electromagnetic calorimeters) with at least $20~\GeV$ of \pt~\cite{PERF-2016-01,ATLAS-CONF-2016-024}. Similarly, `VeryLoose' quality and `Loose' isolation requirements are placed on muons, which must be reconstructed within $|\eta|<2.4$ with $\pt>20~\GeV$~\cite{PERF-2015-10}. The lepton pair must have opposite charge and be kinematically consistent with the decay of a $Z$ boson, requiring the invariant mass of the lepton pair to satisfy $66 < m_{\ell^+\ell^-} < 116~\GeV$. \LargeR jets studied here are calibrated with the simulation calibration and $\eta$-intercalibration described in Sections~\ref{sec:objects}~and~\ref{sec:etacal}.
 
The direct balance method used here closely follows the methodology outlined in Ref.~\cite{PERF-2016-04}. The average momentum balance between the \largeR jet and $Z$ boson is
\begin{equation}
\RDB = \left \langle \frac{\pt^\mathrm{J}}{\ptref} \right \rangle, \label{eq:zjets:B}
\end{equation}
where $\pt^\mathrm{J}$ is the \largeR jet \pt and $\ptref = \pt^{Z}\,\big|\cos\left(\Delta \phi\right)\big|$ is the component of the reference momentum collinear with the jet, with $\Delta\phi$ being the azimuthal angle between the \largeR jet and reference $Z$ boson.
The average value is determined using a Gaussian fit.
 
Even with an ideal detector, the momentum balance \RDB of~Eq.~\ref{eq:zjets:B} will only equal unity for an ideal $2 \rightarrow 2$ process. In practice, there tends to be more QCD radiation in the hemisphere opposite to the colour-neutral $Z$~boson, and therefore \RDB tends to be below unity. The event selection imposes a veto on the \pt of additional sub-leading jets. A minimum requirement is also imposed on the angular separation $\Delta\phi$ of the \largeR jet and reference $Z$ boson. Any mismodelling in the jet energy scale may be evaluated using the balance double ratio of \RDB in data and simulation $\RDB^{\mathrm{data}} / \RDB^{\mathrm{MC}}$. If the event selection criteria are met and the reference object is well measured and correctly modelled in simulation, any deviation from unity in the double ratio can be attributed to a mismodelling of the jet response in simulation and may be taken as an \insitu correction.
 
Calibrated \antikt $R=0.4$ jets constructed from electromagnetic-scale topo-clusters are used to veto additional radiation. These jets are required to be $\Delta R > 1.4$ from the \largeR jet whose response is being probed ($\mathrm{J}_1$), which ensures that there is no overlap. Such small-$R$ jets with $\pt < 60$~\GeV~must also satisfy a requirement on the jet vertex tagger (JVT)~\cite{PERF-2014-03}, which is designed to reject additional jets produced by pile-up interactions using information from the inner detector. The $2 \rightarrow 2$ topology selection only accepts events in which any small-$R$ jet is reconstructed with a $\pt < \max(0.1 \, \ptref,15~\GeV)$ and the $\Delta\phi$ between the \largeR jet and $Z$ boson is greater than $2.8$. A summary of the event selection is presented in Table~\ref{tab:zjetsevsel}. This table also reports variations associated with each criterion, performed by redoing the full analysis for each such variation and taking the difference between the varied and nominal results as the systematic uncertainty.
 
\begin{table}[h!]
\centering
\caption{\label{tab:zjetsevsel} Summary of the $2 \rightarrow 2$ topology selection and systematic variations considered for the $Z$+jet direct balance analysis. The labels J$_{i}$ refer to the $i$th leading \largeR jet, and j$_i$ to the $i$th leading small-$R$ jet that fulfils $\DeltaR(\mathrm{J}_1,\mathrm{j}_i) > 1.4$.}
\begin{tabular}{l c c c}
\hline
Variable & Nominal Selection & Up Variation & Down Variation \\ \hline
$\pt^{\mathrm{j_1}} $ & $\max(0.1\,\ptref,15~\GeV)$ & $\max(0.15\,\ptref,20~\GeV)$ & $\max(0.05\,\ptref,10~\GeV)$ \\
$\Delta\phi(Z,\mathrm{J_1})$ & $>2.8$ & $>2.9$ & $>2.7$ \\
Small-$R$ jet JVT  & $>0.59$ & $>0.91$ & $>0.11$ \\
\hline
\end{tabular}
\end{table}
 
Measurements of \RDB are carried out separately in the electron and muon channels. They are found to be consistent and thus combined to provide a single measurement of the JES. The average momentum balance in $Z$+jet events after this combination is shown in Figure~\ref{fig:zjet_balance}. The balance is found to be consistently below unity as a function of \ptref. The ratio of the predicted balance to the measured balance is consistently 1--4\% above unity. The uncertainties associated with this measurement are shown in Figure~\ref{fig:zjet_uncertainties}, where modelling systematic and statistical uncertainties are the dominant source of error over the \pt range considered.
 
\begin{figure}[tbh]
\begin{center}
\includegraphics[width=0.66\textwidth]{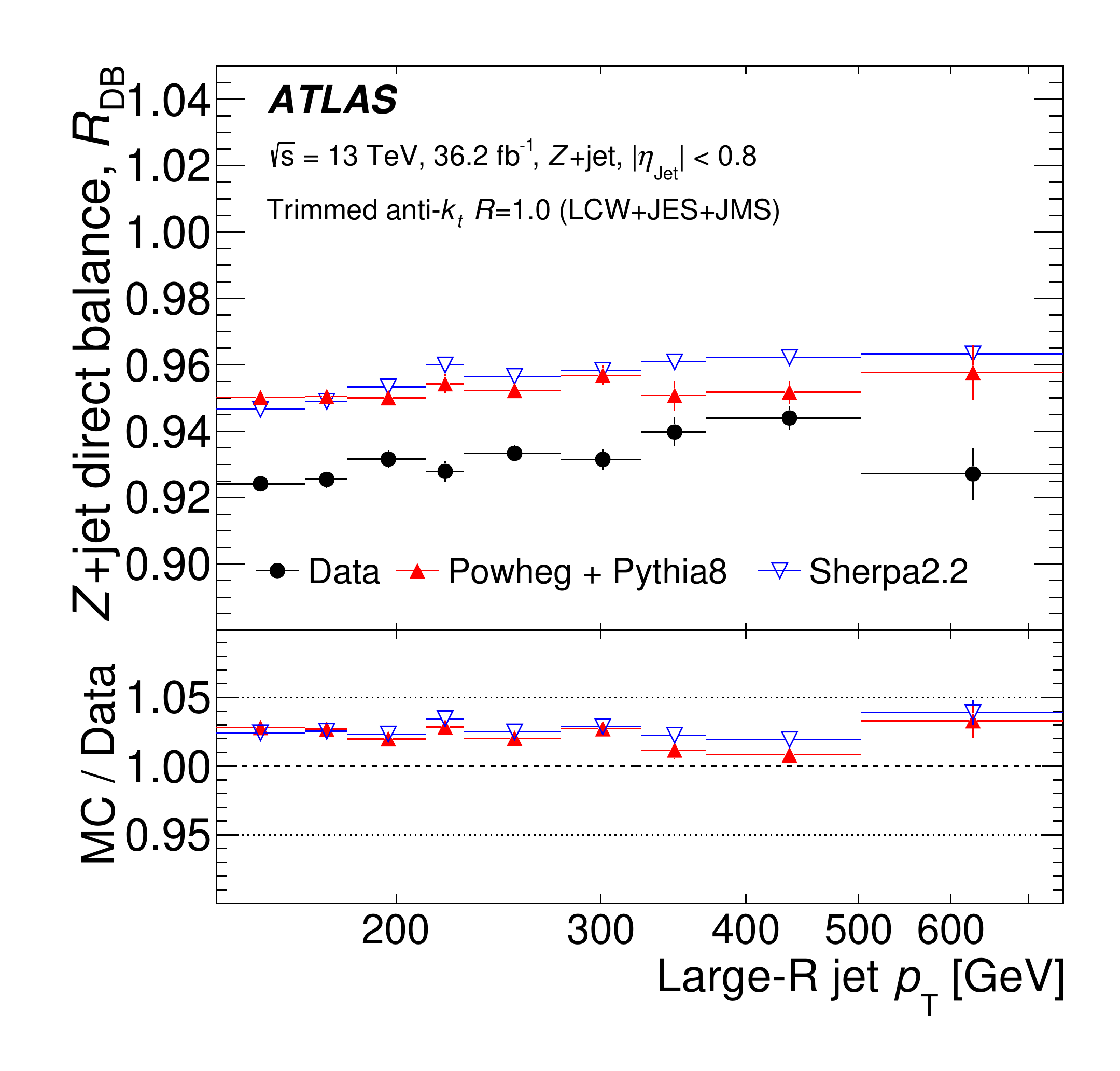}
\caption{\label{fig:zjet_balance} The momentum balance \RDB as a function of the \largeR jet transverse momentum \pt in $Z$+jet events for the combined $e^+e^-$ and $\mu^+\mu^-$ channels. Only statistical uncertainties are shown. For each \pTref{} bin, the measured \RDB{} is plotted against the average jet \pt{} of the bin. The horizontal error bars gives an indication of the the width of the associated \pTref{} bin.}
\end{center}
\end{figure}
 
\begin{figure}[tbh]
\begin{center}
\includegraphics[width=0.66\textwidth]{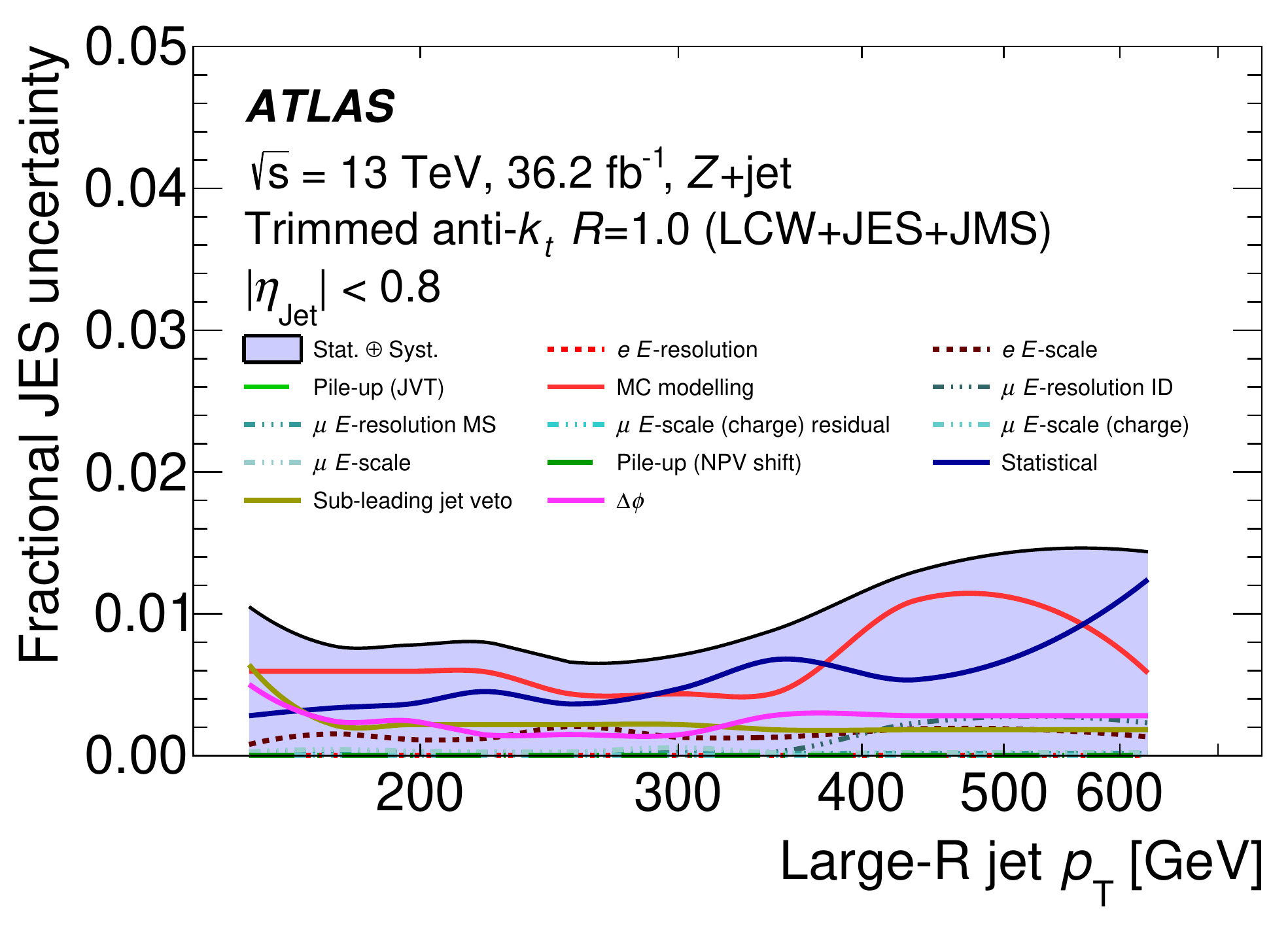}
\caption{\label{fig:zjet_uncertainties} Breakdown of the uncertainties in the JES measurement with the $Z$+jet direct balance method as a function of the \largeR jet transverse momentum \pt. The sources include the statistical uncertainty, variations of the generator (simulation modelling), variations of the event selection (pile-up (JVT), sub-leading jet veto, $\Delta \phi$), the uncertainties in the energy scale and resolution of electrons ($e$ $E$-scale and $e$ $E$-resolution) and muons ($\mu$ $E$-scale and $\mu$ $E$-resolution), and the uncertainty in the pile-up conditions ($N_\text{PV}$ shift). These uncertainties are also discussed in the context of \smallR jets in Ref.~\cite{PERF-2016-04}. The lines are obtained by smoothing a binned representation of these uncertainties using a sliding Gaussian kernel.}
\end{center}
\end{figure}
\clearpage
 
\subsection{$\gamma$+jet balance} \label{sec:gjetbalance}
The \largeR jet energy scale can be measured using the $\gamma$+jet final state
(Figure~\ref{fig:balance_fig}(b)). This method exploits the fact that the energy of photons is
measured more precisely than that of jets. As cross-section for this process is larger than that for $Z$+jets production, this balance technique probes higher \largeR jet \pt. The $\gamma$+jet method is based on the balance between photons and \largeR jets, using the ratio \RDB defined in Eq.~(\ref{eq:zjets:B}), where the reference momentum $\ptref = \pt^{\gamma} \big|\cos\left(\Delta \phi\right)\big|$ is the component of $\pt^{\gamma}$ collinear with the jet.
 
The double ratio of $\RDB^\mathrm{data} / \RDB^{\mathrm{MC}}$ measures
any residual modelling effects in the jet energy scale calibration. If the reference photon is well measured
experimentally and the $\gamma$+jet events are correctly modelled in simulation,
any deviation from unity in the double ratio can be attributed to
a mismodelling of the jet response in the Monte Carlo simulation.
 
Events are selected using the lowest unprescaled single-photon trigger. The offline selection requires the presence of a photon satisfying the `tight' identification and isolation requirements~\cite{ATL-PHYS-PUB-2016-014,PERF-2013-04} with at least 140~\GeV~of $E_{\mathrm{T}}$. This criterion ensures full trigger efficiency. As in the case of $Z$+jet balance (Section~\ref{sec:zjetbalance}), the presence of significant additional radiation in the event invalidates the assumption of a balanced topology. Events are therefore vetoed if a reconstructed, calibrated $R=0.4$ jet built from electromagnetic-scale topo-clusters has a \pt which satisfies $\pt>\max(0.1 \, \ptref,15~\GeV)$. Small-$R$ jets with $\pt < 60$~\GeV~must also satisfy a JVT requirement. Photons must be separated from reconstructed \largeR jets by at least $\Delta \phi(\mathrm{J},\gamma)>2.8$. The simulation calibration and $\eta$-intercalibration described in Sections~\ref{sec:objects}~and~\ref{sec:etacal} are applied to the \largeR jets studied here.
 
A photon purity correction is applied to the mean balance results in data to correct for contamination from misidentified jets or electrons that may skew the nominal \pt balance. The contamination of the photon sample by fakes is derived from data using the double-sideband, or ABCD, method~\cite{STDM-2010-08,STDM-2016-08} in the plane spanned by the photon isolation\footnote{The calorimeter isolation variable $E_{\mathrm{T}}^{\mathrm{iso}}$ is defined as the sum of the $E_{\mathrm{T}}$ of topological clusters deposited in a cone of size $\Delta R = 0.4$ around the photon candidate, excluding an area of size $\Delta \eta \times \Delta \phi = 0.125 \times 0.175$ centred on the photon cluster and subtracting the expected photon energy deposit outside of the excluded area. Fluctuations in the ambient transverse energy of the event are corrected for; the typical size of this correction is 2~\GeV~in the central region.} and the photon identification measure.\footnote{The photon identification decision is based on a set of shower shape variables computed from energy depositions in the first and second layers of the electromagnetic calorimeter and from leakage in the hadronic calorimeter.} The purity correction results in a shift of the relative \RDB value between data and simulation of about 2\%.
 
In Figure~\ref{fig:gamma_jet_response} the result is shown
as a function of the reference \pt for \largeR jets in the region $|\eta|<0.8$. The ratio of the predicted response in the simulation to the measured response is shown in the inset below the main panel. As already observed in Section~\ref{sec:zjetbalance}, the ratio of simulation to data is above unity over the whole \pt range. These results are included in the \insitu calibration that corrects the jet energy response in data.
 
The uniformity of the \largeR jet response across the detector geometry is shown in Figure~\ref{fig:gamma_jet_eta}, as a validation of the $\eta$-intercalibration procedure (Section~\ref{sec:etacal}). The relative response across the detector is constant and well behaved.
 
\begin{figure}[tbh]\centering
\includegraphics[width=0.66\textwidth]{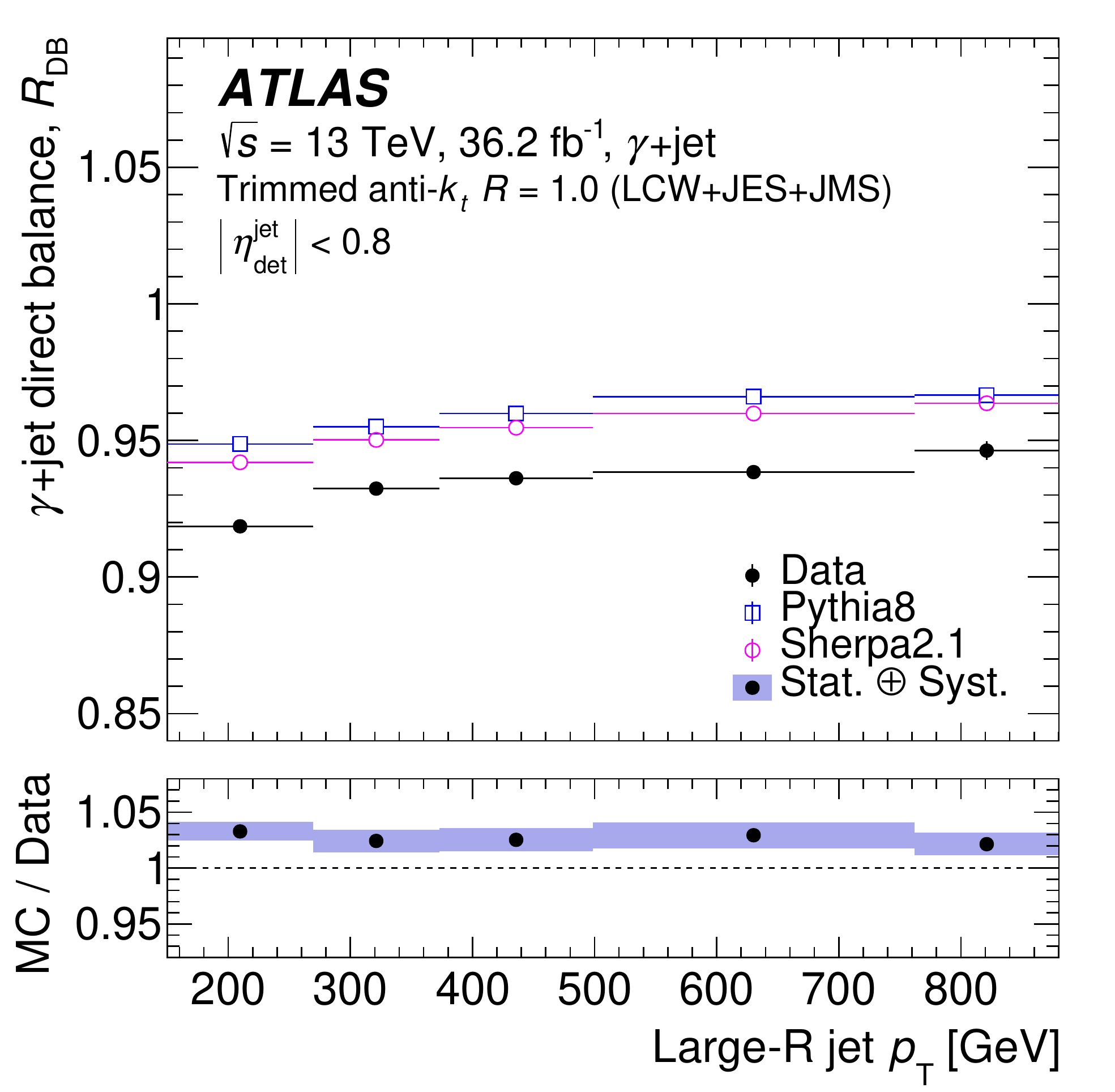}
\caption{The momentum balance \RDB extracted from $\gamma$+jet events in data and simulations as a function of the transverse momentum \pt\ of the \largeR jet. The ratio of the results obtained from the nominal \PYTHIA simulation and from data is shown in the bottom panel. The ratio of \PYTHIA to \SHERPA results, taken as a systematic uncertainty associated with modelling, is included in the shaded band in the ratio panel, which also includes statistical and systematic uncertainties from other sources. For each \pTref{} bin, the measured \RDB{} is plotted against the average jet \pt{} of the bin. The horizontal error bars gives an indication of the the width of the associated \pTref{} bin.}
\label{fig:gamma_jet_response}
\end{figure}
 
\begin{figure}[tbh]\centering
\includegraphics[width=0.66\textwidth]{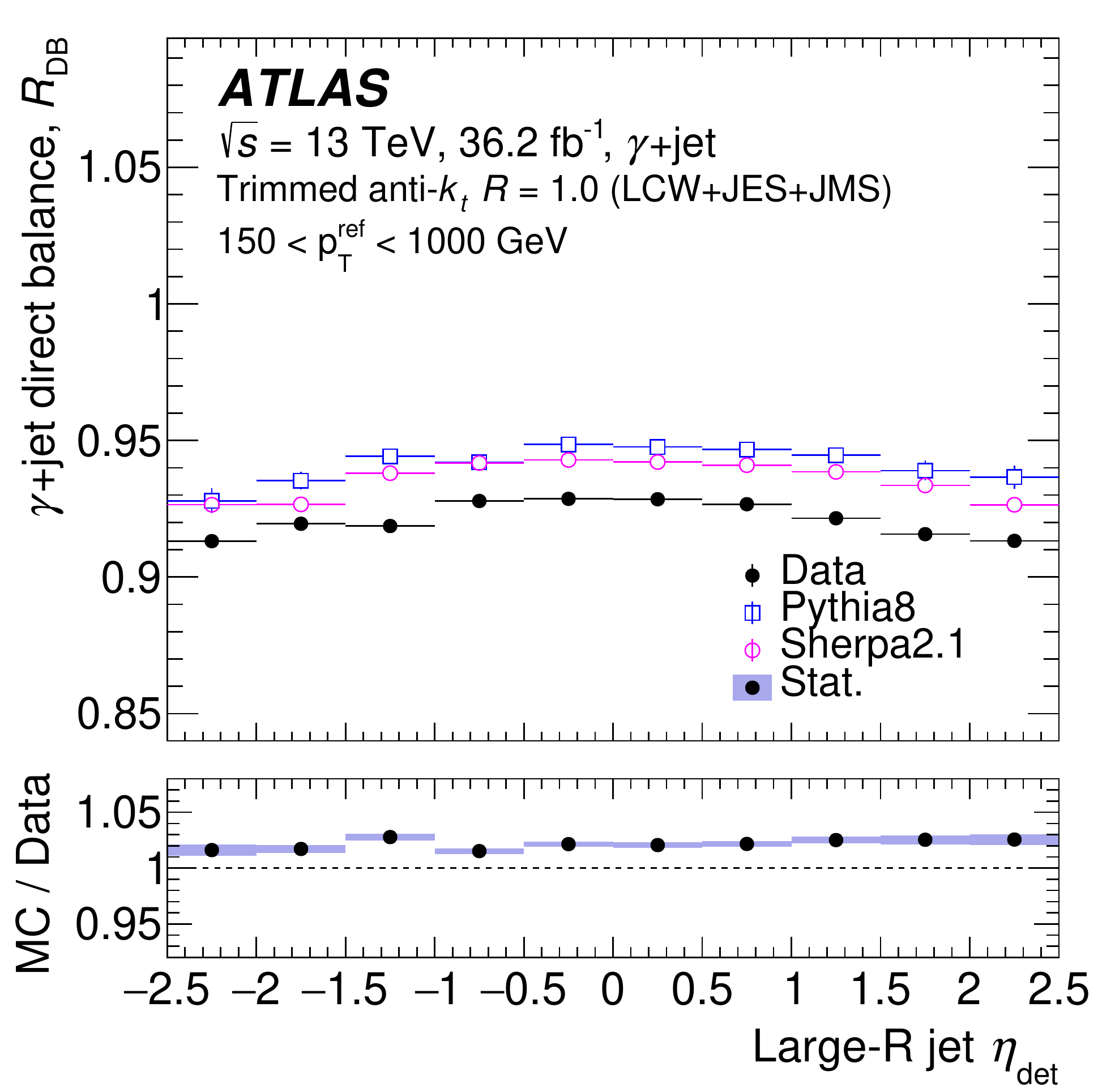}
\caption{The momentum balance \RDB extracted from $\gamma$+jet balance distributions in data and simulation as a function of the \largeR jet detector pseudorapidity $\etadet$. The ratio of the results obtained from the nominal \PYTHIA simulation to the results from data is shown in the bottom panel. The ratio of \PYTHIA to \SHERPA results, taken as a systematic uncertainty associated with modelling, is included in the shaded band in the ratio panel, which also includes statistical and systematic uncertainties from other sources.}
\label{fig:gamma_jet_eta}
\end{figure}
 
There are three main categories of systematic uncertainties in the \RDB measurement: those related to the modelling of additional QCD radiation which affects the balance, uncertainties associated with the photons~\cite{ATL-PHYS-PUB-2016-015,PERF-2013-05}, and effects due to the presence of pile-up jets. The effects of extra radiation on the balance are assessed by varying the topological selections and the overlap removal as described in Table~\ref{tab:gjetvars}. Repeating the analysis separately using $\Delta\phi(\mathrm{J},\mathrm{j})>1.2$ and $\Delta\phi(\mathrm{J},\mathrm{j})>1.6$ produces a negligible systematic shift relative to the nominal result. The effects of the photon measurement are assessed by varying the energy scale and resolution of the photon calibration, as well as by varying the measured photon purity in the purity correction. The effects of pile-up jets on the calibration are estimated by varying the JVT selection threshold for the small-$R$ jets. Lastly, the analysis is repeated with \SHERPA2.1 MC samples, in place of the nominal \PYTHIA8 samples, to assess the modelling uncertainty. As shown in Figure~\ref{fig:gamma_jet_uncertainties}, the overall combined systematic and statistical uncertainty is approximately 1\% for the \pt range from 150 to 880~\GeV. The photon energy scale uncertainty is the dominant source over the entire \pt range.
 
\begin{table}[h!]
\centering
\caption{\label{tab:gjetvars} Summary of the selection and systematic variations considered for the $\gamma$+jet direct balance analysis. The labels J$_{1}$ refers to the leading \largeR jet and j$_1$ to the leading small-$R$ jet that fulfils $\DeltaR(\mathrm{J}_1,\mathrm{j}) > 1.4$.} 
\begin{tabular}{l c c c}
\hline
Variable & Nominal Selection & Up Variation & Down Variation \\ \hline
$\pt^{\,\mathrm{j}_{1}} $ & $\max(0.1\,\ptref,15~\GeV)$ & $\max(0.15\,\ptref,20~\GeV)$ & $\max(0.05\,\ptref,10~\GeV)$ \\
$\Delta\phi(\mathrm{J}_1,\gamma)$ & $>2.8$ & $>2.9$ & $>2.7$ \\
Small-$R$ jet JVT & $>0.59$ & $>0.91$ & $>0.11$ \\
\hline
\end{tabular}
\end{table}
 
\begin{figure}[tbh]\centering
\includegraphics[width=0.80\textwidth]{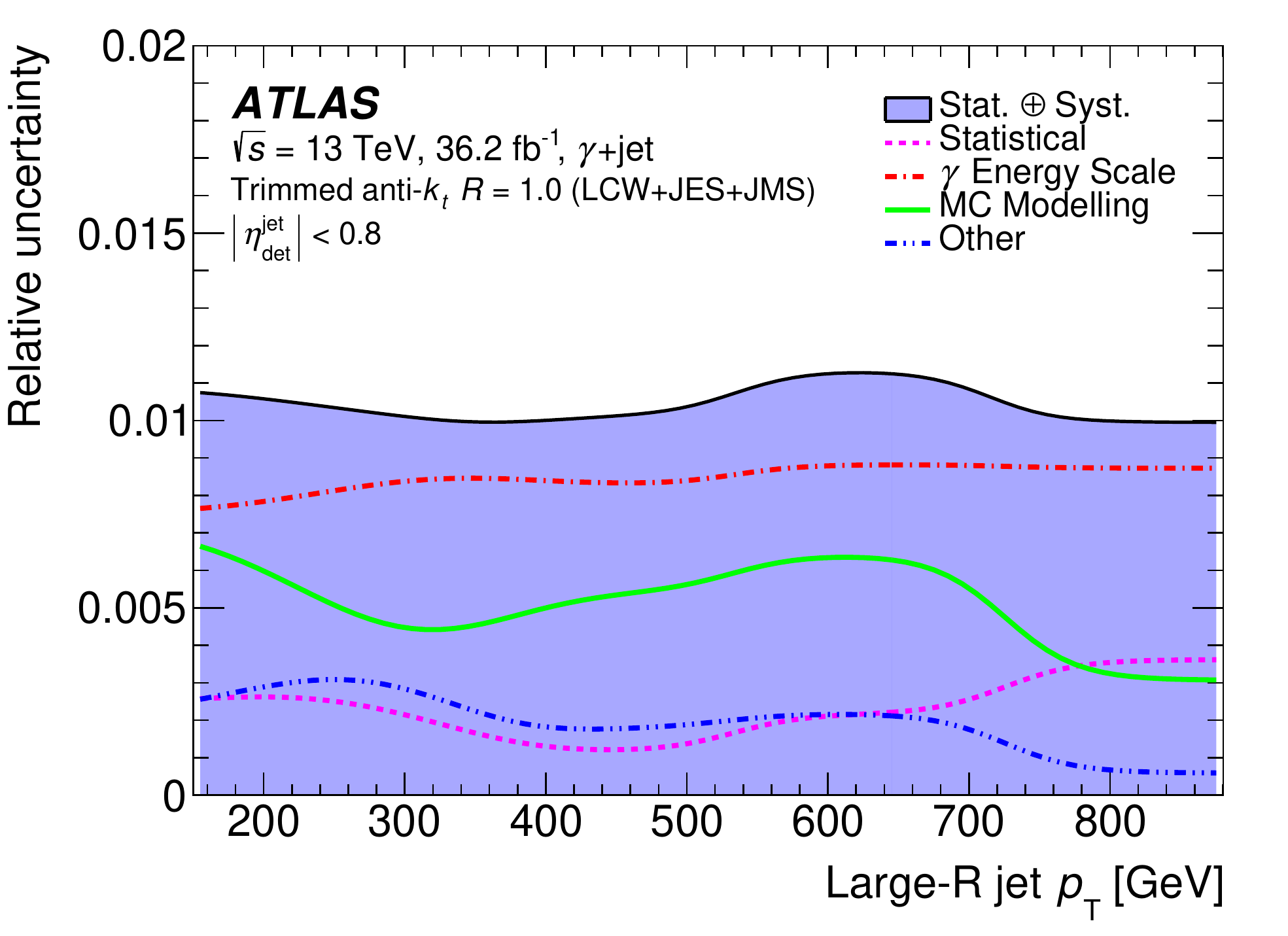}
\caption{Systematic uncertainties in the \insitu measurement of the jet energy scale obtained with the $\gamma$+jet method as a function of the \largeR jet transverse momentum \pt. The lines shown are obtained by smoothing a binned representation of these uncertainties using a sliding Gaussian kernel.}
\label{fig:gamma_jet_uncertainties}
\end{figure}
\clearpage
 
\subsection{Multijet balance} \label{sec:mjbalance}
The $Z$+jet and $\gamma$+jet techniques provide precise constraints on the jet energy scale for jets with \pt{} up to 1~\tev. The energy scale of higher-$\pt$ \largeR jets is measured using multijet events. A schematic representation of the event topology used in this method is shown in Figure~\ref{fig:balance_fig}(c). The multijet balance (MJB) method takes advantage of events where an energetic \largeR jet is balanced against a system that consists of multiple lower-\pt{} jets.
 
For the calibration of \largeR jets the reference $\pt^{\mathrm{recoil}}$ is obtained as the four-vector sum of calibrated \smallR \akt jets. The transverse momentum balance is
\begin{equation*}
\RMJB = \left \langle \frac{\pt^{\mathrm{J}}}{\pt^{\mathrm{recoil}}} \right \rangle,
\end{equation*}
where $\pt^\mathrm{J}$ is the transverse momentum of the leading
\largeR jet and $\pt^{\mathrm{recoil}}$ is the magnitude of the vectorial sum of the
transverse momenta of the recoil system of small-$R$ jets. The average value of the ratio is taken to be the mean value of a Gaussian fit. The value of \RMJB is measured
in data and determined in simulation in several bins of $\pt^{\mathrm{recoil}}$. The data-to-simulation double ratio $\RMJB^{\mathrm{data}}/\RMJB^{\mathrm{MC}}$ allows estimation of the response for high-$\pt$ jets.
 
Events are selected using single small-$R$ jet triggers. Bins of $\pt^{\mathrm{recoil}}$ are defined to correspond to a given fully efficient single small-$R$ jet trigger. The triggers used for 200~\GeV~$<$ $\pt^{\mathrm{recoil}} <$~550~\GeV~are
prescaled, whereas an unprescaled jet trigger is used for $\pt^{\mathrm{recoil}} >$ 550~\GeV.
 
The event selection is summarized in Table~\ref{tab:selection_multijet}.
For small-$R$ jets with $\pt{} < 60$~\GeV~within $|\eta|<2.4$, the JVT selection is applied to suppress pile-up jets.
The \largeR probe jet is required to have $|\eta_{\mathrm{det}}| <$ 0.8,
while the small-$R$ jets that constitute the recoil system are required to have
$|\eta_{\mathrm{det}}| <$ 2.8 and $\pt>25$~\GeV.
To select events with multijet recoil systems, the leading jet in the recoil system
($\mathrm{j_1}$) is allowed to have no more than 80\% of the total
transverse momentum of the recoil system. This selection ensures that the recoil system consists of
several jets with lower \pt than the \largeR jet, which are each well-calibrated by small-$R$ jet \insitu techniques~\cite{PERF-2016-04}.
The angle $\alpha$ in the azimuthal plane between the leading \largeR jet and the vector defining the recoil system is required to
satisfy $|\alpha-\pi| <$ 0.3. The $\Delta R$ distance $\beta$ between the leading \largeR jet and the nearest small-$R$ jet from
the recoil system is required to be greater than 1.5. The simulation calibration and $\eta$-intercalibration described in Sections~\ref{sec:objects}~and~\ref{sec:etacal} are applied to the \largeR jets studied using this technique.

\begin{table}[h!]
\centering
\caption{\label{tab:selection_multijet} Summary of the event selection and systematic variations considered for the multijet direct balance analysis. The label j$_i$ refers to the $i$th leading small-$R$ jet.}
\begin{tabular}{l c c c}
\hline
Variable & Nominal Selection & Up Variation & Down Variation \\ \hline
Separation angle ($\alpha$) & $|\alpha-\pi| <$ 0.3 & $|\alpha-\pi| <$ 0.4 & $|\alpha-\pi| <$ 0.2 \\
$\Delta R$ separation ($\beta$) &  $ > $1.5 &  $ > $1.9 &  $ > $1.1 \\
$\pt^{\mathrm{j_1}}/\pt^{\mathrm{recoil}}$ & $< 0.8$ & $< 0.9$ &	$< 0.7$  \\
Recoil system minimum \pt & 25~\GeV~& 30~\GeV~& 20~\GeV~\\
\hline
\end{tabular}
\end{table}
 
Figure~\ref{fig-MJB-recoilR4} shows the distribution of \RMJB as a function of the \largeR jet \pt. The balance in data decreases from approximately 1.01 at \pt = 300~\GeV~to about $0.99$ for jets with \pt = 2~\tev. The simulation shows a similar downward trend. The response in simulations is 2\% higher than in data, consistent with the findings of the other methods where they overlap.
 
The total uncertainty in the \RMJB measurement is approximately $\pm2\%$ or
lower for \pt $<$ 2~\tev.
The uncertainty in the energy scale of the jets of the recoil \insitu procedure is propagated through the \largeR MJB procedure. Uncertainties associated with high-\pt jets in the recoil system which lie beyond the region covered by the $R=0.4$ \insitu analyses are derived from measurements of the calorimeter response to isolated single charged particles, which are also propagated through this \largeR jet analysis to provide coverage at the highest values of jet \pt (> 1~\TeV)~\cite{PERF-2015-05}. No assumption is made about the flavour of the recoil jets (originating from a gluon, a light quark, or a heavy-flavour quark). This lack of knowledge is a source of systematic uncertainty. The uncertainty in the multijet-balance observable due to the jet flavour response is evaluated using a correlated propagation of the small-$R$ jet flavour response uncertainties, i.e.\ all jets are shifted simultaneously. 
 
In addition to the jet calibration and uncertainties in the reference scale, the event selection criteria and the modelling in the event generators directly affect the \pt balance used to obtain the multijet-balance results. The impact of the event selection criteria is investigated by shifting each event selection criterion up and down by a specified amount and observing the change in the multijet-balance variable. Using an approach to systematic uncertainties similar to that in the small-$R$ \insitu analysis, the transverse momentum threshold for recoil jets is shifted by $\pm$ 5~\GeV, the $\pt^{\mathrm{j1}}/\pt^{\mathrm{recoil}}$ is shifted by $\pm$ 0.1, the angle $\alpha$ is shifted by $\pm$ 0.1, and $\beta$ is shifted by $\pm$ 0.4. The uncertainty due to modelling of multijet events in simulations is estimated from the largest difference between the multijet-balance results obtained from the nominal \PYTHIA8 simulation and those obtained from \SHERPA v2.1 and \HERWIG 7. Figure~\ref{fig-systematics-recoilR4} shows the breakdown of the fractional uncertainties in the jet energy scale derived from this method. Various uncertainties propagated from the reference jet system dominate the measurement across the entire \pt range.
 
\begin{figure}[tbh]
\centering
\includegraphics[width=0.66\textwidth]{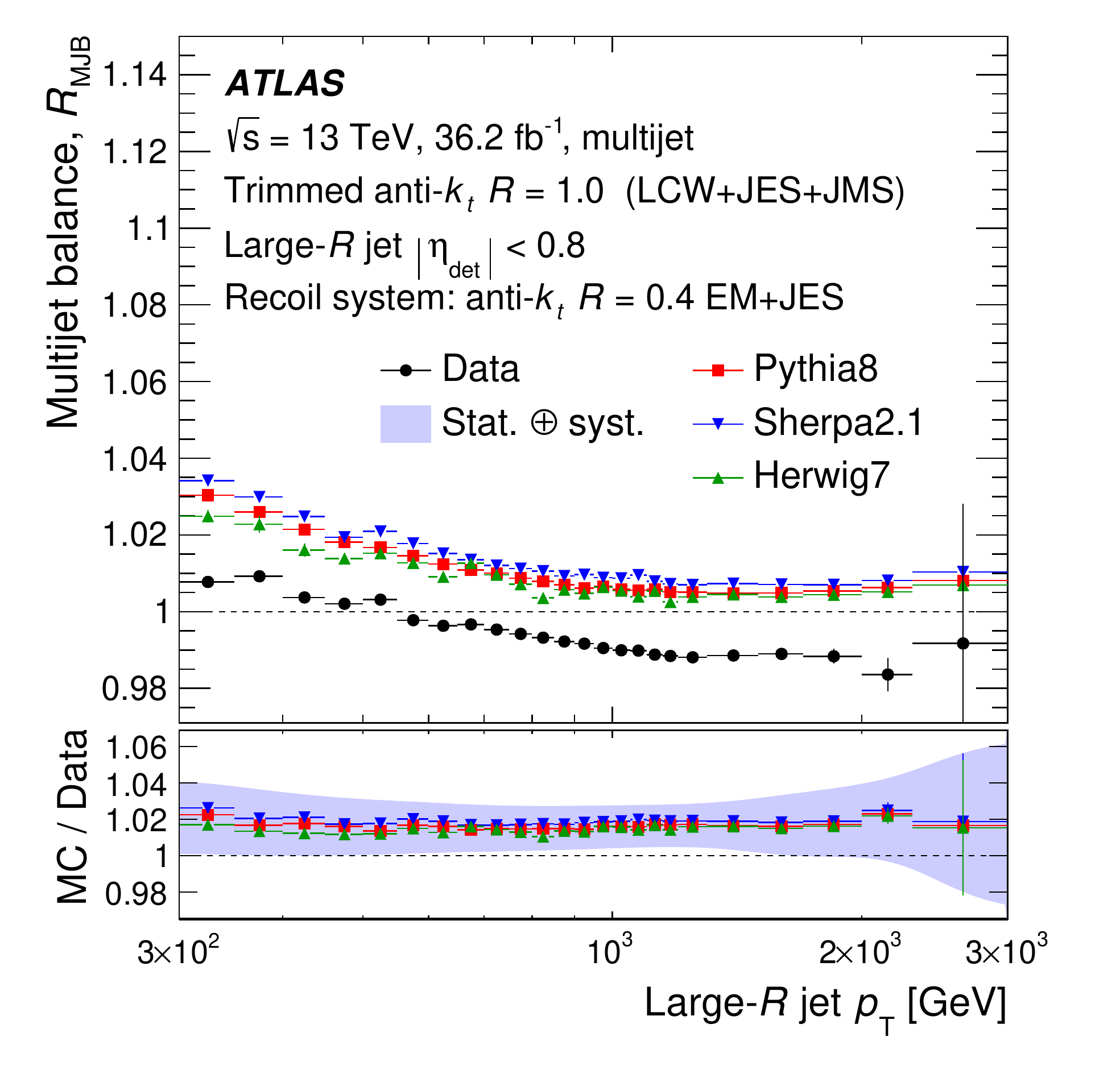}
\caption{
Mean transverse momentum balance \RMJB for leading-$\pt{}$ \largeR jets ($|\eta| <$ 0.8) balanced against a system of at least two small-$R$ jets ($\pt{} \geq 25$~\GeV, $|\eta| <$ 2.8) as a function of the \largeR jet transverse momentum \pt. The measured balance is compared with the prediction of Monte Carlo simulations based on the event generators \PYTHIA8, \SHERPA2.1, and \HERWIG 7. Below, the ratio of response measurements in data and simulation is presented. The shaded band indicates the total uncertainty of the measurement, described in detail in the text. For each \pTref{} bin, the measured \RDB{} is plotted against the average jet \pt{} of the bin. The horizontal error bars gives an indication of the the width of the associated \pTref{} bin.}
\label{fig-MJB-recoilR4}
\end{figure}
 
\begin{figure}[tbh]
\centering
\includegraphics[width=0.66\textwidth]{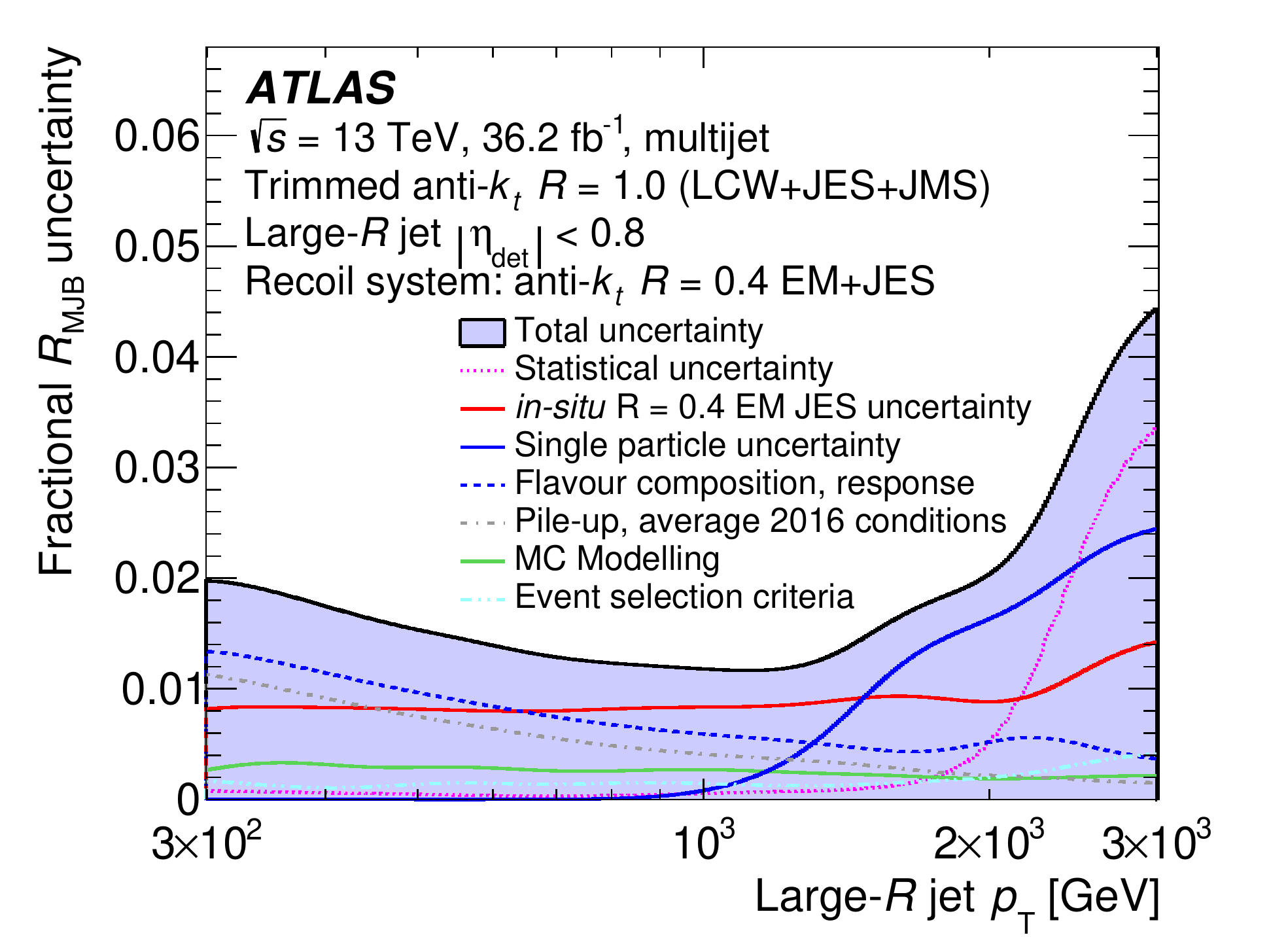}
\caption{
The fractional uncertainty in \RMJB as a function of the \largeR jet transverse momentum \pt. The lines shown are obtained by smoothing a binned representation of these uncertainties using a sliding Gaussian kernel.
}
\label{fig-systematics-recoilR4}
\end{figure}
\clearpage

\section{\Insitu jet mass calibration} 
\label{sec:insitumass}
 
In this section, two methods to derive an \insitu calibration for the \largeR jet mass are presented. The first method, known as the \Rtrk method, relies on the tracker to provide an independent measurement of the jet mass scale and its associated uncertainty. The second method, known as forward folding, fits the mass peaks and jet mass response of the $W$ boson and top quark to measure the relative energy and mass scales and resolutions between data and simulations. Both measurements are performed after applying the \insitu calibration for the energy scale, which also affects the jet mass scale. The results in this section are combined into a global jet mass calibration, detailed in Section~\ref{sec:combination}.
 
\subsection{Calorimeter-to-tracker response ratios} \label{sec:rtrk}
The calorimeter-to-tracker response double-ratio method (or \Rtrk method) is built around the fact that the ATLAS detector provides two independent measurements of the properties of the same jet from the calorimeter and the tracker~\cite{PERF-2012-02}. Jets formed from inner detector tracks only take into account the hits from their charged-particle constituents. Calibrated jets formed from energy depositions within the calorimeter provide a measure of the properties from the full shower. The average calorimeter-to-track jet response
\begin{equation*}
\Rtrk = \left \langle \frac{\pt^{\mathrm{calo}}}{\pt^{\mathrm{track}}}\right \rangle
\end{equation*}
is proportional to the average calorimeter-to-truth jet response. Therefore, a comparison of the double ratio of \Rtrk in simulations and data provides a way to validate the modelling of \largeR jet properties \insitu. The ratio of $\Rtrk$ values determined in data and simulations should be equal to unity for well-modelled observables. Any deviation from this expectation can be taken as a scale uncertainty in the measurement. This method is versatile and allows the determination of uncertainties for several variables, such as the \pt, mass, and substructure information of \largeR jets. Moreover, the dijets process provides a very large sample, such that the analysis can be performed in a large number of \pt{} and mass or $m/\pt$ regions.
 
Figure~\ref{fig:rtrk_pt} shows \Rtrk as a function of the \largeR jet \pt in dijet events for data and several simulation samples. The maximum spread between the two generators and three tracking variations that assume three different types of mismodelling (resolution~\cite{ATL-PHYS-PUB-2015-018}, efficiency within dense environments~\cite{PERF-2015-08}, and alignment~\cite{ATL-PHYS-PUB-2018-003}) is about 8\%. A steady increase in the calorimeter-to-track jet response \Rtrk with increasing \largeR jet \pt is observed, going well beyond the expected ratio of the total and charged transverse momenta of a jet, caused by inefficiencies in the tracker response at high jet \pt.
 
Figure~\ref{fig:rtrk_pt_unc} shows a breakdown of the uncertainties in the \largeR jet \pt derived from this method for the transverse momentum for \largeR jets with values of $m/\pt \approx 0.2$. The main source of uncertainty across the entire \pt range originates from differences between data and the nominal Monte Carlo generator considered in this study. As this uncertainty was expected to be large, the \Rtrk method is neither included in the \insitu JES combination nor used as a source of systematic uncertainty for the JES of \largeR jets. Rather, the \Rtrk \pt results are used as an independent cross-check to validate the JES calibration techniques.
 
\begin{figure}[tbh]\centering
\includegraphics[width=0.75\textwidth]{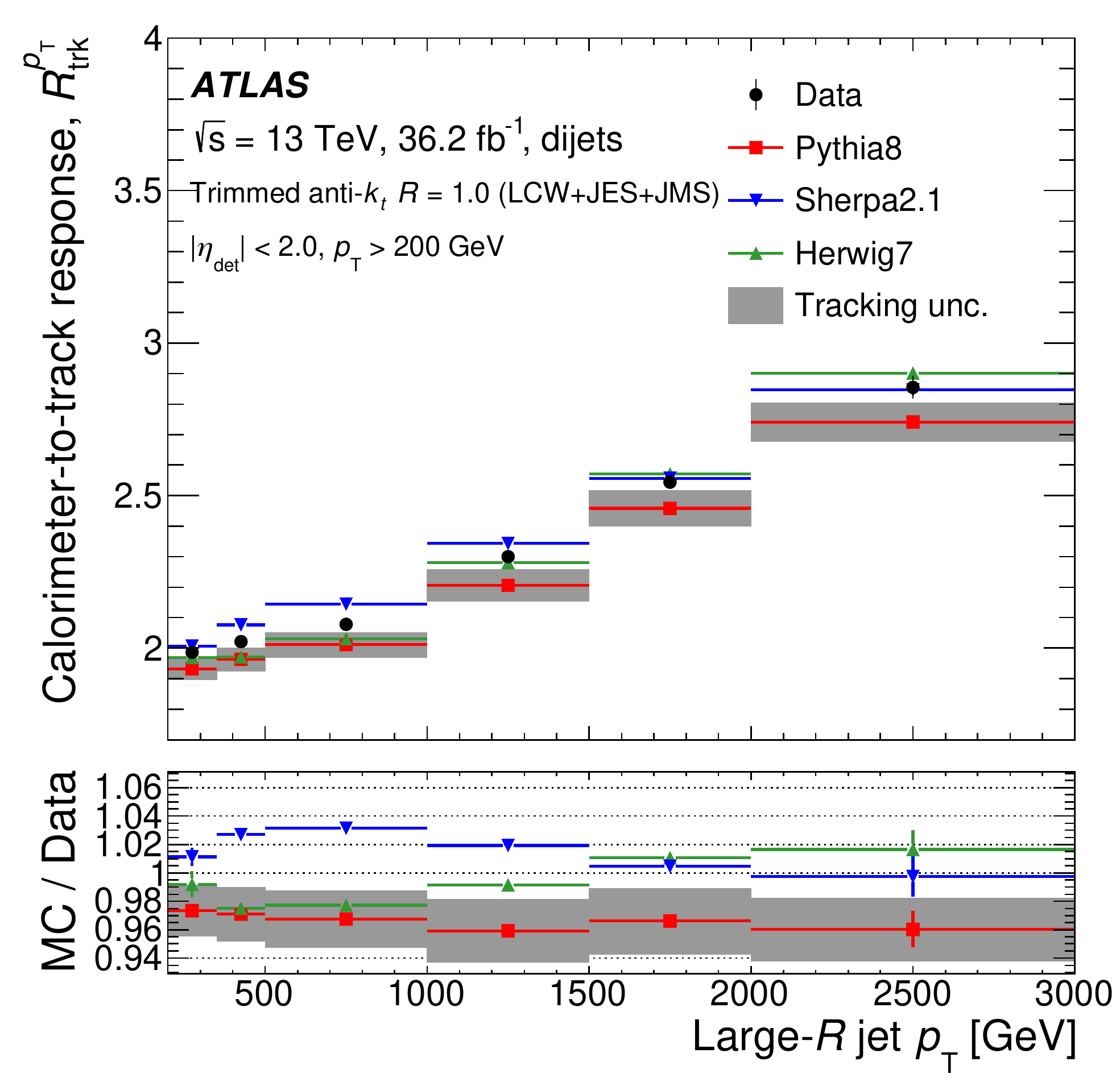} 
\caption{Measurement of $R_{\mathrm{trk}}^{\pt}$ as a function of the \largeR jet transverse momentum \pt for \largeR jets with $m/\pt=0.2$. The \largeR jet \pt is corrected using the simulation calibration, $\eta$-intercalibration, and a combination of \insitu direct balance techniques. Data are compared with three generators and with three tracking variations for the default generator \PYTHIA8 (shown as a band around these points). The double ratio of $R_{\mathrm{trk}}^{\pt}$ measured in simulations and data is shown in the lower panel.}
\label{fig:rtrk_pt}
\end{figure}
 
\begin{figure}[tbh]\centering
\includegraphics[width=0.75\textwidth]{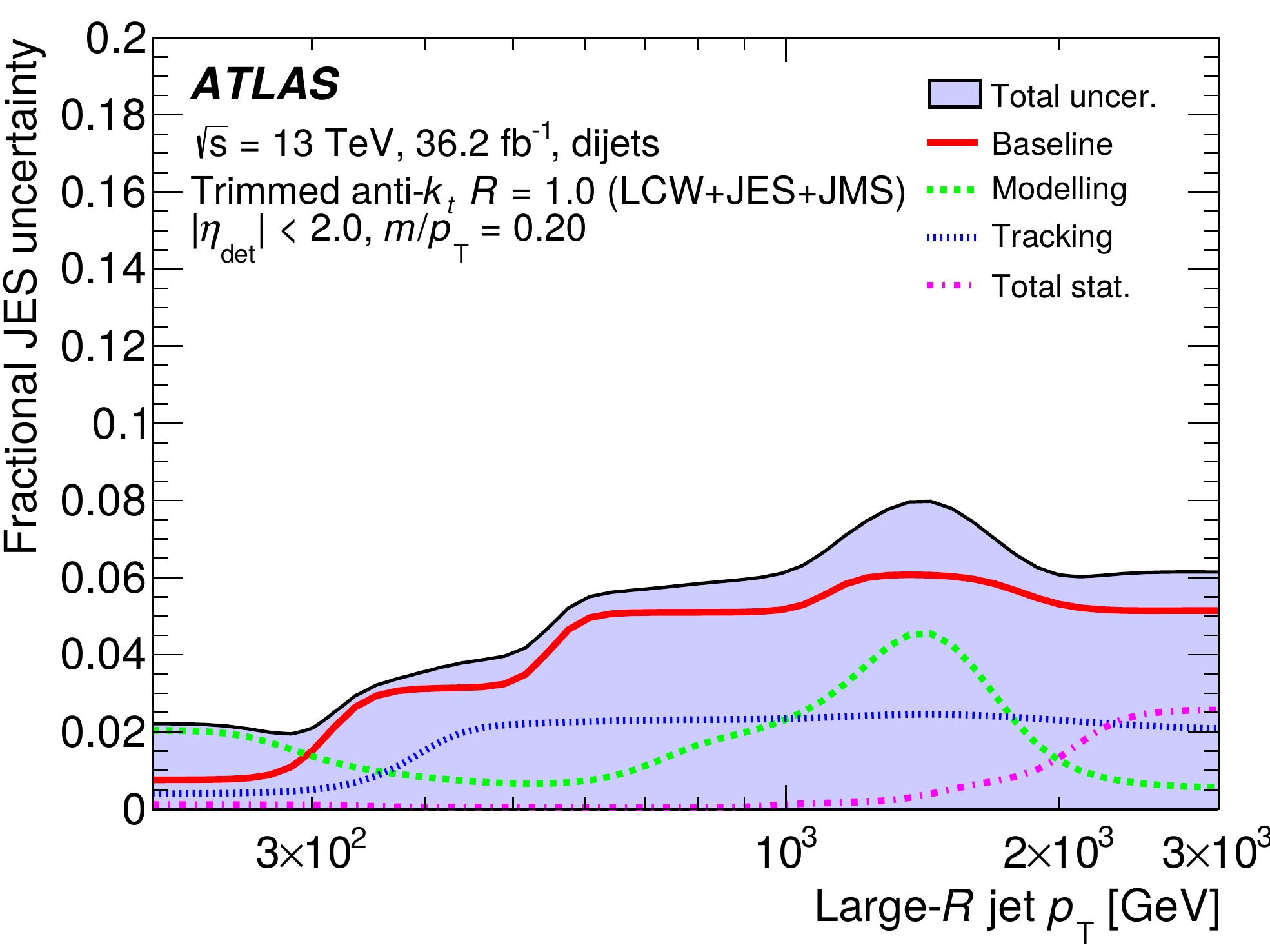}
\caption{The total uncertainty in the relative jet energy scale in data and simulations associated with the $R_{\mathrm{trk}}$ method is plotted as a function of jet transverse momentum \pt. The \largeR jet \pt is corrected using the simulation calibration, $\eta$-intercalibration, and a combination of \insitu direct balance techniques. The contributions from several sources are indicated. The baseline uncertainty represents the deviation of the double ratio from unity for the baseline simulations. The lines shown are obtained by smoothing a binned representation of these uncertainties using a sliding Gaussian kernel.}
\label{fig:rtrk_pt_unc}
\end{figure}
 
The same method is also applied to the \largeR jet calorimeter mass, and is shown in Figure~\ref{fig:rtrk_m}. The largest difference between the considered generators is $\sim$2-3\%. Figure \ref{fig:rtrk_m_unc} shows the various uncertainties in the \largeR jet mass derived from the \Rtrk mass response for \largeR jets with $m/\pt=0.2$. Again, the main source of uncertainty originates from differences between data and the nominal simulation.
 
\begin{figure}[tbh]\centering
\includegraphics[width=0.75\textwidth]{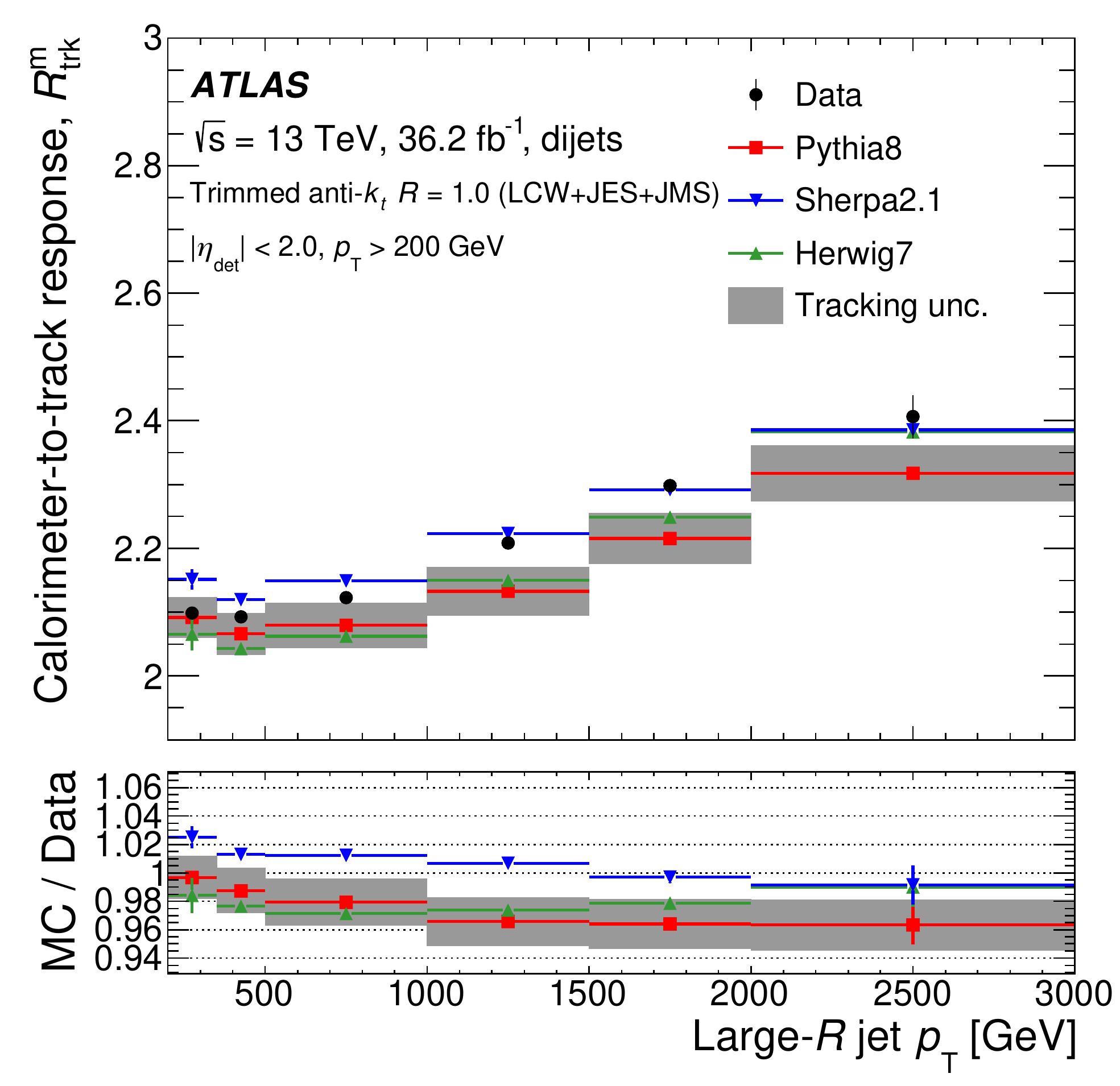} 
\caption{Measurement of $R_{\mathrm{trk}}^m$ as a function of the \largeR jet transverse momentum \pt for \largeR jets with $m/\pt=0.2$. The \largeR jet \pt is corrected using the simulation calibration, $\eta$-intercalibration, and a combination of \insitu direct balance techniques. Data are compared with three generators and with three tracking variations for the default generator \PYTHIA8 (shown as a band around these points). The double ratio of $R_{\mathrm{trk}}^m$ measured in simulations and data is shown in the lower panel.}
\label{fig:rtrk_m}
\end{figure}
 
\begin{figure}[tbh]\centering
\includegraphics[width=0.75\textwidth]{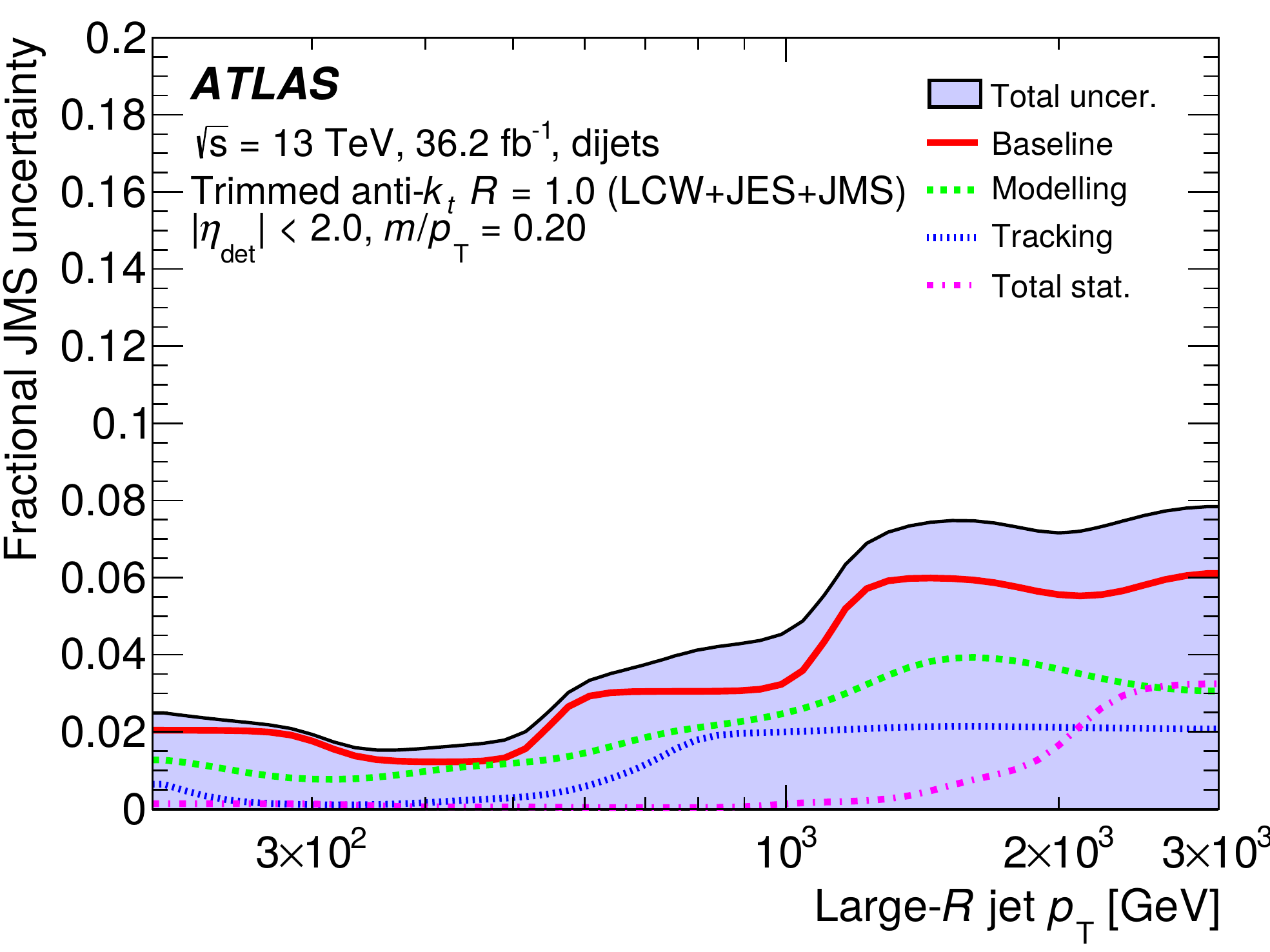}
\caption{The total uncertainty in the relative jet mass scale between data and simulation associated with the $R_{\mathrm{trk}}$ method is plotted as a function of jet transverse momentum \pt for \largeR jets with $m/\pt=0.2$. The \largeR jet \pt is corrected using the simulation calibration, $\eta$-intercalibration, and a combination of \insitu direct balance techniques. The contributions from several sources are indicated. The baseline uncertainty represents the deviation of the double ratio from unity for the baseline simulations. The lines shown are obtained by smoothing a binned representation of these uncertainties using a sliding Gaussian kernel.}
\label{fig:rtrk_m_unc}
\end{figure}
 
The $R_{\mathrm{trk}}$ method can also be used to study the topology dependence of the response modelling. The double ratio is constructed in two event samples, with different jet flavours (jets originating from light quarks or gluons and jets containining a hadronic top quark decay). The dijet sample used for Figure~\ref{fig:rtrk_pt} is dominated by gluon jets at low transverse momenta, while at higher momenta the fraction of light-quark jets in the sample increases. The $t\bar{t}$ sample of Section~\ref{sec:forwardfolding} is enriched in \largeR jets that contain a complete high-\pt object's decay (either a top quark or $W$ boson). In Figure~\ref{fig:rtrk_topology} the double ratios of the two samples are compared for jet \pt and jet mass. The jets in the samples correspond to the same pseudorapidity range $|\eta|<$ 2.0 and the same \pt and jet mass intervals. In both samples, the double ratio is constructed with the nominal simulation events, which rely on \PYTHIA 8 for hadronization. As systematic uncertainties are expected to partially cancel out, only statistical uncertainties are shown.
 
\begin{figure}[tbh]\centering
\subfigure[]{
\includegraphics[width=0.48\textwidth]{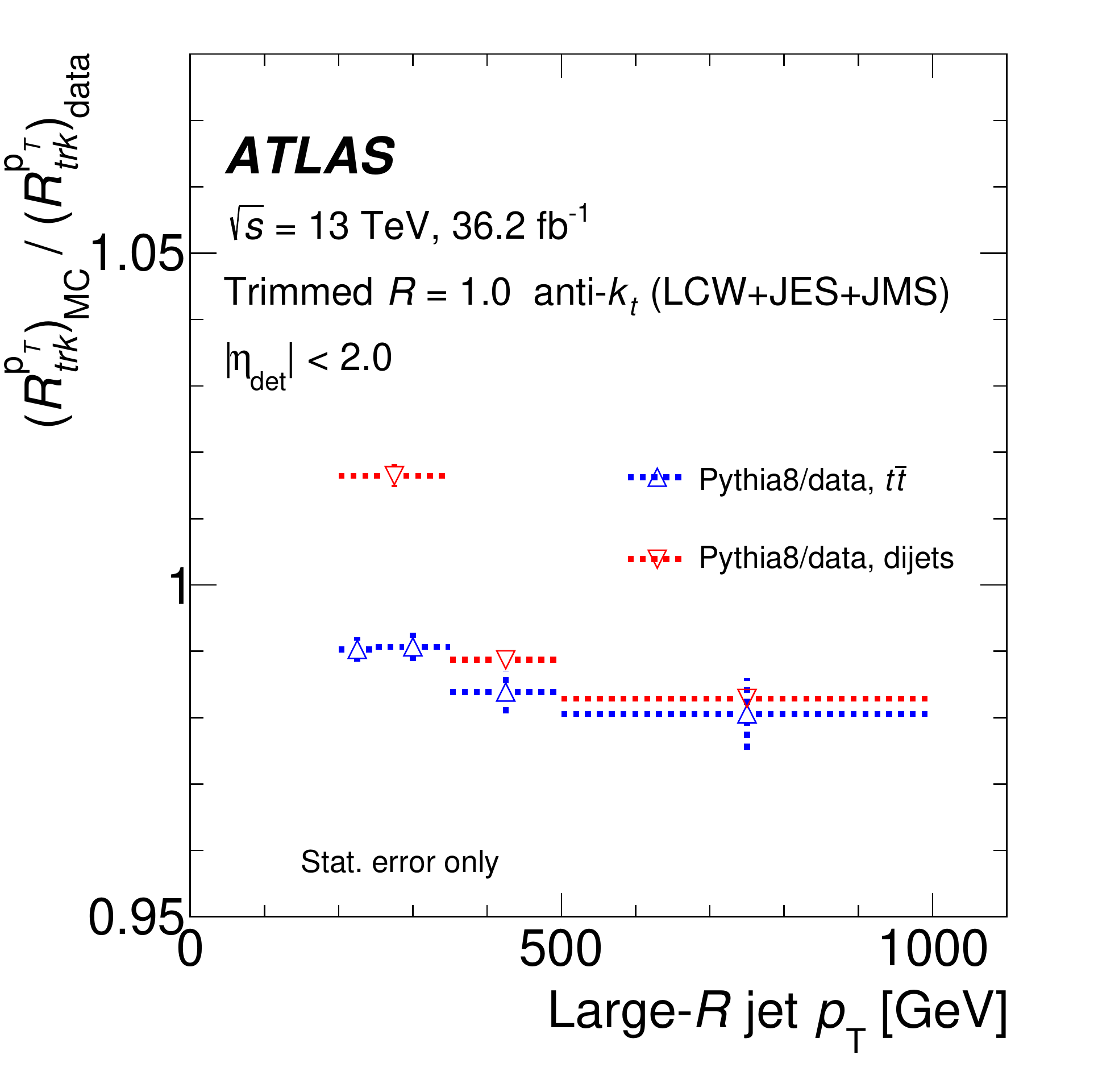} }
\subfigure[]{
\includegraphics[width=0.48\textwidth]{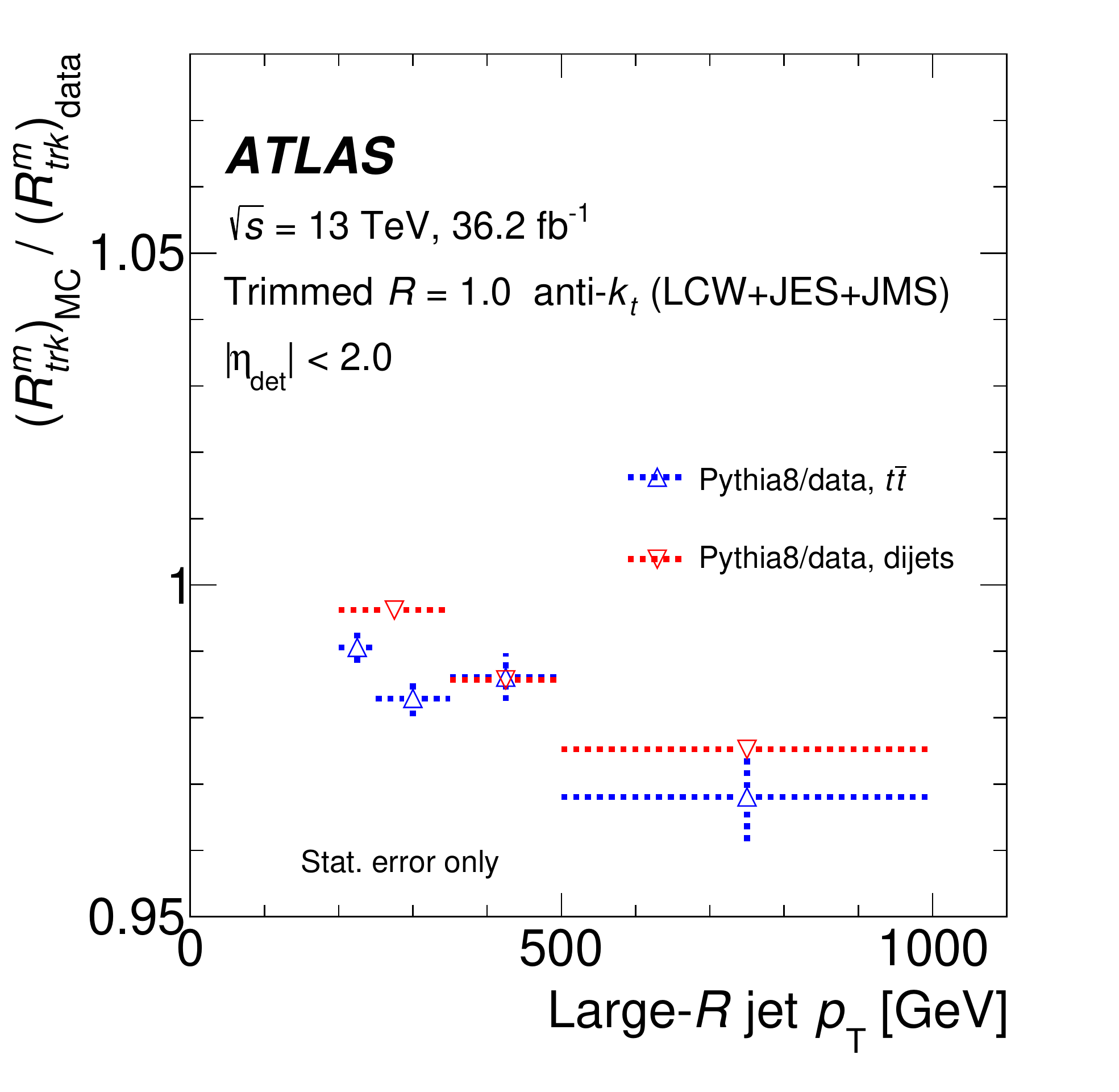} }
\caption{The simulation/data ratio of $R_{\mathrm{trk}}$ for (a) \largeR jet \pt and (b) calorimeter mass as a function of the \largeR jet transverse momentum \pt. Two sets of results are derived from a dijet sample, dominated by light-quark and gluon jets, and a \ttbar sample, where the \largeR jets contain a boosted $W$ boson or top quark. The \largeR jet \pt is corrected using the simulation calibration, $\eta$-intercalibration, and a combination of \insitu direct balance techniques. The jets in both samples correspond to the same pseudorapidity range $|\eta|<$ 2.0 and the same \pt{} and jet mass intervals. The double ratio is constructed with the nominal \PYTHIA 8 samples for dijet events and \POWHEG+\PYTHIA 8 samples for the \ttbar sample. The error bars indicate statistical uncertainties.}
\label{fig:rtrk_topology}
\end{figure}
 
There is a mild tension between the double-ratio results from the two samples. The double ratio in the \ttbar sample is systematically somewhat higher than the equivalent result in the dijet sample. The difference is typically 1\% or less, except in the first bin of the double ratio for jet mass. This is significant compared to the statistical uncertainties but is small in comparison with the modelling uncertainties of the $R_{\mathrm{trk}}$ method. Some properties of these two jet populations differ, such as the distribution of their $m/\pt$ and their flavour composition, and so it is not expected that the modelling uncertainties will cancel out exactly. No additional uncertainty is assigned to account for the topology dependence.
\clearpage
 
\subsection{Forward folding} \label{sec:forwardfolding}
A high-purity signal sample of \largeR jets with high-\pt, hadronically decaying $W$ bosons and top
quarks is obtained by selecting \ttbar events in the lepton+jets final state, where a
hadronically decaying top quark balances one which decays to a leptonically decaying $W$ boson and $b$-quark.
This sample is used to measure the response for jets in signal-like topologies
which contain jets consisting of multiple regions of high energy density~\cite{PERF-2015-03,PERF-2015-04}.
The jet mass response is determined by fits to the $W$ boson and top quark mass peaks in the
\largeR jet invariant mass distribution of the hadronically decaying top quark candidate.
 
The event selection is based on the ATLAS search for $t\bar{t}$
resonances~\cite{EXOT-2015-04} and is summarized in
Table~\ref{tab:ttbar_selection}. It requires a central high-\pt,
isolated muon, and significant missing transverse momentum (\MET)~\cite{PERF-2016-07}. The $W$ boson transverse
mass obtained from \mbox{$m_{\text{T}}^2=2p_{\text{T}}^{\text{lep}}E_{\text{T}}^{\text{miss}}(1-\cos(\Delta\phi))$}, where
$\Delta\phi$ is the azimuthal angle between the charged lepton and the
direction of the missing transverse momentum, must be greater
than 60~\GeV. A multivariate $b$-tagging algorithm is used to identify $R=0.4$ jets which originate
from the decays of $b$-quarks based on information about the impact parameters of inner detector tracks matched to the jet, the presence of displaced secondary vertices, and the reconstructed flight paths of $b$- and $c$-hadrons inside the jet; the 70\% signal tagging efficiency working point is used here~\cite{PERF-2016-05}.
 
\begin{table}[h]
\centering
\caption{Summary of the event selection for the top quark events decaying
into lepton+jets, to be used for top and $W$ mass calibration. The \pt bins into which
events are divided are also shown.}
\begin{tabular}{ c c c }
\hline
Object & Selection & Description \\
\hline
 
Muon ($\mu$) & Single-muon triggers & Trigger \\
& $\pt^{\mu} > 25$ GeV, $\left|\eta \right| < 2.5$ & Preselection \\
& \texttt{tight} muon ID                           & Identification \\
& $\Delta R_{\mu, \mathrm{jet}} > 0.4$      &  Isolation \\
\hline
\MET       & \MET $>$ 20~\GeV,  $m_\text{T} > $ 60~\GeV~ \\
\hline
Small-$R$ Jets j & \pt $>$ 25~\GeV~& Jet selection \\
& at least one jet with $\Delta R (j,\mu) <$ 1.5 &  Boosted top decay \\
& at least one $b$-tagged jet & Flavour tagging   \\
\hline
Large-$R$ Jets & $\Delta R_{\mathrm{j},\mathrm{J}} > $ 2  & Opposite hemispheres \\
\hline
\multirow{2}{*}{\pt~$\in~[200, 250]~\GeV,~[250, 350]~\GeV$ } & $\Delta R_{b,\mathrm{J}} > 1$ & Boosted $W$ sample,  \\
& $\Delta R_{b,\mathrm{J}} > 1$ &  $b$-jet veto \\
\hline
\multirow{2}{*}{\pt~$\in~[350, 500]~\GeV,~[500, 1000]~\GeV$}  & $\Delta R_{b,\mathrm{J}} < 1$ & Boosted top,   \\
& $\Delta R_{b,\mathrm{J}} < 1$ &  $b$-jet matched \\
 
\hline
\end{tabular}
\label{tab:ttbar_selection}
\end{table}

The \largeR jet mass distribution of the highest-\pt \largeR jet in the hemisphere opposite to the charged
lepton is shown in Figure~\ref{fig:ttbar} for two categories of events, and for both the calorimeter-only and track-assisted jet masses.
For \largeR jets with intermediate \pt (200~\GeV~$ < \pt < $~350~\GeV), in
Figures~\ref{fig:ttbar}(a) and \ref{fig:ttbar}(c), the decay products of the hadronic $W$ boson
are captured in a single \largeR jet. For high-\pt jets with \pt $>$
350~\GeV, in Figures~\ref{fig:ttbar}(b) and \ref{fig:ttbar}(d), the complete hadronic top decay
is captured in the main \largeR jet. The high-\pt $W$ boson and top quark
topologies are confirmed by, respectively, vetoing or requiring a $b$-tagged
small-$R$ jet that overlaps with the \largeR jet.
 
\begin{figure}[tbh]
\centering
\subfigure[]{
\includegraphics[width=0.45\textwidth]{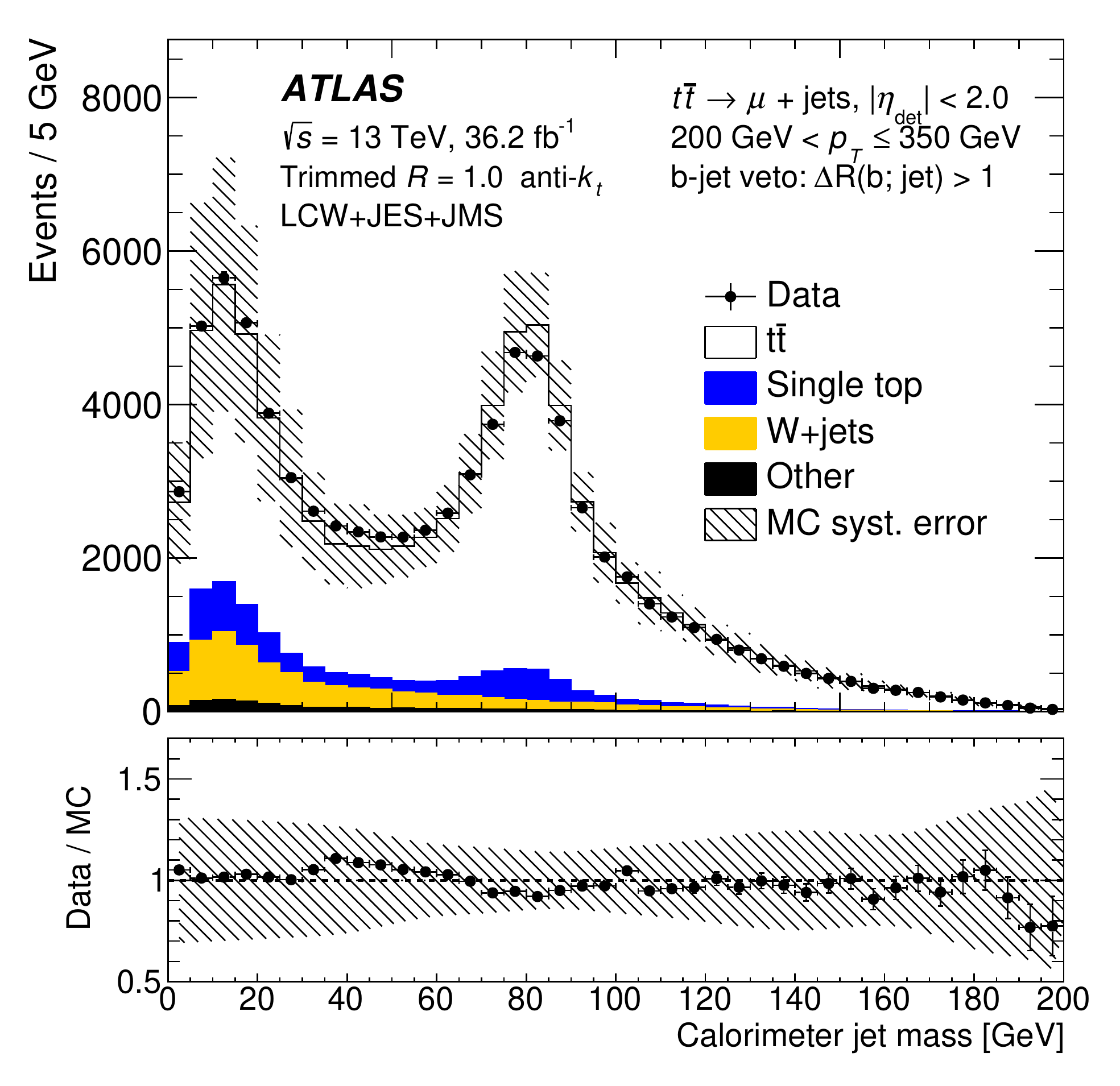}
}
\subfigure[]{
\includegraphics[width=0.45\textwidth]{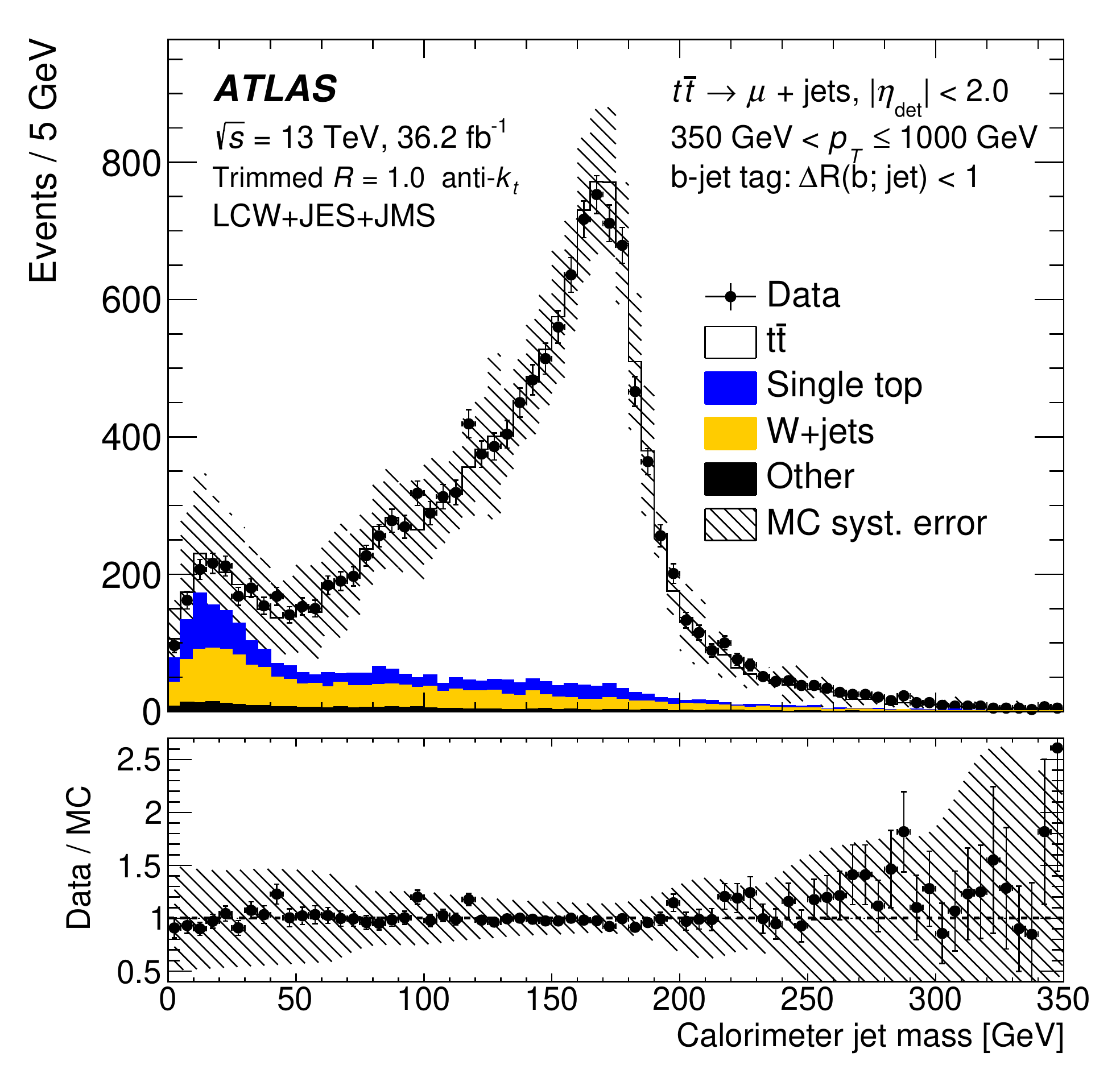}
}
\subfigure[]{
\includegraphics[width=0.45\textwidth]{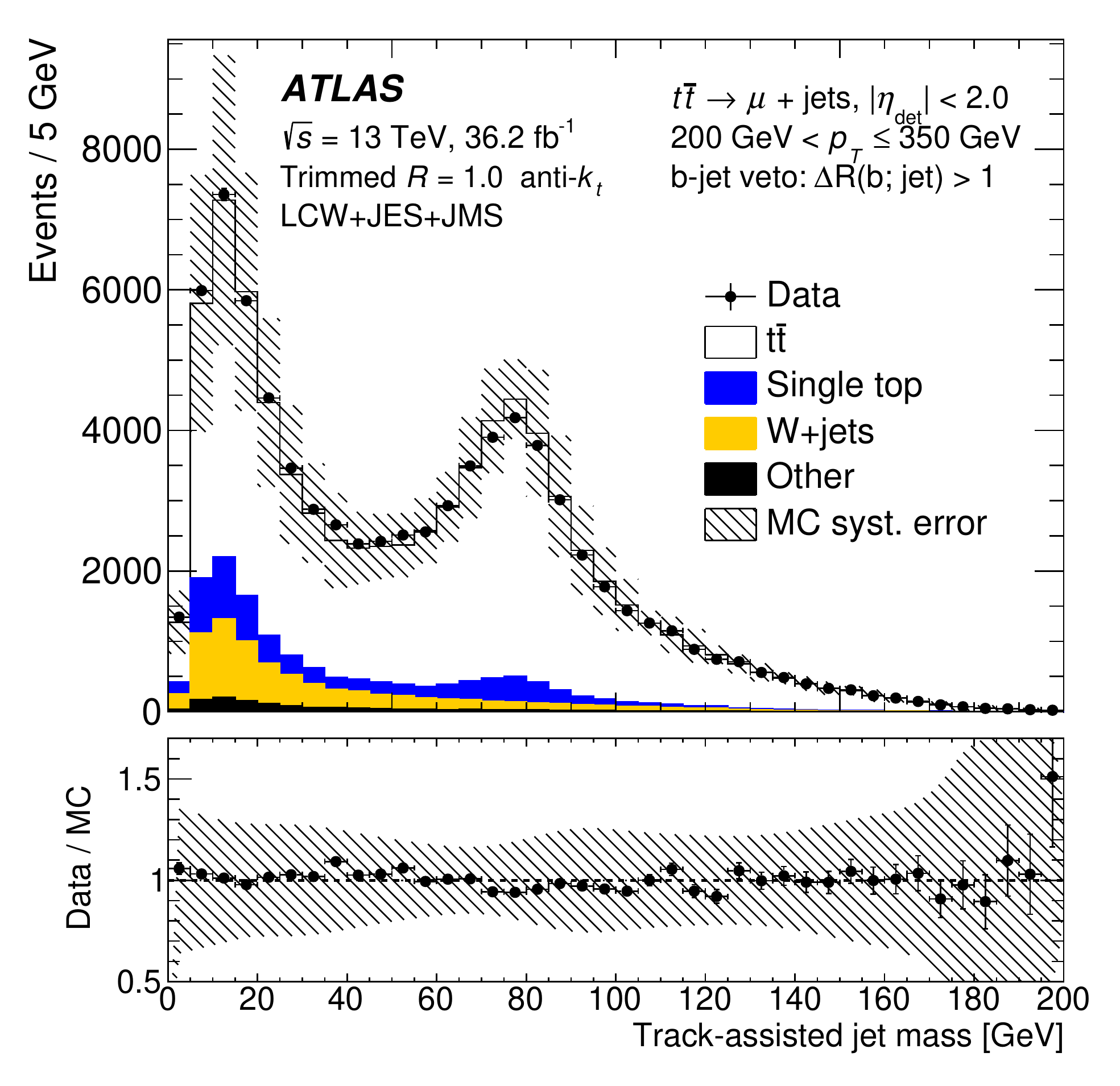}
}
\subfigure[]{
\includegraphics[width=0.45\textwidth]{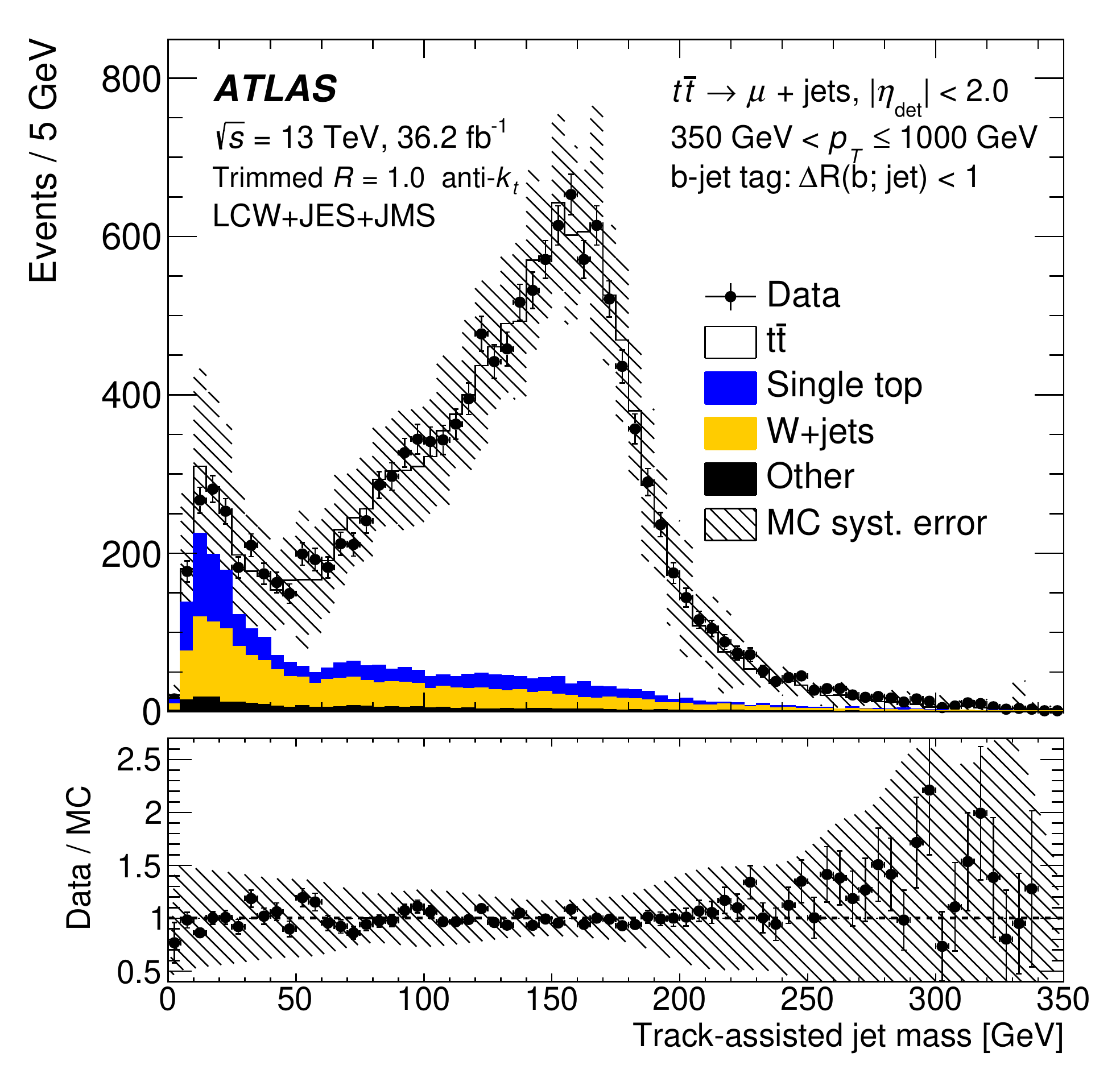}
}
\caption{The distributions of the jet invariant mass for \largeR jets in samples enriched in (a,c) boosted $W$ bosons and (b,d) boosted top quarks. The distribution of the calorimeter mass is shown in~(a)~and(b), and the distribution of the track-assisted mass of the same jets is shown in~(c)~and~(d). The \largeR jet transverse momentum \pt is corrected using the simulation calibration, $\eta$-intercalibration, and a combination of \insitu direct balance techniques. The template estimated from simulations is rescaled to match the observed yield. The lower panels display the data-to-simulation ratio. The error bars on the data represent the statistical uncertainty. The dashed uncertainty band on the simulation template includes the systematic uncertainties due to signal and detector modelling.}
\label{fig:ttbar}
\end{figure}
 
The track-assisted mass (Eq.~(\ref{eq:track_assisted})) is obtained
by scaling the invariant mass of the charged-particle jet
by the ratio of the \pt of the calorimeter and charged-particle jets.
The resulting jet mass distributions in the $W$ boson and top quark \largeR jet samples
are presented in Figures~\ref{fig:ttbar}(c) and \ref{fig:ttbar}(d). The selection for
this second set of plots is entirely based on the properties of the
matched calorimeter jet, such that plots~(a)~and~(c) and plots~(b)~and~(d) are populated by the same jets.
The track-assisted mass peaks in (c)~and~(d) are slightly broader than the
calorimeter-based mass peaks in (a)~and~(b) for
\largeR jets with a
large invariant mass and relatively low \pt.
 
The position and shape of the mass peaks provide information about the \largeR jet mass scale and resolution.
Values for the ratio of the response in data and simulations ($s = R^{m}_{\mathrm{data}}/R^{m}_{\mathrm{MC}}$) and the ratio
of the resolution in data and simulations ($r = \sigma^{m}_{\mathrm{data}}/\sigma^{m}_{\mathrm{MC}}$) are extracted from
the jet mass spectrum. These two parameters are extracted simultaneously in a fit referred to as forward folding~\cite{ATLAS-CONF-2016-008}. This method produces simulation-based predictions of the jet mass spectrum with variable response and resolution. This is achieved by folding particle-level jets with a response function. The default response function
is taken from the nominal simulations. The predicted detector-level jet mass spectrum
for arbitrary values of $s$ and $r$ is obtained by modifying
the response function by
\begin{equation*}
m^{\mathrm{fold}} = s \, m^{\mathrm{reco}} + \left( m^{\mathrm{reco}} -  m^\text{truth} \, R_m(m^\text{truth},\pt^\text{truth}) \right ) \,(r - s),
\end{equation*}
where $m^{\mathrm{reco}}$ is the detector-level \largeR jet mass and $R_m$ is the \largeR jet mass response. The value of $R_m$ is obtained from simulations, as discussed in Section~\ref{sec:objects}. Typical values of $R_m$ are in the range 0.8--1.5, depending on jet \pt{} and mass.
The forward-folding procedure does not require the response to be Gaussian. The scale factors $s$ and $r$ also modify
the non-Gaussian tails of the response function, if these are present in the simulations.
 
The prediction from simulation is fit to the data by minimizing the $\chi^2$ built with the predicted and
observed distributions. The best-fit values for $s$ and $r$ are taken as the data-to-simulation scale factors
for the \largeR jet mass response and jet mass resolution. This method has the advantage that
the response for the \ttbar{} events and events from other Standard Model processes is varied
consistently. It was first applied to 2012 data~\cite{ATLAS-CONF-2016-008}.
Further details of the forward-folding procedure are in
Refs.~\cite{ATLAS-CONF-2016-035,ATLAS-CONF-2017-063}.
 
The results of the fits are shown in Figure~\ref{fig:ttbar_response}.
The data sample is divided in several \pt bins. The $W$ boson peak is
fitted in two intervals: 200~\GeV~$<~\pt < $~250~\GeV~and 250~\GeV~$<~\pt < $
350~\GeV. The top quark peak is fitted for \pt between 350 and 500~\GeV\
and between 500~\GeV~and~1~\TeV. The small error bars on the points
represent the statistical uncertainty, and the larger error bars represent the total
uncertainty. The dominant systematic effect is expected to be due to
the modelling of top quark pair production, estimated by repeating
the analysis with \POWHEG + \HERWIG7, \SHERPA, and several
variations of the generator settings that regulate the probability of
hard initial- and final-state radiation.
 
\begin{figure}[tbh]\centering
\includegraphics[width=0.9\textwidth]{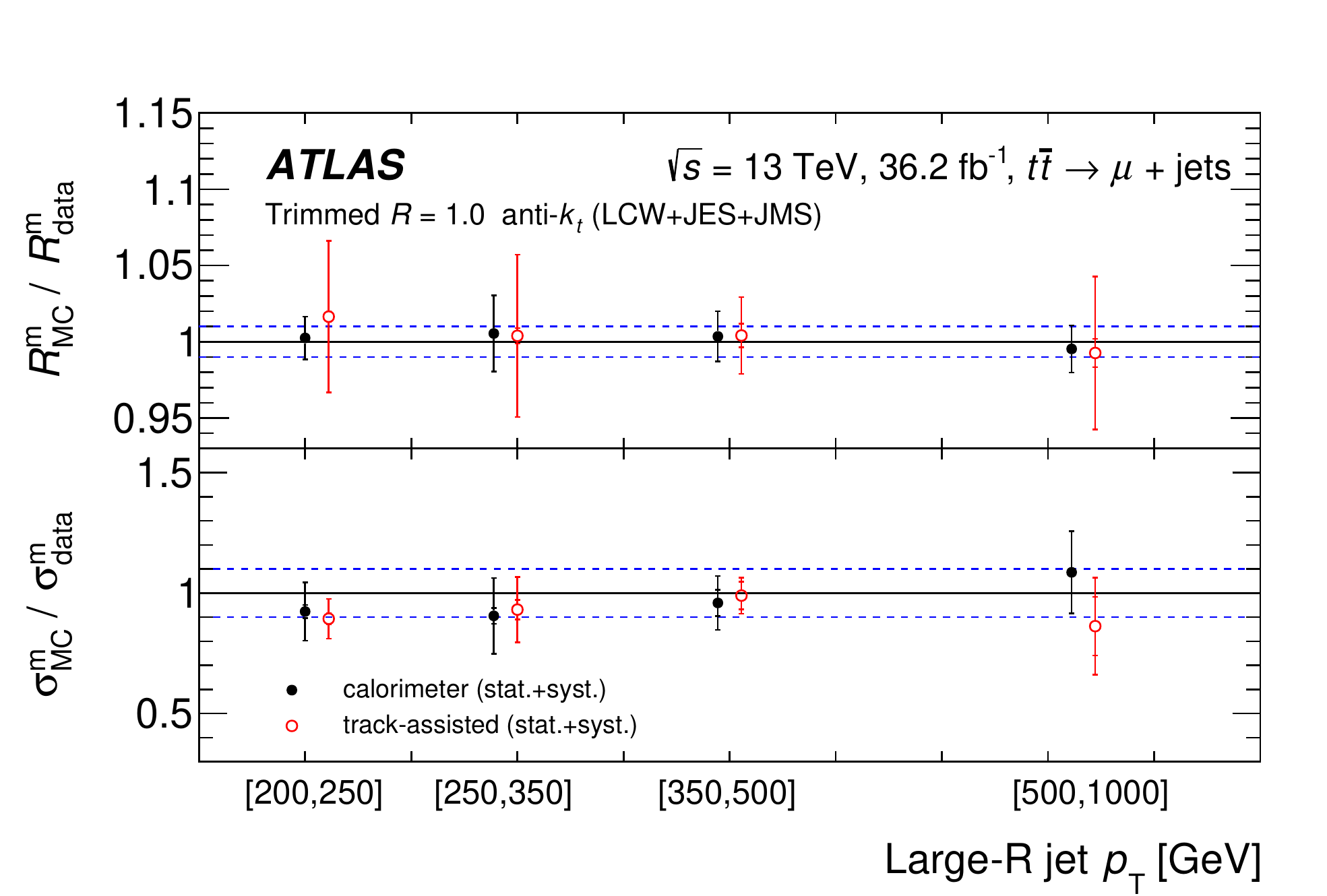}
\caption{Summary of the \insitu measurements of the \largeR jet mass response in \ttbar{} events with a lepton+jets final state as a function of the \largeR jet transverse momentum \pt. The \largeR jet \pt is corrected using the simulation calibration, $\eta$-intercalibration, and a combination of \insitu direct balance techniques. The closed circles correspond to the JMS and JMR of trimmed \largeR jets reconstructed from calorimeter clusters. The open circles represent the equivalent result for the track-assisted mass. The dashed lines, corresponding to $\pm$1\% for the JMS and $\pm$10\% for the JMR, are drawn for reference. The results in the first two \pt bins (200~\GeV~$ < \pt < $~250~\GeV~and 250~\GeV~$ < \pt < $~350~\GeV) correspond to a sample of high-\pt $W$ bosons, and the highest two bins (350~\GeV~$<~\pt < $~500~\GeV~and 0.5~\TeV~$<~\pt < $~1~\tev) correspond to high-\pt top quarks. In each subsample, the JMS and JMR are extracted simultaneously in a two-parameter fit to the mass distribution. The statistical and total uncertainties are indicated with the small and large error bars on the data points, respectively. }
\label{fig:ttbar_response}
\end{figure}
 
An \insitu calibration is also derived for the track-assisted mass in a
completely analogous fashion. The JMS and JMR results are shown with
open circles in Figure~\ref{fig:ttbar_response}. The statistical and systematic
uncertainties are indicated on the data points. The systematic
uncertainties are dominated by modelling uncertainties and are expected to be
strongly correlated between the two measurements.
The \insitu scales of the two mass measurements are found to be within
1\% for all points and within 0.5\% for three out of four. As the
track-assisted mass is primarily sensitive to the \pt response of the
calorimeter, this level of agreement implies that the \pt and mass
scales are closely connected for these high-mass jets with relatively low \pt.
 
Measurements of the \pt response of high-\pt $W$ bosons or top quarks can be obtained directly by fitting
the balance distribution of the two top quark candidates. This provides a cross-check of
the direct balance methods discussed previously in Sections~\ref{sec:etacal}--\ref{sec:mjbalance} in
a topology with a very different radiation pattern. The reference system is formed by the $b$-jet, the
charged lepton, and the neutrino from the semileptonic top quark decay. It is reconstructed by adding the
four-vectors of the charged lepton, the leading (and possibly $b$-tagged) \smallR jet in a cone of size $\Delta R =$ 1.5
around the charged lepton, and the neutrino~\cite{TOPQ-2012-18}. The transverse momentum of the neutrino is inferred by
assigning the \MET to the neutrino \pt, and its $p_z$ can be reconstructed using a $W$-mass constraint (but does not
affect the balance measurement). The resulting balance distribution of the probe jet \pt and the recoiling semileptonic
top quark decay system has a  distinctive peak around 1. The peak position is sensitive to
the \largeR jet energy scale, and its width is sensitive to the resolution. Measurements of the relative jet mass scale and resolution obtained by fitting the balance distribution with the same forward-folding technique are shown in Figure~\ref{fig:ttbar_pt_response}, after the application of the \insitu JES calibration derived from light quark and gluon jets (Section~\ref{sec:jes}). The results are compatible with unit JES within the precision of the measurement. This provides another confirmation that the Monte Carlo modelling of the response of high-\pt, hadronically decaying $W$ bosons or top quarks is adequate within 2--3\%, and that a calibration derived from jets without hard substructure is applicable to topologies with hard substructure.
 
\begin{figure}[tbh]\centering
\includegraphics[width=0.9\textwidth]{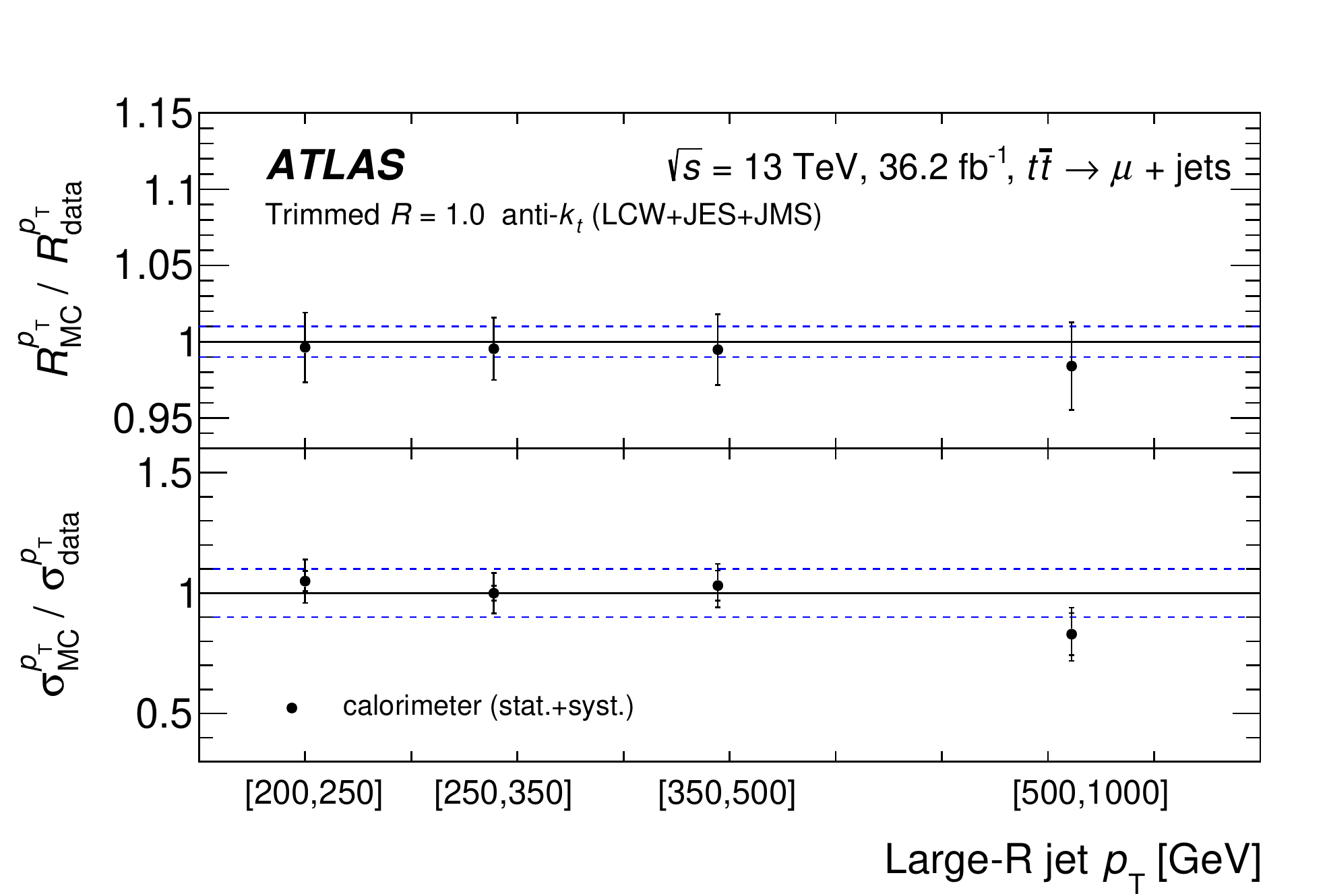}
\caption{Summary of the \insitu measurements of the \largeR jet response in \ttbar{} events with a lepton+jets final state as a function of the \largeR jet tranverse momentum \pt. The closed circles correspond to the JES and JER of trimmed \largeR jets reconstructed from calorimeter clusters. The \largeR jet \pt is corrected using the simulation calibration, $\eta$-intercalibration, and a combination of \insitu direct balance techniques. The error bars represent the total uncertainty. Statistical uncertainties are indicated with the inner error bar (only visible on some of the points). The dashed lines, corresponding to $\pm$1\% for the JES and $\pm$10\% for the JER, are drawn for reference. The results in the first two \pt bins (200 $ < \pt < $ 250~\GeV~and 250 $ < \pt < $ 350~\GeV) correspond to a sample of high-\pt $W$ bosons, and the highest two bins correspond to high-\pt top quarks.}
\label{fig:ttbar_pt_response}
\end{figure}
\clearpage
 
\section{Measurement of the \largeR jet \pt resolution} \label{sec:djjer}
 
The \insitu measurement of the ATLAS jet \pt{} resolution\footnote{The relative resolution $\sigma(\pt)/\pt$ is equal to the relative energy resolution $\sigma(E)/E$ to a good approximation, and the term jet energy resolution is used to refer to both quantities.} relies on a measurement that exploits the momentum balance between the leading and sub-leading \largeR jets in dijet events. This measurement follows the event selection criteria outlined for the $\eta$-intercalibration provided in Section~\ref{sec:etacal}, including the trigger strategy. The simulation calibration and $\eta$-intercalibration described in Sections~\ref{sec:objects}~and~\ref{sec:etacal} are applied to the \largeR jets studied here, and the \largeR jet \pt is also corrected using the combination of the \insitu direct balance techniques discussed in Sections~\ref{sec:zjetbalance},~\ref{sec:gjetbalance},~and~\ref{sec:mjbalance}, which is presented in Section~\ref{sec:combination}.
 
The asymmetry distribution of Section~(\ref{eq:asymmetry_dijet}) is studied in dijet events in bins of the dijet system \ptavg and the probe \largeR jet $\etadet$. The width of the asymmetry distribution depends on the resolution of the jet \pt measurement and on the intrinsic particle-level width, which arises due to balance fluctuations and out-of-cone effects. Since the latter effect is uncorrelated with the detector response, the component of the asymmetry width due to the detector resolution can be determined by subtracting in quadrature the asymmetry width of particle-level (`truth-level') jets from that of reconstructed jets, giving
\begin{eqnarray*} 
\sigma_{\mathcal{A},\text{det}} &= \sqrt{\sigma_{\mathcal{A},\text{reco}}^2 - \sigma_{\mathcal{A},\text{truth}}^2}.
\end{eqnarray*}

The jet energy resolution is measured in two \etadet bins: the central reference region $|\etadet| < 0.8$, denoted ``ref'', and a forward region $0.8 < |\etadet| < 2.0$, denoted ``fwd''. If both \largeR jets are within the central reference region, they have the same \pt resolution. In this case, the determination of the probe jet is arbitrary, and the assignment proceeds using a random-number generator. Since both jets contribute the same amount to the asymmetry distribution, the relative jet-\pt resolution of the reference region is defined by
\begin{equation*}
\left(\frac{\sigma_{\pt}}{\pt}\right)_\text{ref} = \frac{\sigma_{\mathcal{A},\text{det}}^\text{ref}}{\sqrt{2}}.
\end{equation*}
 
The resolution of forward jets is extracted from the width of the asymmetry distribution in events where a central reference jet balances a forward probe jet (in the region $0.8 < |\etadet| < $~2.0). The result is corrected for the resolution
of central jets by subtracting the asymmetry of central dijet systems, giving
\begin{equation}
\left(\frac{\sigma_{\pt}}{\pt}\right)_\text{fwd} = \sqrt{(\sigma_{\mathcal{A},\text{det}}^\text{fwd})^2 - \frac{(\sigma_{\mathcal{A},\text{det}}^\text{ref})^2}{2}}.
\label{eq:resol}
\end{equation}
 
Figure~\ref{fig:jer_nosub} shows $\sigma_{\mathcal{A}}$ for reconstructed- and truth-level dijet systems as a function of \ptavg in two \etadet bins, as well as for data. For each of the event generators, the width of the detector-level asymmetry is shown as a solid line, while the particle-level asymmetry is indicated by a dashed line. For forward jets, the additional correction shown in Eq.~(\ref{eq:resol}) is applied to account for the effect of the resolution of the \largeR jet within the central reference region.
 
\begin{figure}[tbh!]
\centering
\subfigure[]{\includegraphics[width=0.48\textwidth]{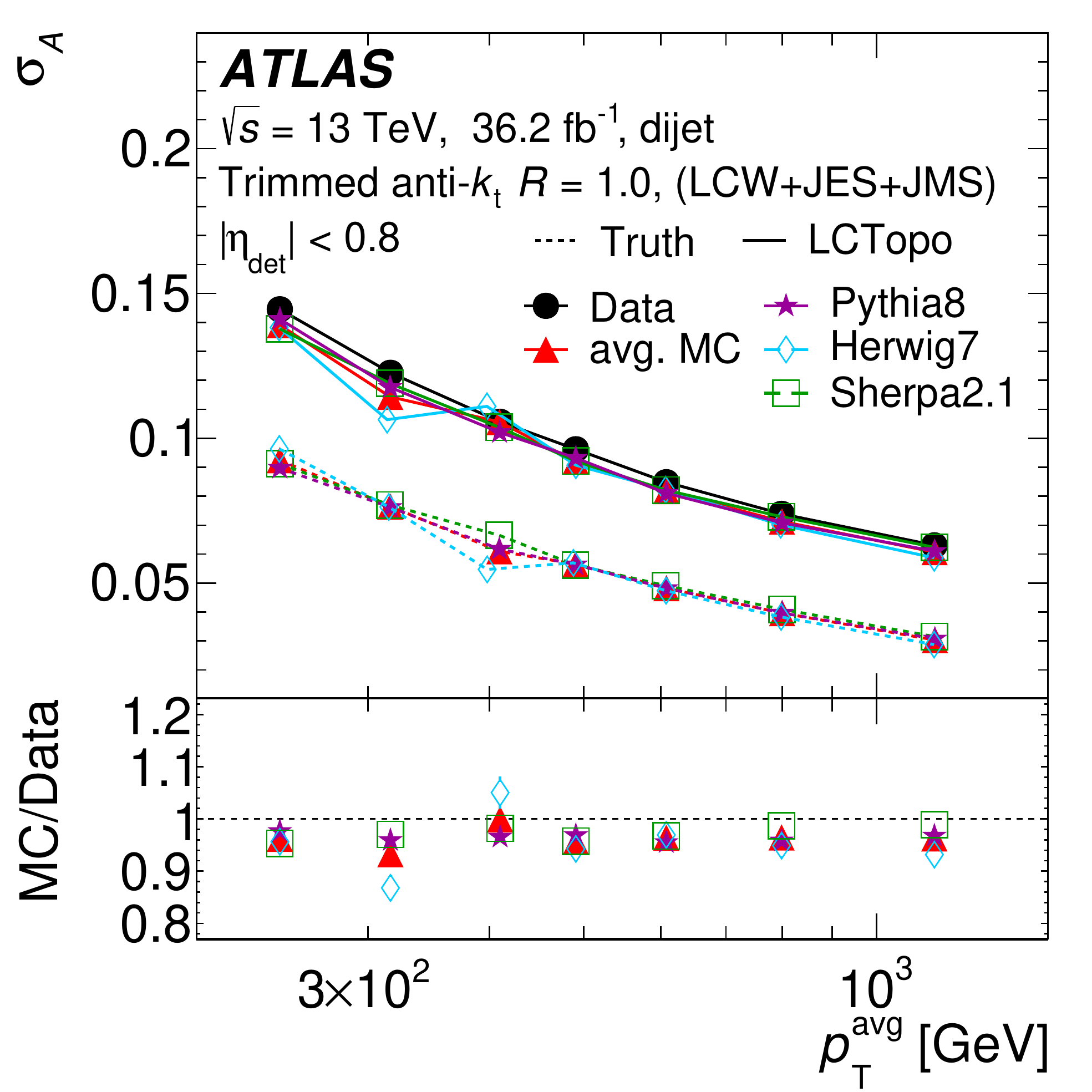}}
\subfigure[]{\includegraphics[width=0.48\textwidth]{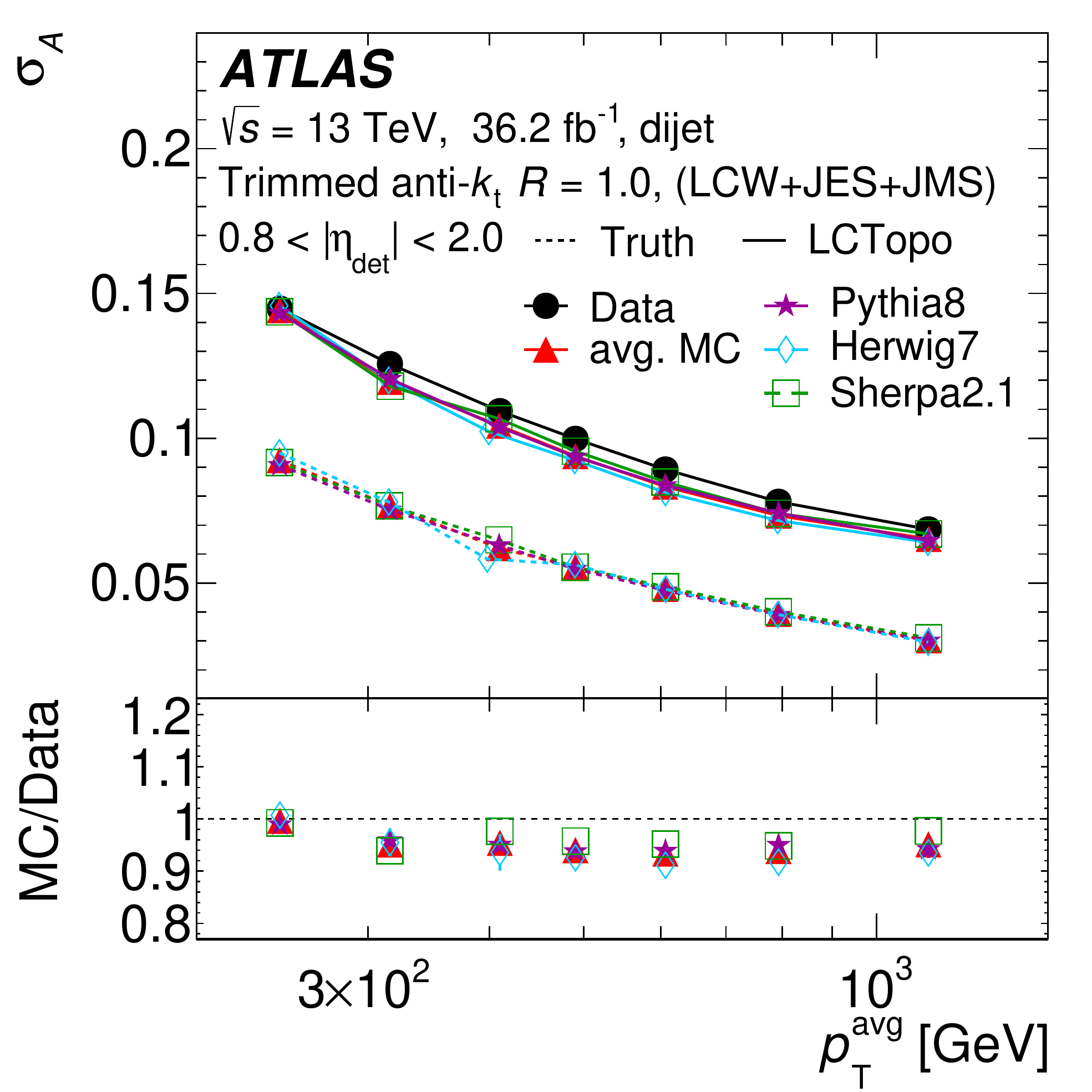}}
\caption{Width of the dijet asymmetry distribution obtained using reconstructed ($\sigma_{\mathcal{A},\mathrm{reco}}$) and particle-level jets ($\sigma_{\mathcal{A},\mathrm{truth}}$) as a function of the average jet transverse momentum $\pt^\mathrm{avg}$. Results are shown (a) for events where both jets have detector pseudorapidity in the range $|\etadet| < 0.8$ and (b) for events where the probe jet has $0.8 < |\etadet| < 2.0$, and the reference jet is still within $|\etadet| < 0.8$. The measurement is compared with the prediction from simulations based on the three generators \PYTHIA8, \HERWIG7, and \SHERPA2.1. Also an unweighted average of the three is shown. The \largeR jet \pt is corrected using the simulation calibration, $\eta$-intercalibration, and a combination of \insitu direct balance techniques. Statistical errors are usually smaller than the size of the marker. The resolution at the particle level is also shown as a dashed line.}\label{fig:jer_nosub}
\end{figure}
 
Following the correction for the particle-level width, the results of a fit to the asymmetry distribution obtained in data and from several event generators (\PYTHIA8, \HERWIG7, and \SHERPA 2.1) are shown in Figure~\ref{fig:JERfromdijets}, where the measured relative resolution $ \sigma ( \pt{} ) / \pt $ is plotted as a function of the average \pt{} of the two jets, $\pt^{\mathrm{avg}}$. The correction for the particle-level resolution is estimated using the \PYTHIA sample. The measured resolution in the central region is in fair agreement with the predicted resolution. The resolution of forward jets in data and simulations is compatible within the observed uncertainties. The choice of event generator has a small effect on the resolution.

In Figure~\ref{fig:JERfromdijets}, the relative \pt resolution, $\pt^{\mathrm{reco}} / \pt^{\mathrm{true}}$, as predicted by the simulations is compared with the result of the extraction of the resolution from the asymmetry in simulated events. The difference between the two indicates a bias in the method that is taken as an additional uncertainty (labelled non-closure).
 
\begin{figure}[tbh]
\begin{center}
\subfigure[]{
\includegraphics[width=0.48\textwidth]{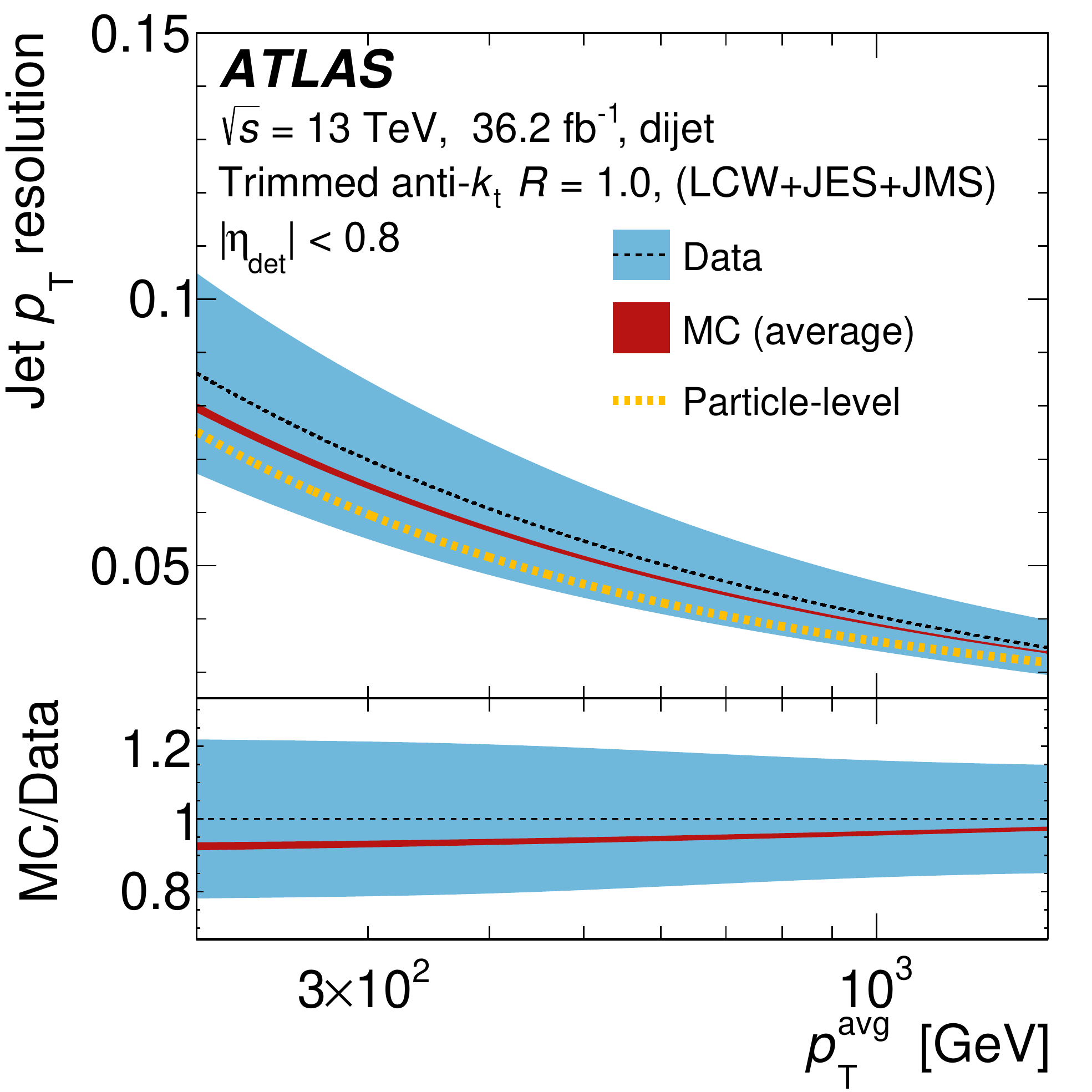} 
}
\subfigure[]{
\includegraphics[width=0.48\textwidth]{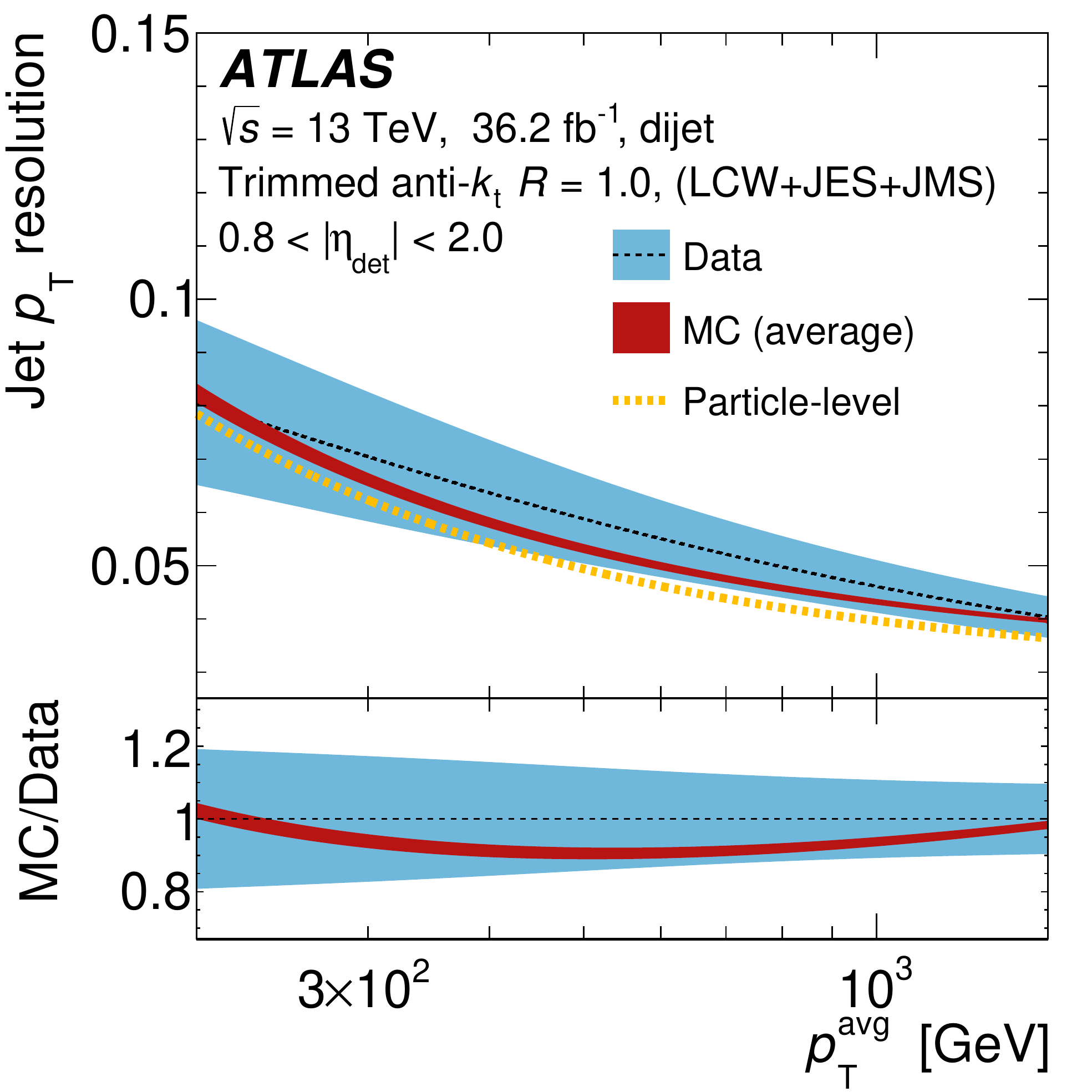} 
}
\caption{\label{fig:JERfromdijets} Comparison of the measured jet \pt{} resolution with the resolution determined in simulation, averaged between different generators as a function of the average jet $\pt$ and in two bins of  detector pseudorapidity $\etadet$ from (a)~$|\etadet| < 0.8$ and from (b)~$0.8 < |\etadet| < 2.0$. The \largeR jet \pt is corrected using the simulation calibration, $\eta$-intercalibration, and a combination of \insitu direct balance techniques. The error band, drawn as a light band, represents the statistical and systematic uncertainties added in quadrature. The determination of the relative resolution using the \insitu technique for an average of three simulations and their envelope is also shown as a dark band. Inconsistencies between the resolution determined using the \insitu technique and of the resolution determined from the response in simulation by matching particle-level jets to reco-level jets (light dotted line) are taken as an additional uncertainty in the measurement. The lines shown are obtained by smoothing a binned representation of these uncertainties using a sliding Gaussian kernel.}
\end{center}
\end{figure}
 
The total uncertainty in the determination of the JER is shown in Figure~\ref{fig:JERfromdijets_uncertainties} as a function of the
average \pt{} and in the two \etadet regions. A breakdown of the uncertainties into individual sources is presented. The \largeR jet energy scale is varied according to its uncertainty, leading to a 10--15\% variation in the measured resolution due to its impact on the asymmetry (labelled as `JES uncertainty'). The non-closure uncertainty is found to be a nearly constant 10\% effect in the central region and to be 5--10\% in the forward region. The $\Delta\phi$ requirement is also varied by $\pm$0.5, which has a small effect primarily for low-\pt jets. The modelling uncertainty is estimated as the variation of the result when using different generators for the particle-level momentum imbalance, where \PYTHIA8 is chosen as a nominal sample and \HERWIG7 and \SHERPA2.1 are chosen as the variations.
 
\begin{figure}[tbh]
\begin{center}
\subfigure[]{
\includegraphics[width=0.48\textwidth]{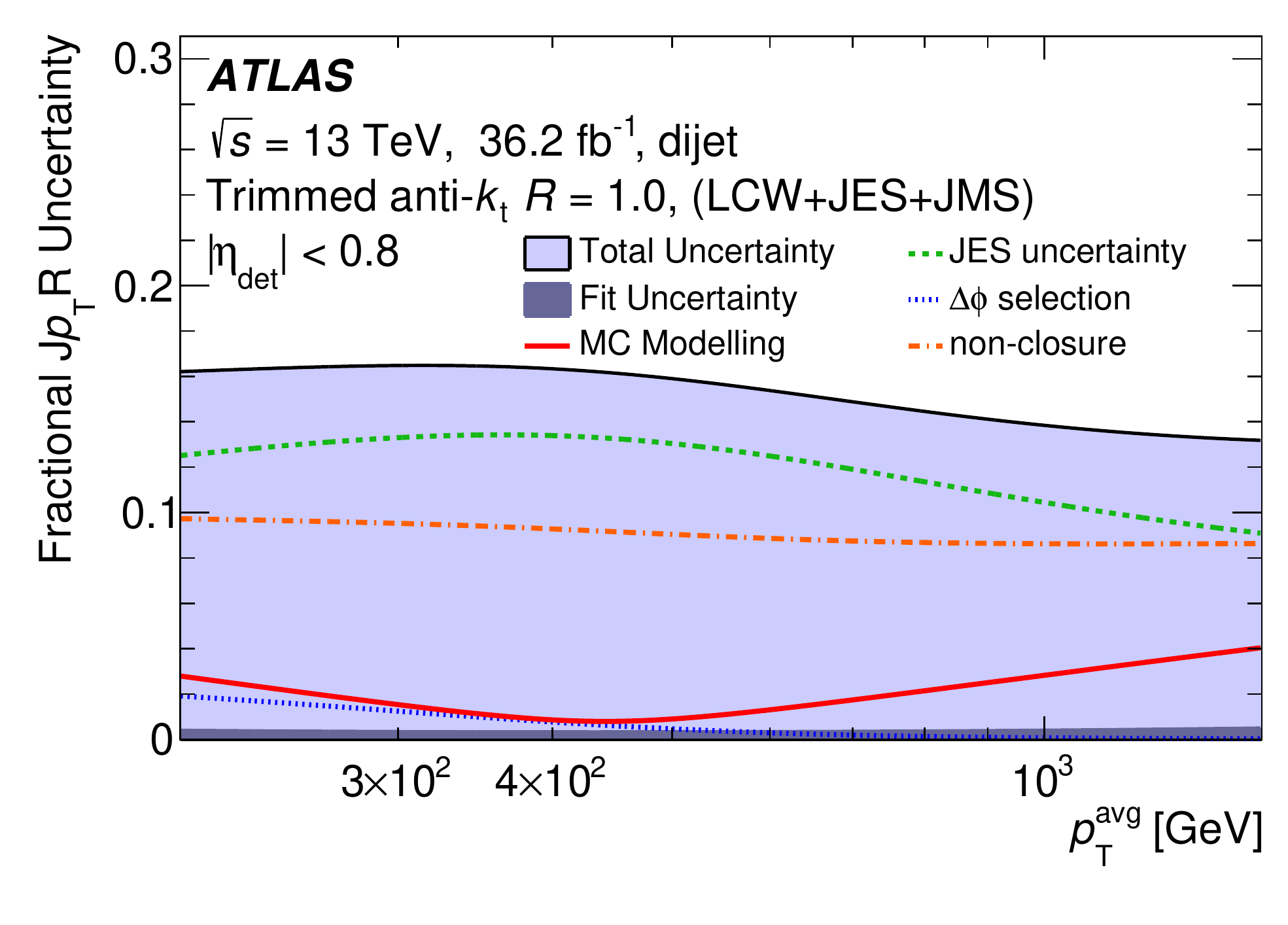} 
}
\subfigure[]{
\includegraphics[width=0.48\textwidth]{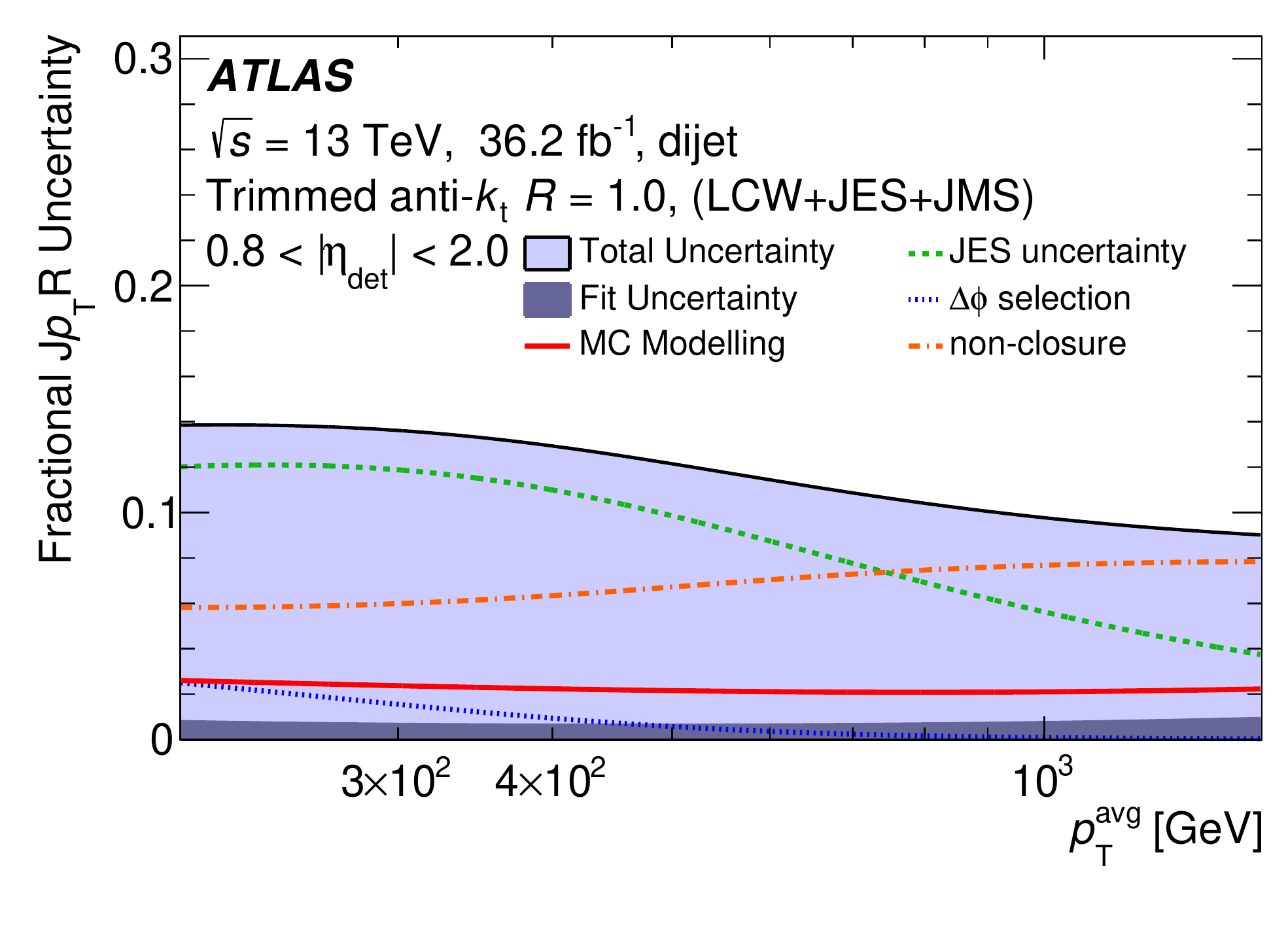}
}
\caption{\label{fig:JERfromdijets_uncertainties} The relative uncertainty in the ratio of the jet transverse momentum \pt{} resolution measured in dijet events and in simulations as a function of the average jet $\pt$ in pseudorapidity $\eta$ bins (a)~$|\eta| < 0.8$ and (b)~$0.8 < |\eta| < 2.0$. The \largeR jet \pt is corrected using the simulation calibration, $\eta$-intercalibration, and a combination of \insitu direct balance techniques. Contributions from three sources are estimated separately by propagating the uncertainty in the energy scale to the measurement, by varying the $\Delta\phi$ selection, and by varying the event generator. The lines shown are obtained by smoothing a binned representation of these uncertainties using a sliding Gaussian kernel.}
\end{center}
\end{figure}
\clearpage
 
\section{Combined \largeR jet calibration results} \label{sec:combination}
The measurements of the trimmed \largeR jet response relative to simulation obtained using the different \insitu methods presented in Sections~\ref{sec:balancept} and~\ref{sec:insitumass} are combined to determine the relative jet energy and mass scales over a broad range of jet transverse momenta.  The combination procedure is described in detail in Ref.~\cite{PERF-2011-03}.
 
The data-to-simulation response ratios obtained from the $\gamma$+jet, $Z$+jet, and multijet balance methods are combined to produce a jet \pt-dependent calibration curve. The uncertainties in the \pt calibration are obtained by error propagation of the uncertainties associated with the \insitu methods.  A jet mass calibration is derived analogously using the jet mass response measurements provided by the forward-folding and \Rtrk methods.
 
The measurements of the \pt response are performed in bins of the jet transverse momentum (the $\pt^{\mathrm{ref}}$ values are translated to jet \pt) and evaluated inclusively in mass. The jet mass response combination is performed in bins of the jet transverse momentum and in two bins of the jet mass. The combination proceeds in three steps which take into account correlations between uncertainties and possible inconsistencies between the \insitu methods:\\
\begin{itemize}
\item \textbf{Simple Monte Carlo method:} Pseudo-experiments are created that
represent the ensemble of measurements and contain the full data-treatment chain including interpolation and
averaging (described in the following steps). These pseudo-experiments
are used to consistently propagate all uncertainties into the evaluation of the average. They are generated taking into account all known correlations by coherently shifting all correction factors by one
standard deviation. The difference between the shifted-correction result and the nominal
result provides an estimate of the propagated systematic uncertainty.
\item \textbf{Interpolation:} The relative \pt (mass) response is defined
in fine $\pt$ bins, separately for each \insitu method using interpolating
splines based on first- or second-order polynomials.
\item \textbf{Averaging:} The actual combination is carried out using a
weighted average of the \insitu measurements based on a
$\chi^2$-minimization. The weights take into account the statistical
and systematic uncertainties, as well as correlations and differing
bin sizes. The local $\chi^2$ is also useful to define the level of
agreement between \insitu measurements where they overlap.
\end{itemize}
 
The uncertainty sources are treated according to the Hessian formalism: each uncertainty source is fully correlated across kinematic regions (i.e.\ as a function of \pt and  $\eta$) but is uncorrelated with other sources. Sources of uncertainty that affect both the \smallR and \largeR jet \insitu calibration are treated as fully correlated. The reduced $\chi^2$ is estimated as $\sqrt{\chi^2/N_{\mathrm{dof}}}$, where $N_{\mathrm{dof}}$ is the number of degrees of freedom (in this case, the number of combined measurements contributing to the average in a particular \pt bin). In case of disagreement between different \insitu measurements, i.e.\ when the reduced $\chi^2$ value is larger than 1, the uncertainty sources are rescaled by $\sqrt{\chi^2/N_{\mathrm{dof}}}$.
 
A smoothing procedure using a variable-size sliding interval with a Gaussian kernel is applied to the response ratio and its associated systematic uncertainties. This smoothing removes spikes  due to  statistical  fluctuations in the measurements, as well as discontinuities at the first and last point in a given measurement.
 
In Figure~\ref{fig:pTresponse}, the ratio of the jet \pt response in data
and simulations is shown as a function of the jet transverse momentum.
Data points are shown for the $\gamma$+jet, $Z$+jet, and multijet balance
methods, and the band corresponds to the result of the combination.
 
\begin{figure}[tbh]
\centering
\includegraphics[scale=0.7]{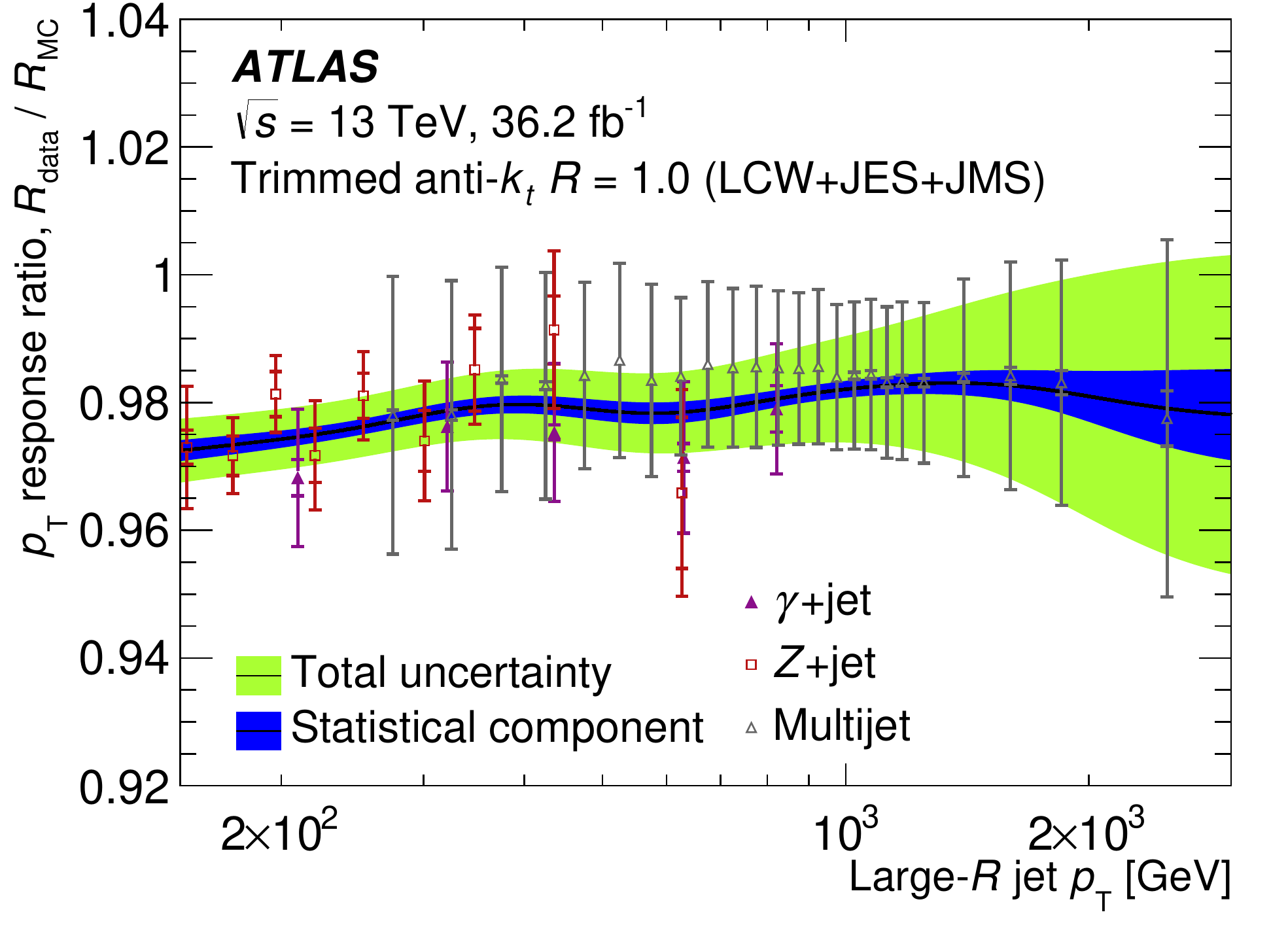}
\caption{Data-to-simulation ratio of the average jet \pt response as a function of \largeR jet $\pt$. The combined result (band) is based on three \insitu techniques: the $Z$+jet balance method (open squares), $\gamma$+jet balance method (closed triangles), and the multijet balance (open triangles). The errors represent the statistical (inner error bars) and the total uncertainty (statistical and systematic uncertainties added in quadrature, outer error bars). The results apply to trimmed anti-\textit{$k_t$} jets with $R=1.0$. The lines shown are smoothed using a sliding Gaussian kernel. }
\label{fig:pTresponse}
\end{figure}
 
The relative weight in the fit of the three methods is shown in Figure~\ref{fig:pTweights}. The $Z$+jet balance makes the largest contribution up to transverse momenta of approximately 500~\GeV. Between 500~\GeV{} and 1~\TeV, the $\gamma$+jet balance recieves the largest weight. At higher \pt, the multijet balance method acquires more weight in the combination. Beyond 1~\tev{}, it provides the only measurement and extends the jet energy scale beyond 2~\tev.
 
The local $\chi^2$ per degree of freedom in Figure~\ref{fig:pTchi2} quantifies the level of agreement between the three sets of measurements. The results of the three methods agree in the whole \pt range $0.1~\TeV < \pt < 1$~\TeV, where all three provide results.
 
\begin{figure}[h!]
\centering
\includegraphics[scale=0.7]{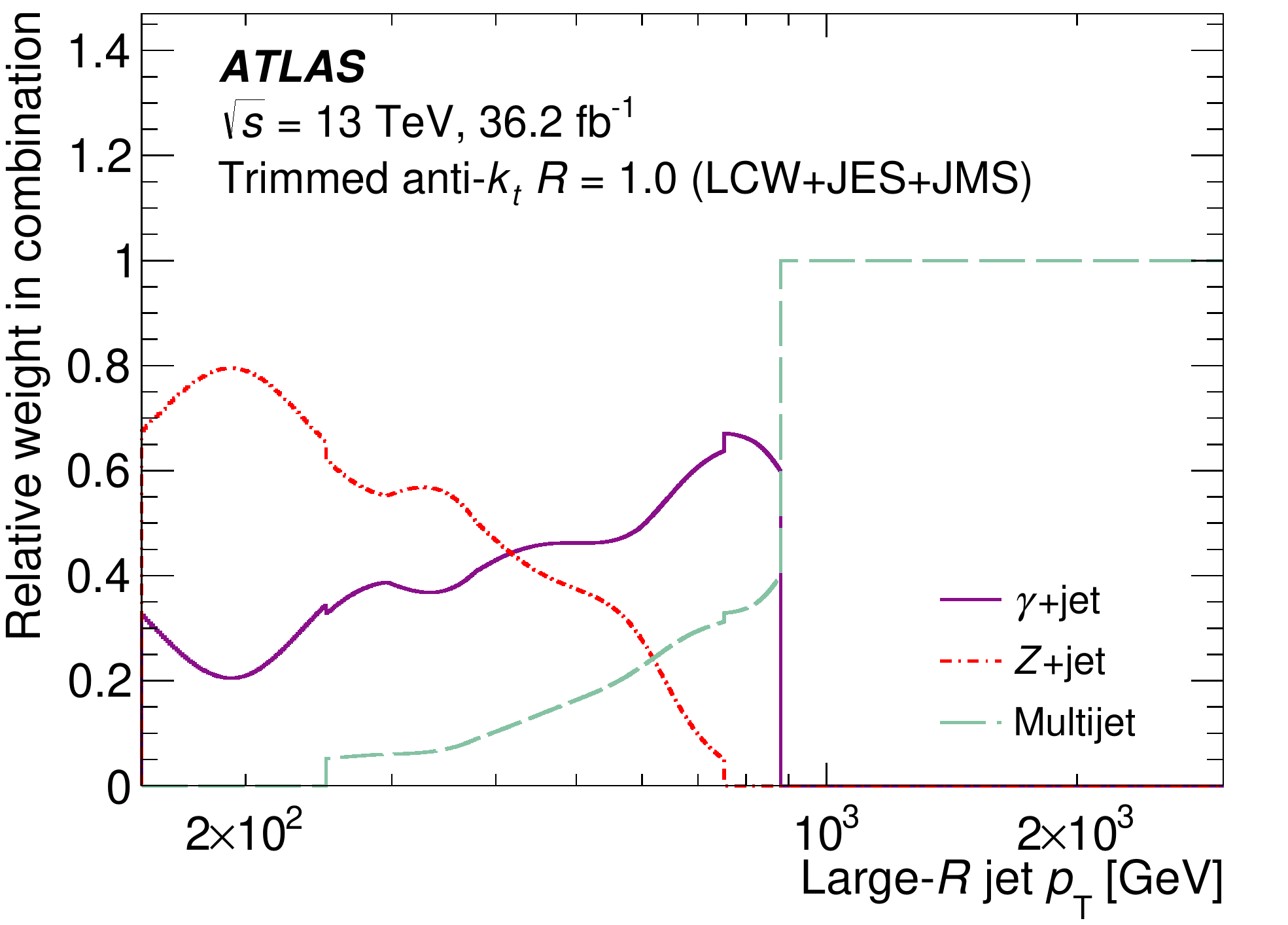}
\caption{The weight assigned to different techniques in the combination of \insitu measurements of the relative \pt response of \largeR jets in data and simulations, as a function of the jet transverse momentum \pt. For each \pt bin, the weight of the $Z$+jet, $\gamma$+jet, and multijet balance methods are shown. The slight discontinuities observed in the weights correspond to the onset of the multijet balance method at $\pt \sim $ 300~\GeV\ and to the upper end of the $Z$+jets method for $\pt \sim $ 800~\GeV. }
\label{fig:pTweights}
\end{figure}
 
The combined \pt response in data is approximately 3\% lower than in the simulation over most of the \pt range. The deviation from unity in the data/MC ratio is significant, as the total uncertainty approaches 1\% in the intermediate \pt region. These observations are consistent with previous \insitu measurements of the $R=0.4$ JES during Run 2~\cite{PERF-2016-04} with similar levels of associated uncertainty. At low \pt, the uncertainty reaches about 1\% at 200~\GeV. Above 1.5~\tev{}, the uncertainty increases, reaching over 2\% at 2.4~\tev.
 
A breakdown of the total JES uncertainty is presented graphically in Figures~\ref{fig:uncertainties_JES_gzjets} and~\ref{fig:uncertainties_JES_multijet}. This includes uncertainties in $\gamma$+jet, $Z$+jet, and multijet balance methods associated with the simulation modelling,
reference system construction and calibration, and the event selection. Furthermore, as the \largeR multijet balance method uses
small-$R$ jets as a reference system, all nuisance parameters from the small-$R$ jet calibration enter as uncertainties in the combination presented here.
 
\begin{figure}[h!]
\centering
\includegraphics[scale=0.7]{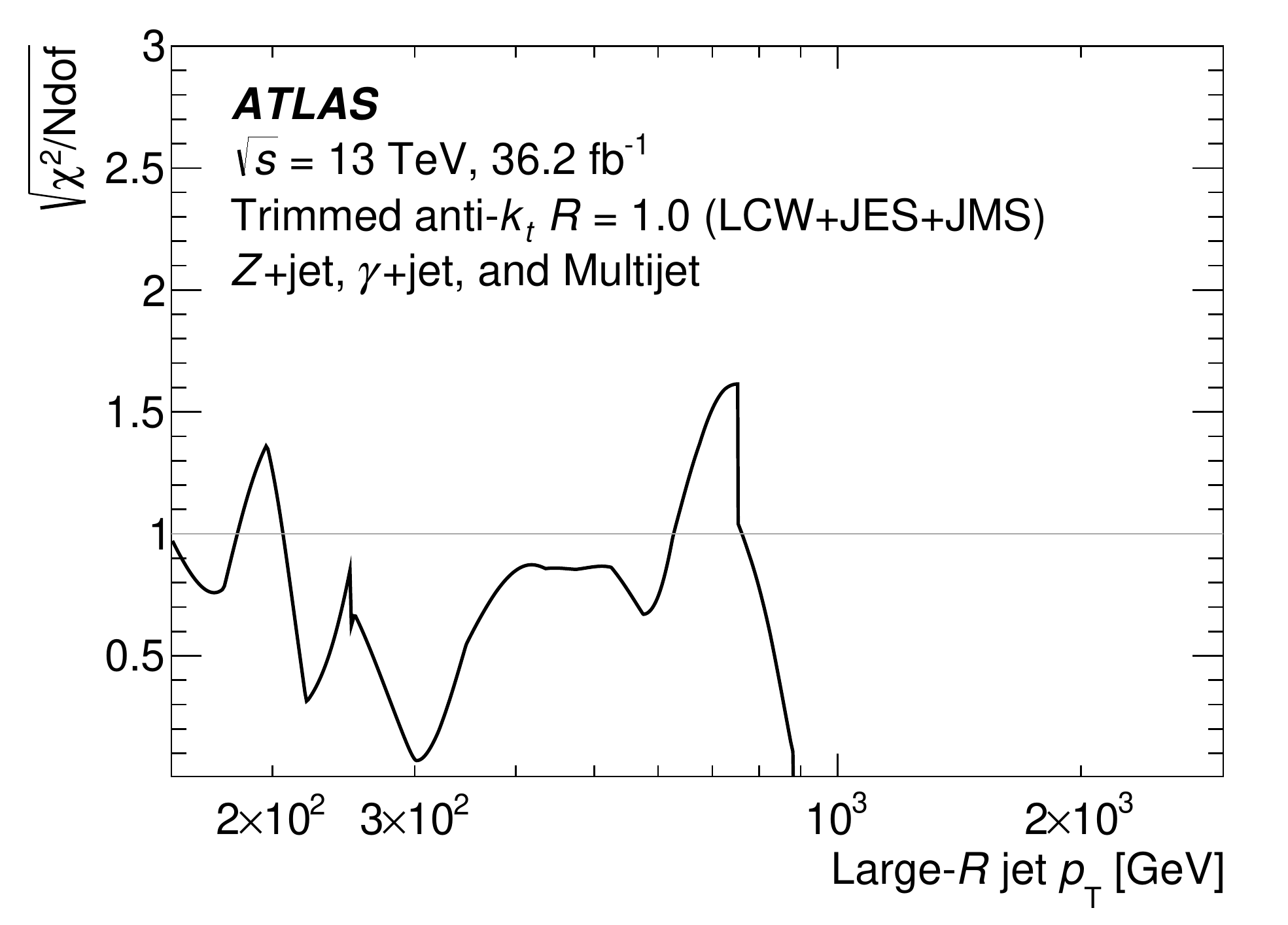}
\caption{The $\chi^2/N_{\mathrm{dof}}$ of the combination of \insitu measurements of the relative jet-\pt response of \largeR jets in data and simulations as a function of the jet transverse momentum \pt. The $\chi^2/N_{\mathrm{dof}}$ indicates the level of tension between the results of the $\gamma$+jet and multijet balance methods in each \pt bin. For transverse momenta beyond 1~\TeV, only one method is available, and the $\chi^2/N_{\mathrm{dof}}$ goes to zero.}
\label{fig:pTchi2}
\end{figure}
 
\begin{figure}[tbh]\centering
\subfigure[]{
\includegraphics[width=0.48\textwidth]{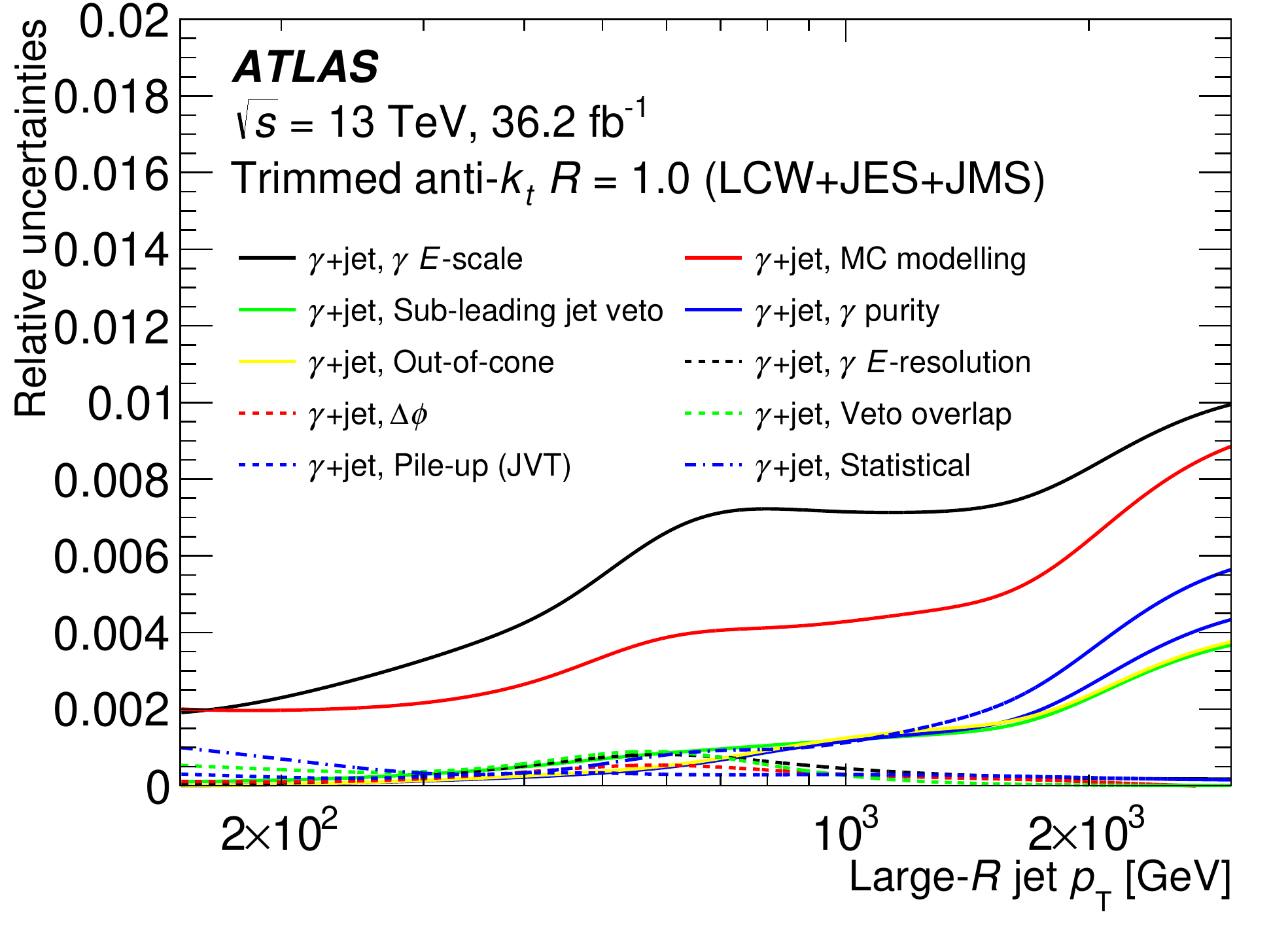} } 
\subfigure[]{
\includegraphics[width=0.48\textwidth]{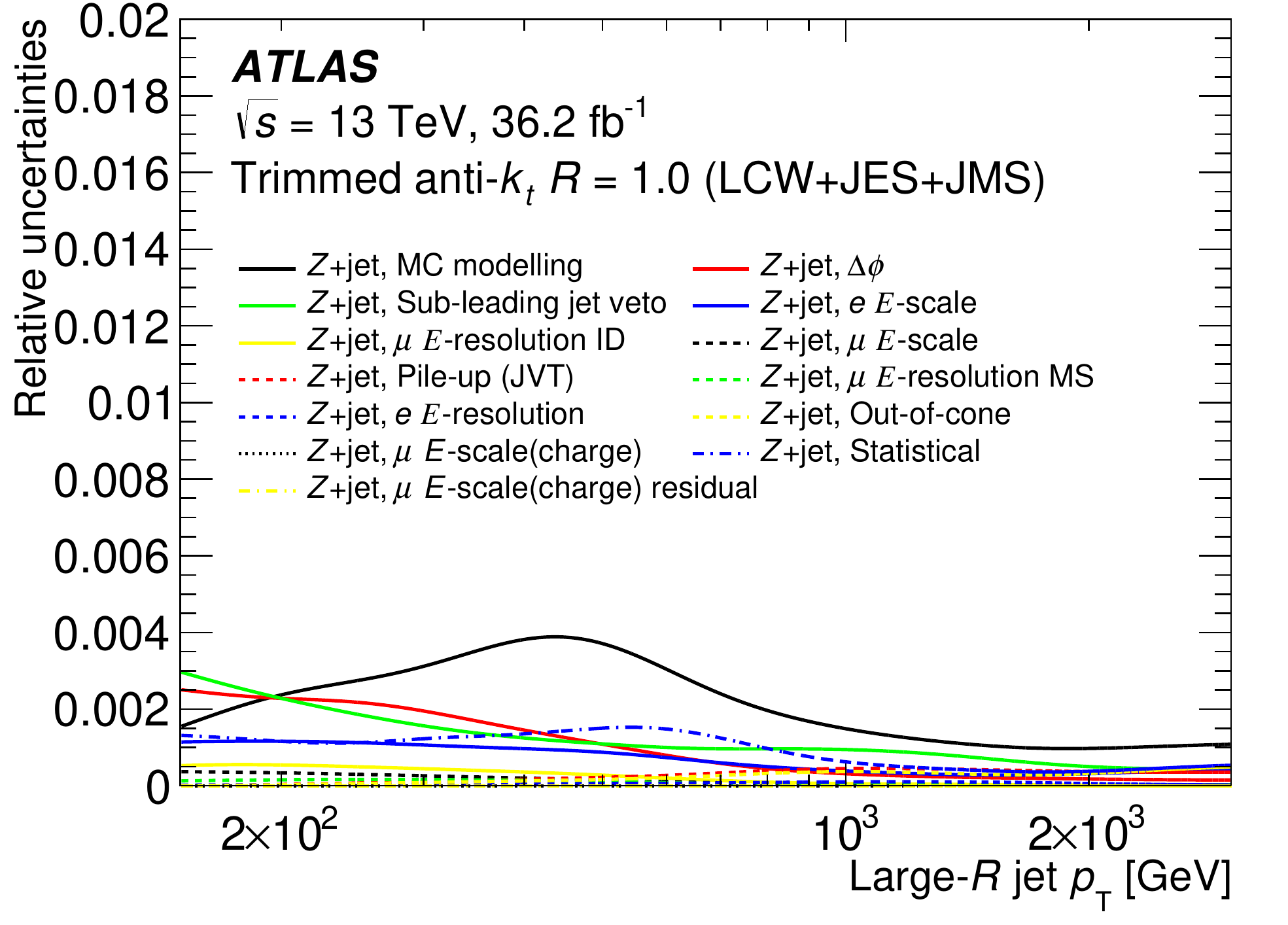} } 
\caption{Breakdown of the combined uncertainty in the \largeR jet \pt response as a function of the jet transverse momentum $\pt$, for the (a) $\gamma$+jet and (b) $Z$+jet analyses. Contributions are shown for each of the nuisance parameters of the $\gamma$+jet and $Z$+jet balance methods. The vertical axis reflects the uncertainty introduced by a given nuisance parameter in combination, incorporating the weight of the method from which it originates. The nuisance parameters related to the $\gamma$+jet method (both directly and through their effect on the multijet balance) are shown in the left panel, and those of the Z+jets method are shown in the right panel. The lines shown are smoothed using a sliding Gaussian kernel.}
\label{fig:uncertainties_JES_gzjets}
\end{figure}
 
\begin{figure}[tbh] \centering
\subfigure[]{
\includegraphics[width=0.48\textwidth]{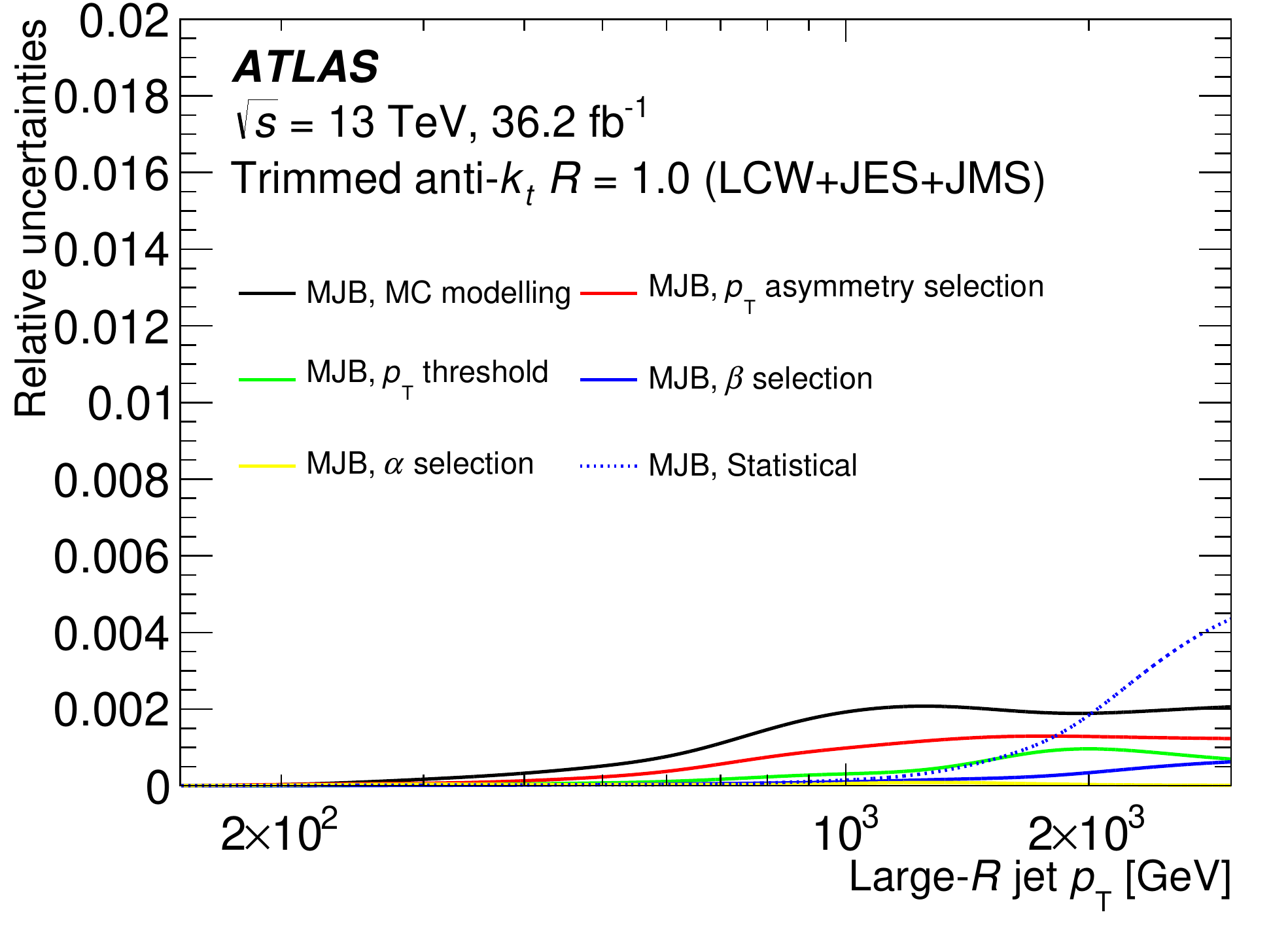} }
\subfigure[]{
\includegraphics[width=0.48\textwidth]{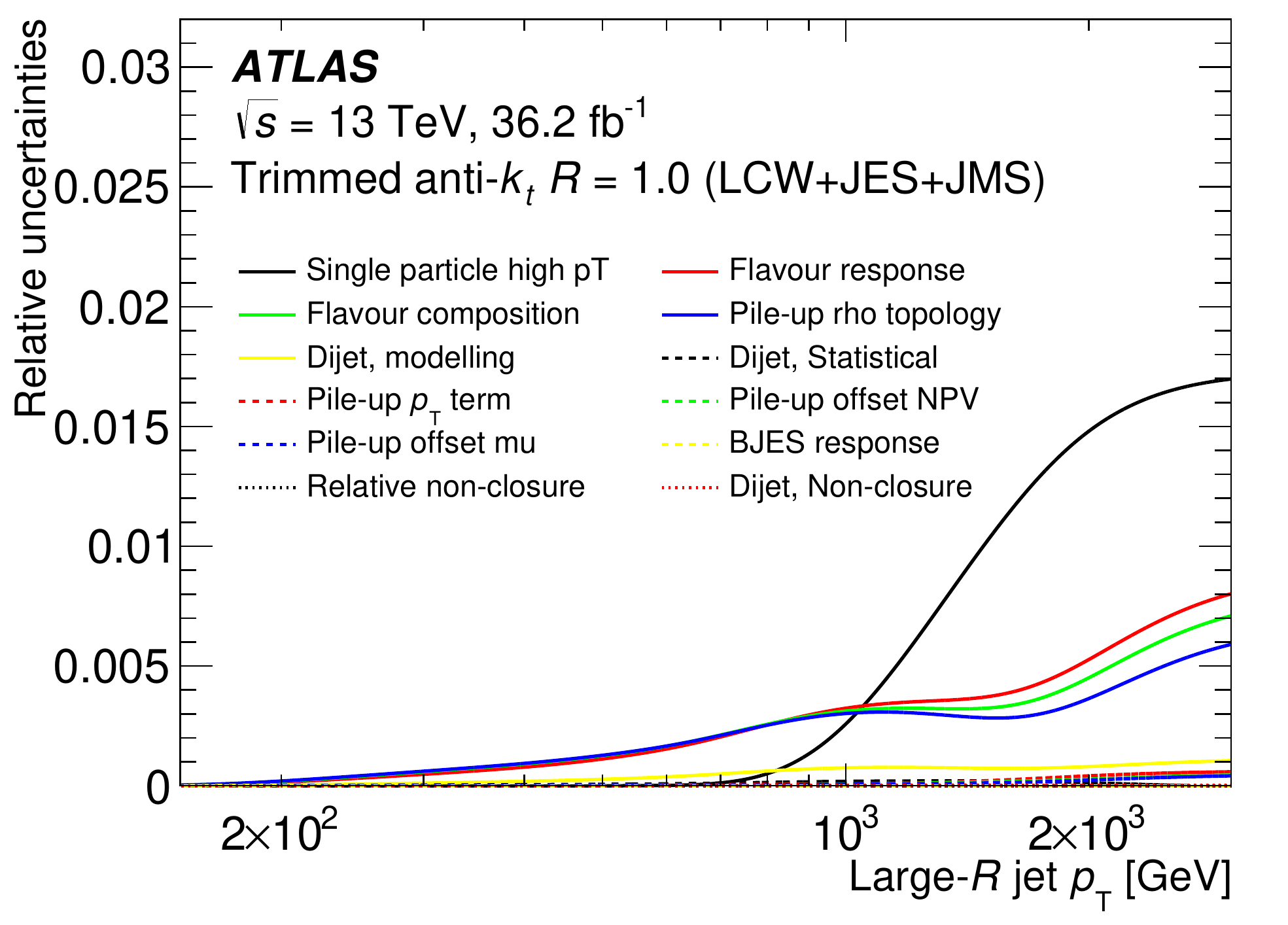} }
\caption{Breakdown of the combined uncertainty in the \largeR jet \pt response as a function of the jet transverse momentum $\pt$. Contributions are shown for nuisance parameters of the multijet balance method for nuisance parameters (a) originating from the MJB selection and (b) propagated from the small-$R$ jets which constitute the recoil system. The vertical axis reflects the uncertainty introduced by a given nuisance parameter in combination, incorporating the weight of the method from which it originates. Since the multijet balance method relies on the small-$R$ jet \pt, nuisance parameters from all associated uncertainties are propagated. The lines shown are smoothed using a sliding Gaussian kernel.}
\label{fig:uncertainties_JES_multijet}
\end{figure}
 
The combination of the jet mass response includes results from two methods.
Forward folding provides four measurements in the \pt range below
1~\TeV. The \Rtrk method takes advantage of a large data sample
and can be finely binned in mass and \pt, extending to over
2~\TeV. The combined result is shown in Figure~\ref{fig:Massresponse}
for two jet mass intervals: the plot in the upper panel
corresponds to the $W$ boson mass window with $50~\GeV < m < 120~\GeV$,
and the lower panel corresponds to the top quark mass window with $120~\GeV < m < 300~\GeV$.
 
The \insitu jet mass calibration factor is defined from the combined mass response shown in Figure~\ref{fig:Massresponse} as $c_m = R_\text{MC}^m/R_\text{data}^m$. It is applied as a scale factor to the jet mass but does not affect the jet momentum vector.
The full calibration applied to \largeR jets in data impacts the reconstructed jet energy, mass, pseudorapidity, and \pt according to
\begin{equation*}
\begin{split}
&E_{\mathrm{reco}} = c_s\sqrt{E_0^2+c_\text{JMS}\,m_0\left(c_m^2-1\right)},~~~m_{\mathrm{reco}} = c_s\,c_\text{JMS}\,c_m\,m_0,~~~\eta_{\mathrm{reco}} = \eta_0+\Delta\eta,\\
&\pt^{\mathrm{reco}} = c_s\,\sqrt{\left(E_0^2-c_\text{JMS}^2\,m_0^2\right)}\cosh{(\eta+\Delta\eta)},
\end{split}
\end{equation*}
where $c_s =c_\text{JES}\,c_\text{abs}\,c_\text{rel}$ is the product of several calibration factors. The factor $c_\text{JES}$ corresponds to the simulation-based JES calibration, $c_\text{rel}$ to the relative \insitu correction obtained from the $\eta$-intercalibration, and $c_\text{abs}$ to the absolute \insitu correction from the balance methods. All $c$-factors and the factor $\Delta \eta$ are smooth functions of the \largeR jet kinematics. The terms $E_0$, $m_0$, $\eta_0$ and $\vec{p}_0$ refer to the jet properties prior to any calibration, as returned by the trimming algorithm.
 
The measured JMS correction is consistent with unity within the precision of the combined measurements. This suggests that the application of an \insitu JES correction is sufficient to correct the JMS of these trimmed \largeR jets in the mass and \pt{} ranges considered here. The level of precision with which the JMS is measured depends on the kinematic region in question.  For \largeR jets in the high-mass bin with \pT between 400~\GeV~and 1~\TeV, the uncertainties are 2--5\%. In other kinematic regions the uncertainty is larger, approaching 10\% at high \pt in both mass bins.
 
\begin{figure}[tbh]
\centering
\includegraphics[scale=0.7]{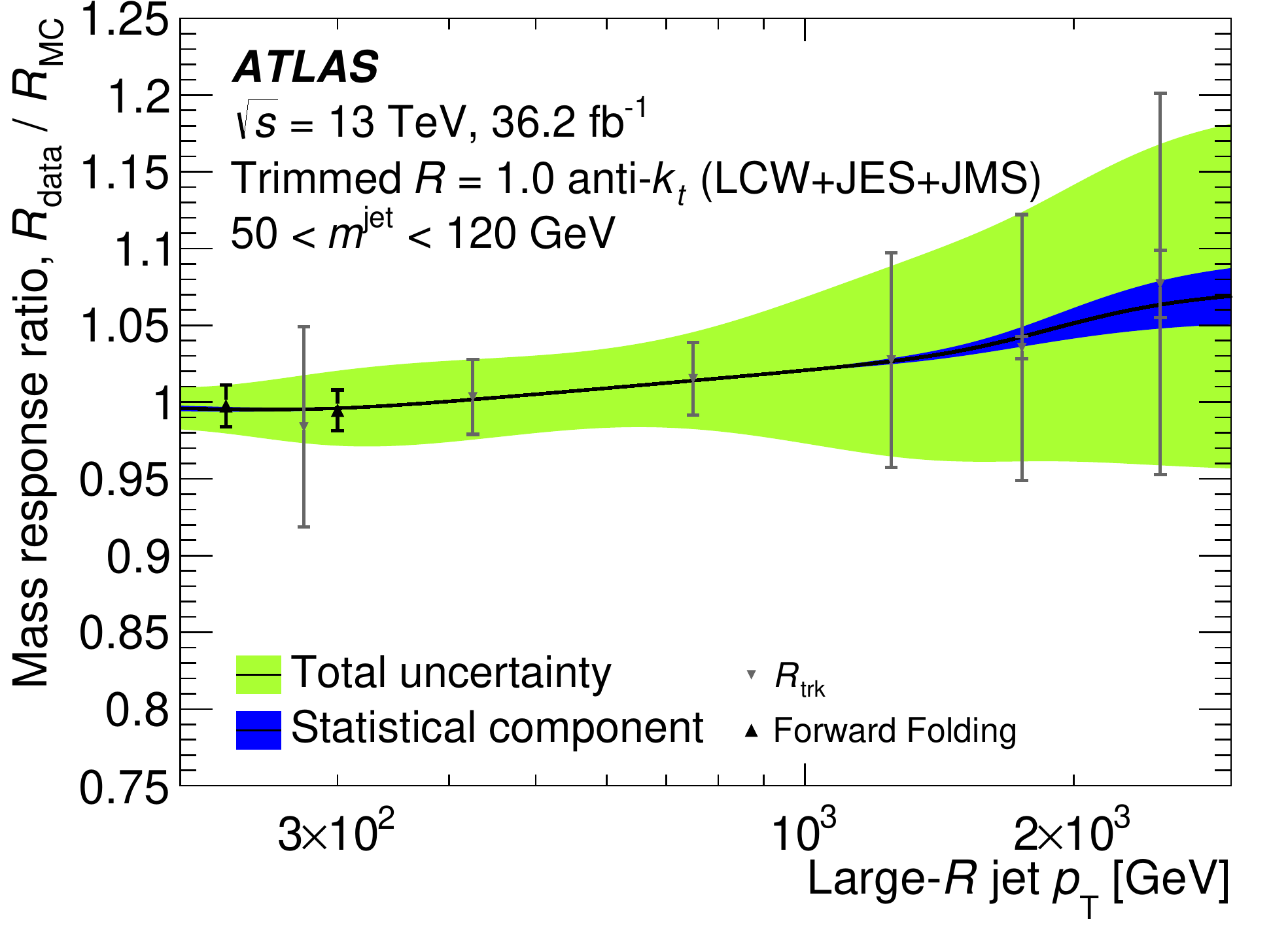} \includegraphics[scale=0.7]{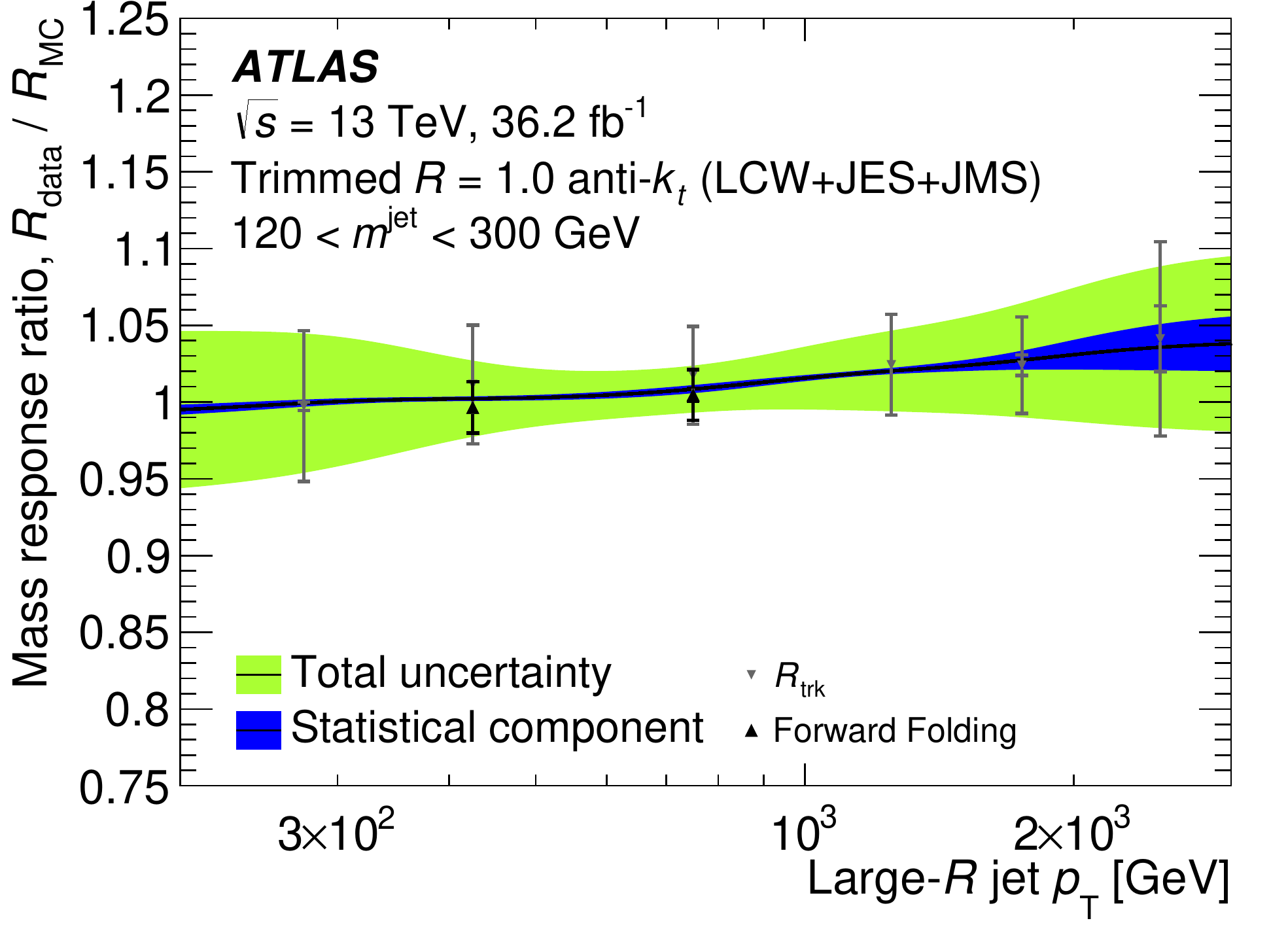}
\caption{Data-to-simulation ratio of the average jet mass response as a function of the \largeR jet $\pt$. Corrections using a combination of two \insitu methods, the \Rtrk and forward-folding approaches, are applied. The fit is performed for \largeR jet mass in the $W$ mass range 50--120~\GeV\ (upper), and the top mass range 120--300~\GeV\ (lower). The error bars represent the statistical and systematic uncertainties added in quadrature. The results apply to \antikt jets with $R=1.0$ calibrated with the LC+JES+JMS scheme. The lines shown are smoothed using a sliding Gaussian kernel.}
\label{fig:Massresponse}
\end{figure}
 
\begin{figure}[tbh] \centering
\subfigure[]{
\includegraphics[width=0.48\textwidth]{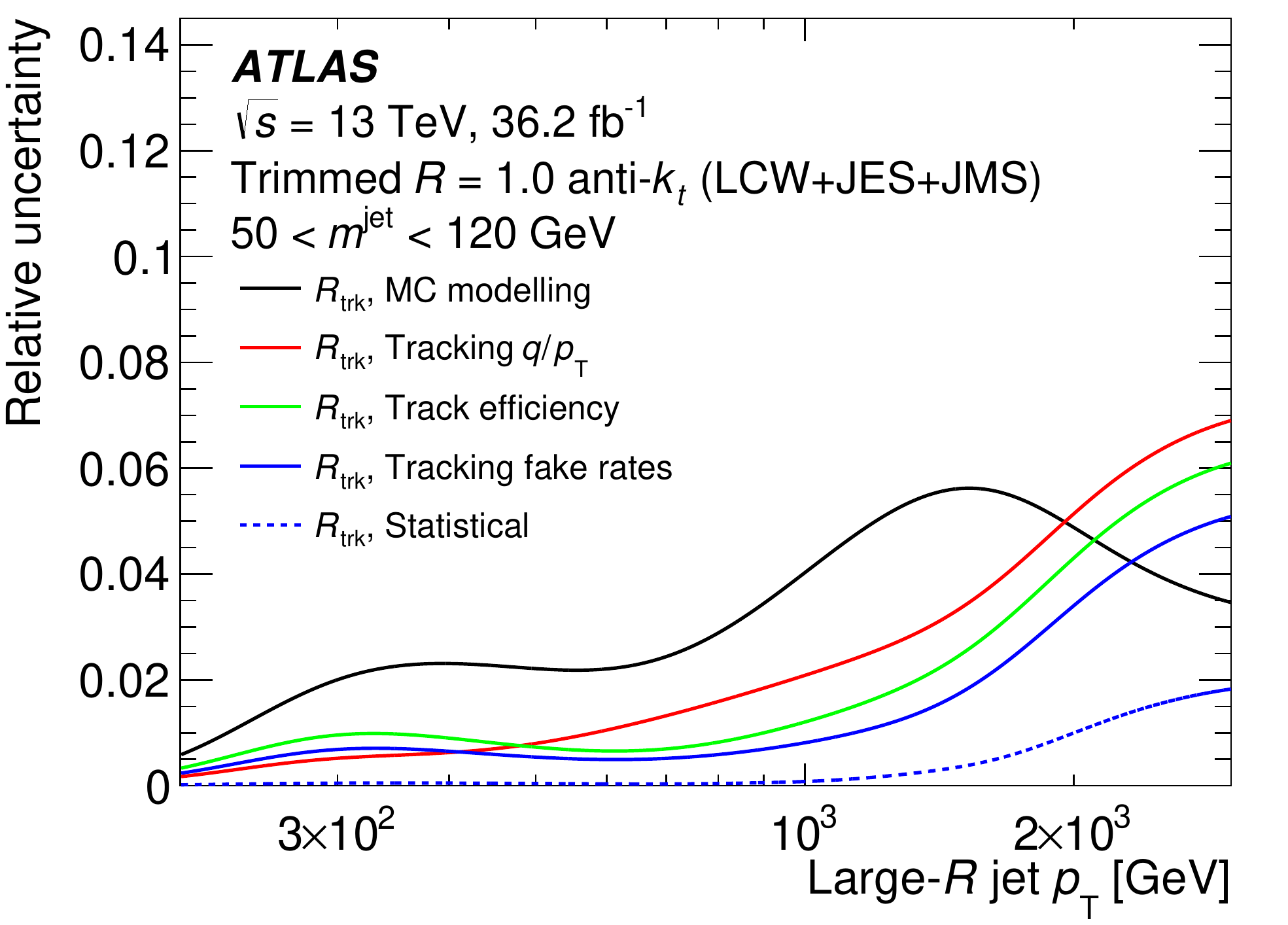} }
\subfigure[]{
\includegraphics[width=0.48\textwidth]{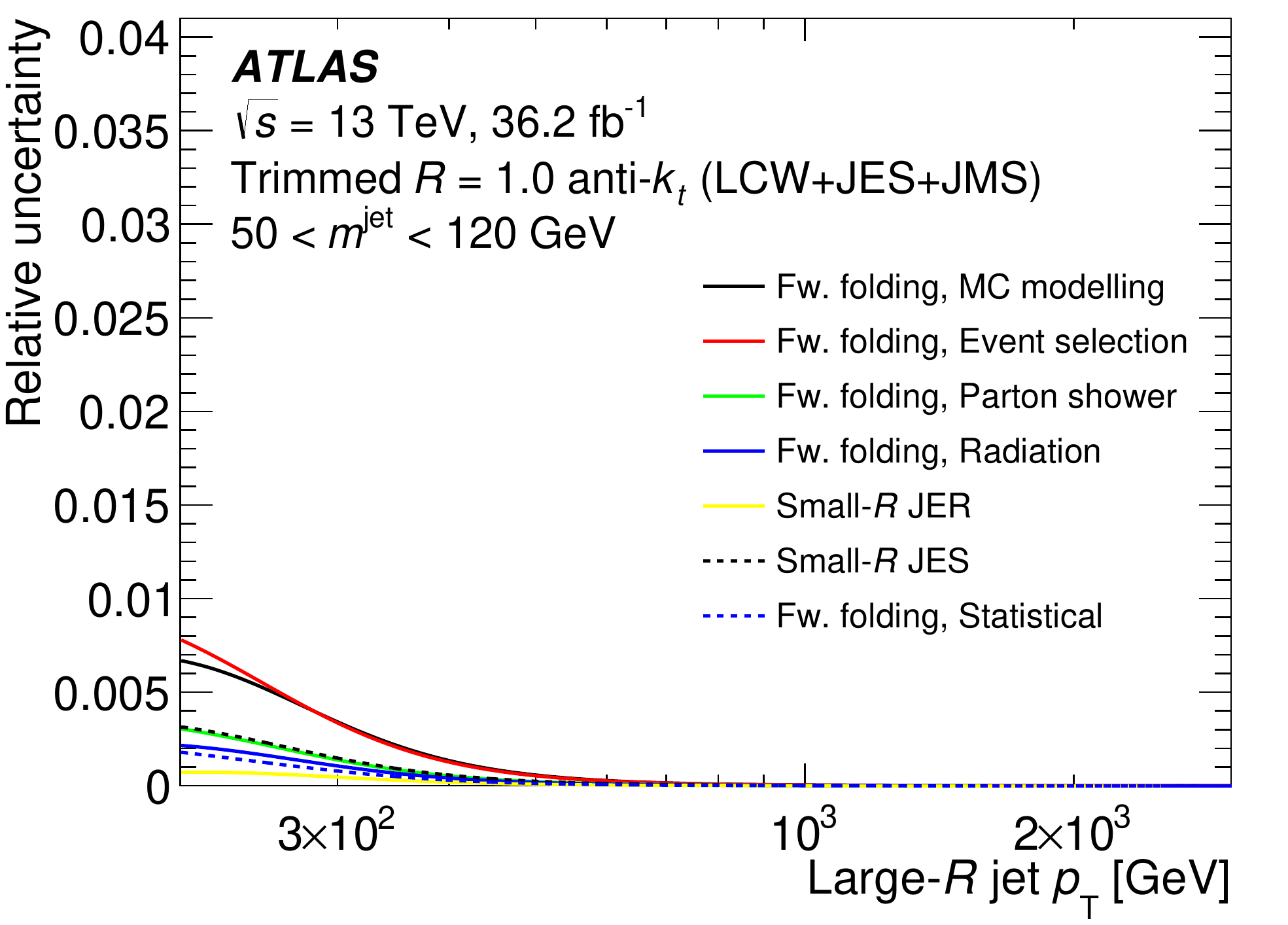} }
\caption{Breakdown of the combined JMS uncertainty shown in Figure~\ref{fig:Massresponse} as a function of jet transverse momentum $\pt$ for the jet mass bin 50--120~\GeV. Contributions are shown for each of the nuisance parameters of the (a) \Rtrk and (b) forward-folding methods. The vertical axis reflects the uncertainty introduced by a given nuisance parameter in combination, incorporating the weight of the method from which it originates. This weight is dominated at high \pt by the \Rtrk method. The lines shown are smoothed using a sliding Gaussian kernel.}
\label{fig:JMS_uncertainty_sources_one}
\end{figure}
 
The contributions of several sources to the uncertainty in the combined jet mass scale are presented in Figures~\ref{fig:JMS_uncertainty_sources_one} and~\ref{fig:JMS_uncertainty_sources_two}. In both the \Rtrk and forward-folding techniques, the leading systematic uncertainties are associated with uncertainties in the event generators across most of the \pt range and for the two mass intervals considered.
 
\begin{figure}[tbh]
\centering
\subfigure[]{
\includegraphics[width=0.48\textwidth]{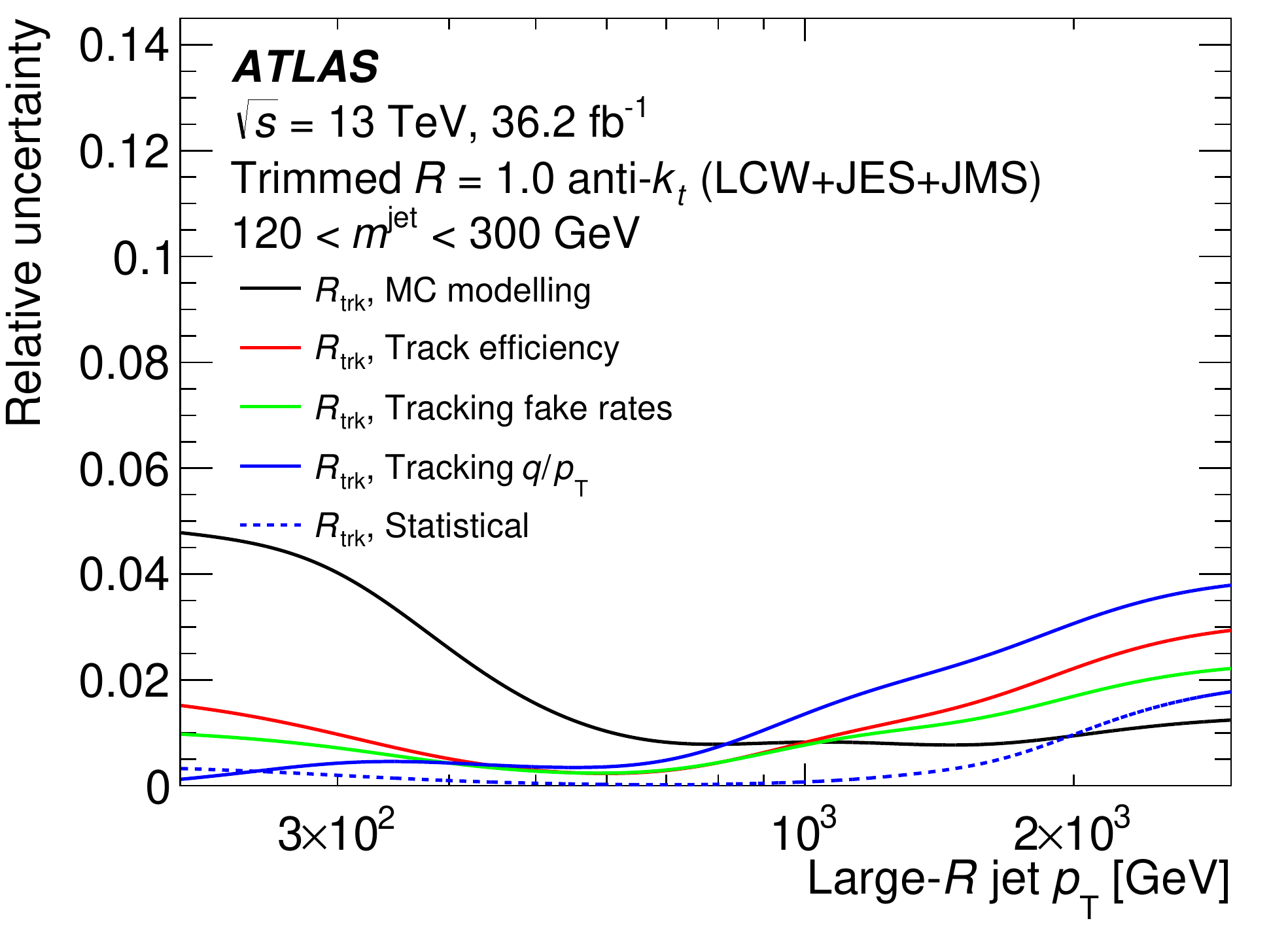} }
\subfigure[]{
\includegraphics[width=0.48\textwidth]{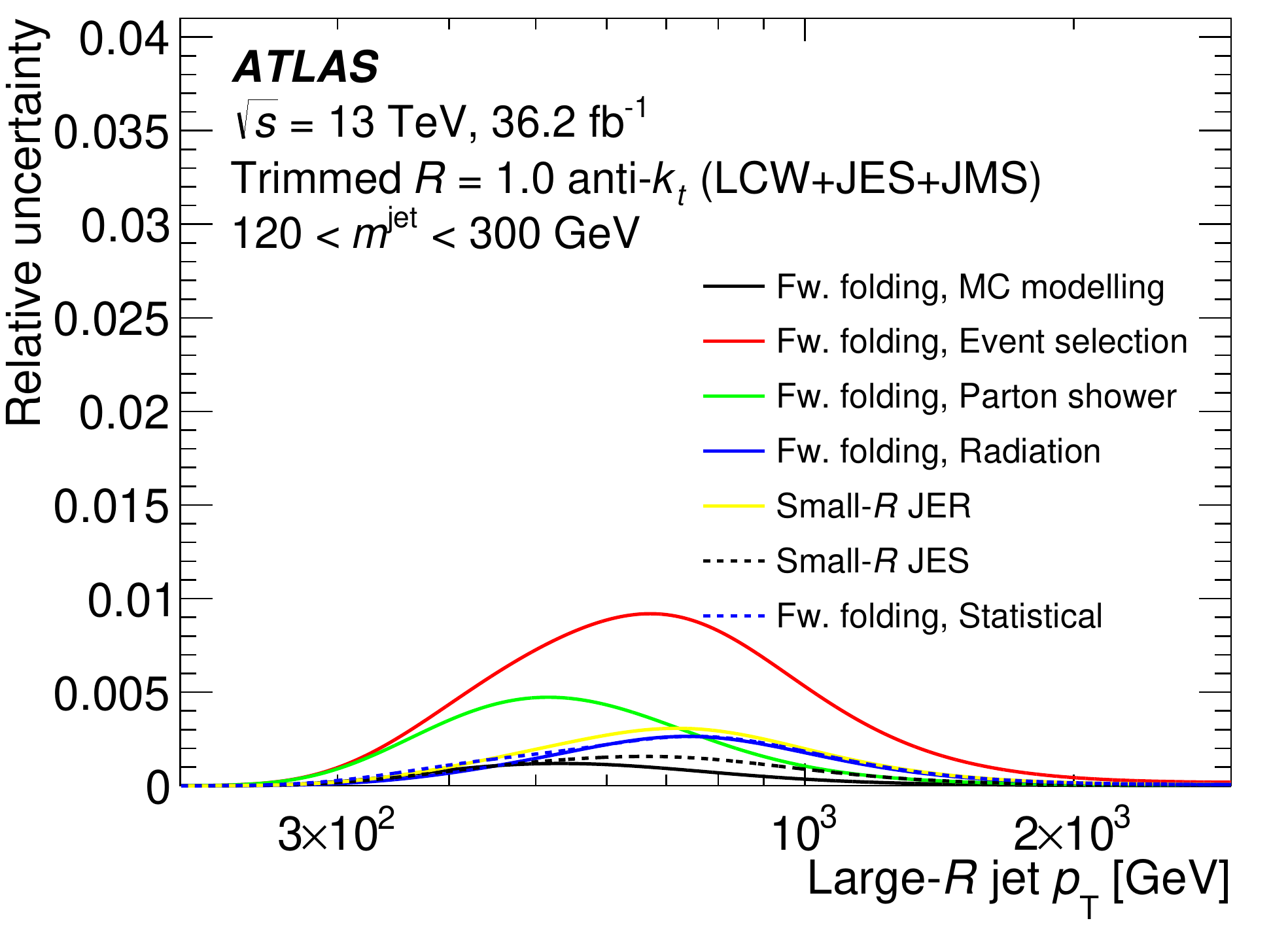} }
\caption{Breakdown of the combined JMS uncertainty shown in Figure~\ref{fig:Massresponse} as a function of jet transverse momentum $\pt$ for the jet mass bin 120--300~\GeV. Contributions are shown for each of the nuisance parameters of the (a) \Rtrk and (b) forward-folding methods. The vertical axis reflects the uncertainty introduced by a given nuisance parameter in combination, incorporating the weight of the method from which it originates. This weight is dominated at high \pt by the \Rtrk method. The lines shown are smoothed using a sliding Gaussian kernel.}
\label{fig:JMS_uncertainty_sources_two}
\end{figure}
\clearpage
 
\FloatBarrier
 
\section{Conclusion} \label{sec:conclusion}
Several \insitu calibration methods are used to measure the response of the ATLAS detector to trimmed \largeR jets using 36.2~\ifb~of $\sqrt{s}=$ 13~\TeV~proton--proton collision data provided by the LHC and collected by the ATLAS experiment during 2015 and 2016. These methods exploit the transverse momentum balance in events where a jet recoils against a reference system with a precisely known energy scale, the independence of measurements performed with different sub-detectors, or the position and width of known mass peaks. With this ensemble of techniques, dedicated jet energy scale and jet mass scale calibrations are derived for \largeR jets. The results of several techniques applied to a variety of final states are consistent within the uncertainties, indicating that after calibration, the simulations model the flavour dependence of the jet \pt{} and mass response to within a few percent.

The results of all methods are combined taking into account  correlations between uncertainties and possible discrepancies  between the results of different \insitu methods. The combined measurement of the ratio of the energy scales in data and simulations are used to derive an \insitu correction to the response, which determines the \largeR{} jet energy and mass scales. The residual uncertainty in the ratio of the energy scales in data and simulations is 1--2\% for transverse momenta from 150~\gev{}~to~2~\tev. The precision of the jet mass scale varies from 2\% to 10\% over the same \pt range. The results of the simulations for jet \pt{} and mass resolution are also validated \insitu and found to agree with the measured resolution within 10--15\%. The \insitu JES calibration, derived from light quark and gluon jets, is found to fully correct the energy and mass scales of high \pt $W$ bosons and top quarks to within the precision of the present measurement (1--3\%).
 
\LargeR jets are a vital ingredient of the ATLAS physics programme. This new \insitu calibration leads to significantly reduced uncertainties in the reconstructed \largeR jet \pt{} and mass, thus increasing the sensitivity of searches and the precision of Standard
Model measurements using \largeR jets.
 
\section*{Acknowledgements}
 
 
We thank CERN for the very successful operation of the LHC, as well as the
support staff from our institutions without whom ATLAS could not be
operated efficiently.
 
We acknowledge the support of ANPCyT, Argentina; YerPhI, Armenia; ARC, Australia; BMWFW and FWF, Austria; ANAS, Azerbaijan; SSTC, Belarus; CNPq and FAPESP, Brazil; NSERC, NRC and CFI, Canada; CERN; CONICYT, Chile; CAS, MOST and NSFC, China; COLCIENCIAS, Colombia; MSMT CR, MPO CR and VSC CR, Czech Republic; DNRF and DNSRC, Denmark; IN2P3-CNRS, CEA-DRF/IRFU, France; SRNSFG, Georgia; BMBF, HGF, and MPG, Germany; GSRT, Greece; RGC, Hong Kong SAR, China; ISF and Benoziyo Center, Israel; INFN, Italy; MEXT and JSPS, Japan; CNRST, Morocco; NWO, Netherlands; RCN, Norway; MNiSW and NCN, Poland; FCT, Portugal; MNE/IFA, Romania; MES of Russia and NRC KI, Russian Federation; JINR; MESTD, Serbia; MSSR, Slovakia; ARRS and MIZ\v{S}, Slovenia; DST/NRF, South Africa; MINECO, Spain; SRC and Wallenberg Foundation, Sweden; SERI, SNSF and Cantons of Bern and Geneva, Switzerland; MOST, Taiwan; TAEK, Turkey; STFC, United Kingdom; DOE and NSF, United States of America. In addition, individual groups and members have received support from BCKDF, CANARIE, CRC and Compute Canada, Canada; COST, ERC, ERDF, Horizon 2020, and Marie Sk{\l}odowska-Curie Actions, European Union; Investissements d' Avenir Labex and Idex, ANR, France; DFG and AvH Foundation, Germany; Herakleitos, Thales and Aristeia programmes co-financed by EU-ESF and the Greek NSRF, Greece; BSF-NSF and GIF, Israel; CERCA Programme Generalitat de Catalunya, Spain; The Royal Society and Leverhulme Trust, United Kingdom.
 
The crucial computing support from all WLCG partners is acknowledged gratefully, in particular from CERN, the ATLAS Tier-1 facilities at TRIUMF (Canada), NDGF (Denmark, Norway, Sweden), CC-IN2P3 (France), KIT/GridKA (Germany), INFN-CNAF (Italy), NL-T1 (Netherlands), PIC (Spain), ASGC (Taiwan), RAL (UK) and BNL (USA), the Tier-2 facilities worldwide and large non-WLCG resource providers. Major contributors of computing resources are listed in Ref.~\cite{ATL-GEN-PUB-2016-002}.
 
 
\printbibliography

\clearpage

\clearpage
\input{atlas_authlist.tex} 
\end{document}

%% file: atlas_authlist.tex
 
\begin{flushleft}
{\Large The ATLAS Collaboration}

\bigskip

M.~Aaboud$^\textrm{\scriptsize 34d}$,    
G.~Aad$^\textrm{\scriptsize 99}$,    
B.~Abbott$^\textrm{\scriptsize 124}$,    
O.~Abdinov$^\textrm{\scriptsize 13,*}$,    
B.~Abeloos$^\textrm{\scriptsize 128}$,    
D.K.~Abhayasinghe$^\textrm{\scriptsize 91}$,    
S.H.~Abidi$^\textrm{\scriptsize 164}$,    
O.S.~AbouZeid$^\textrm{\scriptsize 39}$,    
N.L.~Abraham$^\textrm{\scriptsize 153}$,    
H.~Abramowicz$^\textrm{\scriptsize 158}$
% ...
\bigskip
\\

% ...
$^{13}$Institute of Physics, Azerbaijan Academy of Sciences, Baku; Azerbaijan.\\
% ...
$^{34}$$^{(a)}$Facult\'e des Sciences Ain Chock, R\'eseau Universitaire de Physique des Hautes Energies - Universit\'e Hassan II, Casablanca;$^{(b)}$Centre National de l'Energie des Sciences Techniques Nucleaires (CNESTEN), Rabat;$^{(c)}$Facult\'e des Sciences Semlalia, Universit\'e Cadi Ayyad, LPHEA-Marrakech;$^{(d)}$Facult\'e des Sciences, Universit\'e Mohamed Premier and LPTPM, Oujda;$^{(e)}$Facult\'e des sciences, Universit\'e Mohammed V, Rabat; Morocco.\\
% ...
$^{39}$Niels Bohr Institute, University of Copenhagen, Copenhagen; Denmark.\\
% ...
$^{91}$Department of Physics, Royal Holloway University of London, Egham; United Kingdom.\\
% ...
$^{99}$CPPM, Aix-Marseille Universit\'e, CNRS/IN2P3, Marseille; France.\\
% ...
$^{124}$Homer L. Dodge Department of Physics and Astronomy, University of Oklahoma, Norman OK; United States of America.\\
% ...
$^{128}$LAL, Universit\'e Paris-Sud, CNRS/IN2P3, Universit\'e Paris-Saclay, Orsay; France.\\
% ...
$^{153}$Department of Physics and Astronomy, University of Sussex, Brighton; United Kingdom.\\
% ...
$^{158}$Raymond and Beverly Sackler School of Physics and Astronomy, Tel Aviv University, Tel Aviv; Israel.\\
% ...
$^{164}$Department of Physics, University of Toronto, Toronto ON; Canada.\\
% ...
$^{a}$ Also at Borough of Manhattan Community College, City University of New York, NY; United States of America.\\
$^{b}$ Also at Centre for High Performance Computing, CSIR Campus, Rosebank, Cape Town; South Africa.\\
$^{c}$ Also at CERN, Geneva; Switzerland.\\
$^{d}$ Also at CPPM, Aix-Marseille Universit\'e, CNRS/IN2P3, Marseille; France.\\
$^{e}$ Also at D\'epartement de Physique Nucl\'eaire et Corpusculaire, Universit\'e de Gen\`eve, Gen\`eve; Switzerland.\\
% ...
$^{*}$ Deceased

\end{flushleft}
